\documentstyle[12pt]{article}
\renewcommand{\thefootnote}{\fnsymbol{footnote}}

\begin{document}

\baselineskip 6mm
\begin{flushright}
{\tt TMI-07-1 \\
     July 2007}
\end{flushright}
\thispagestyle{empty}
\vspace*{5mm}
\begin{center}
   {\Large {\bf Composite Model and CP Violation}}
\vspace*{20mm} \\
  {\sc Takeo~~Matsushima}
\footnote[1]  
{   e-mail : mtakeo@toyota-ct.ac.jp  }
\vspace{5mm} \\
  \sl{265 Takaodouji Fuso-Cho Niwa-Gun}\\
  \it{Aichi-Prefecture, Japan}
\end{center}

\vspace{15mm}  

\begin{center}
{\bf Abstract} \\
\vspace{3mm}
\begin{minipage}[t]{120mm}
\baselineskip 5mm
{\small
In an axiomatic way we propose a fermion-boson-type 
Composite Model for quarks and leptons based on the 
gauge theory equipped with {\bf Cartan connectjons}.
 Elementary fields are only one kind of spin-1/2 
and spin-0 {\bf preon}. Both are in the global supersymmetric
pair with the common electric charge of `` ${\bf e/6}$ '' and 
belong to the fundamental representations of 
( {\bf 3 , 2 , 2} ) under the spontaneously {\bf unbroken}  
$SU(3)_C\otimes{SU(2)}_L^{h}\otimes{SU(2)}_R^{h}$
gauge symmetry ( $h$ means hyper-color gauge ). 
Preons are composed into {\bf subquarks} which are 
{\bf intermediate clusters} towards quarks 
and leptons. Weak interactions are residual ones of 
hyper-color gauge interactions. {\bf W}-and {\bf Z}-boson 
are also composite objects of subquarks, 
which introduces the idea of existence of their scalar 
partners ($\;${\bf S}$\;$) by {\bf hyper-fine-splitting}  
whose masses would be around 
$110\sim120$ GeV. The mechanism of making higher 
generations is obtained by adding neutral scalar subquark 
($\;$${\bf y}$$\;$) composed of a preon-antipreon pair. 
Creation or annihilation of ${\bf y}$ inside quarks induces 
the coupling constants of flavor-mixing weak interactions 
which are {\bf all complex} numbers 
(contrary to CKM-matrix elements) and then they 
all become sources of direct and mixing-induced CP violations. 
Exchange of ${\bf y}$ between quark and anti-quark inside 
neutral pseudo-scalar meson ($\;$$P^0$$\;$) gives indirect 
CP violation and mass-difference of $P^0$ and $\overline{P^0}$. 
Current experimental results of CP violation ( Belle, BaBar, 
CLEO, KTeV and NA48 ) are inspected by this Composite Model. 
This model suggests the candidates for `` {\bf Dark Energy} ''
and `` {\bf Dark Matter} ''.  
}
\end{minipage}

%

\end{center}



\newpage
\baselineskip 18pt
\section{Introduction}
\hspace*{\parindent}
  The discovery of the top-quark[1] has finally confirmed the 
 existence of three quark-lepton symmetric generations.
 So far the standard $SU(2)_{L}\otimes{U(1)}$ model
 (denoted by the SM)
 has successfully explained various experimental evidences.
 Nevertheless, as is well known, the SM is not regarded as
 the final theory because it has many arbitrary parameters, e.g.,
 quark and lepton masses, quark-mixing parameters, etc. .
 \par
 Therefore it is meaningful to investigate the 
 origins of these parameters and the relationship among them.
  In order to overcome such problems some attempts have done, e.g.,
 Grand Unification Theory (GUT), Supersymmetry,Super String Theory,
 Composite model, etc. . In the theory except Composite model
 quarks and leptons are elementary fields in general. 
  On the contrary in the composite scenario they are
 literally the composite objects constructed from the elementary
 fields (so called `` preon ''). The lists of various Composite 
 model are in Ref.[2].
 \par
 If quarks and leptons are 
 elementary, in order to solve the above problems it is necessary 
 to introduce some external relationship or symmetries among them. 
  On the other hand the composite models have ability to explain
 the origin of these parameters in terms of the substructure 
 dynamics of quarks and leptons. Further, the composite scenario 
 naturally leads us to the thought that the intermediate vector bosons
 of weak interactions (${\bf W,Z}$) are not elementary gauge fields
 (which is so in the SM)
 but composite objects constructed from preons (same as ${\bf \rho}$
 -meson from quarks). Many studies based on such conception have 
 done after Bjorken's[3] and Hung and Sakurai's[4] suggestions of the 
 alternative way to unified weak-electromagnetic gauge theory[5$\sim$11].
 In this scheme the weak interactions are regarded as the effective
 residual interactions among preons. The fundamental fields for
 intermediate forces are massless gauge fields belonging to
 some gauge groups and they confine preons into singlet states
 to build quarks and leptons and ${\bf W,Z}$.
\par
  The conception of our model is that the fundamental interacting
 forces are all originated from massless gauge fields belonging
 to the adjoint representations of some gauge groups which have 
 nothing to do with the spontaneous breakdown and that the 
 elementary matter fields are only one kind of spin-1/2 preon
 and spin-0 preon carrying common `` $e/6$ '' 
 electric charge ($e>0$). Quarks, leptons and ${\bf W,Z}$ 
 are all composites of them and usual weak interactions are 
 regarded as effective residual interactions. 
 \par
 Based on such scenario various CP-violating phenomena are 
 investigated. The most outstanding point is that CP-violations
 originate from interactions among subquarks inside quarks. 
 \par The outline of this article is as follows. In Section 2
 we introduce brief presentation about the gauge theory 
 inspiring composite quarks and leptons. In Section 3
 we discuss the composite model naturally inherited from
 above mentioned gauge theory. In Section 4 we give the 
 definition of the flavor-mixing matrix elements, which 
 come from subquark dynamics. These correspond 
 to CKM-matrix elements in the SM. In Section 5 we discuss 
 the mass difference (denoted by $\Delta{M_{P}}$) by  
 $P^0$-$\overline{P^0}$ mixing ($P^0$ is pseudo scalar meson).
 This originates from ${\bf y}$-subquark-exchange between 
 quark and anti-quark inside $P^{0}$.
 In Section 6 indirect CP-violations are investigated.
 This is also caused by ${\bf y}$-subquark-exchange between 
 quark and anti-quark inside $P^{0}$. 
 In Section 7 we study direct and 
 mixing-induced CP-violations which originate from 
 flavor-mixing interactions caused by subquark dynamics.
 Lastly we give conclusions in Section 8.

 \section{  Gauge theory inspiring quark-lepton composite scenario}
 \hspace*{\parindent}
 
  In our model the existence of fundamental matter fields (preon)
 are inspired by the gauge theory with Cartan connections[14]. 
 Let us briefly summarize the basic features of that.\par
 Generally 
 gauge fields, including gravity, are considered as geometrical  
 objects, that is, connection coefficients of principal fiber 
 bundles. It is said that there exist some different points between 
 Yang-Mills gauge theories and gravitationary theory, though both 
 theories commonly possess the fiber bundle structures. Namely 
 the latter is equipped with  the fiber bundle essentially related 
 to 4-dimensional space-time  freedoms but the former with the
 fiber bundle belonging to  the internal space which has nothing to 
 do with the space-time  coordinates.\par 
 In case of gravity it is usually considered that 
 there exist ten gauge fields, that is, six spin connection fields 
 in $SO(1,3)$ gauge group and four vierbein fields in $GL(4,R)$ 
 gauge group from which the metric tensor ${\bf g}^{{\mu}{\nu}}$ 
 is constructed in a bilinear function of them. Both altogether 
 belong to  Poincar\'e group $ISO(1,3)=SO(1,3)\otimes{R}^4$ 
 which is semi-direct product. In this scheme spin connection 
 fields and vierbein fields are independent but only if there is 
 no torsion, both come to have some relationship. Seeing this, 
 $ISO(1,3)$ gauge theory seems to  have the logical weak point not to  
 answer how two kinds of gravity fields are related to each other 
 intrinsically.
 \par   
 In the theory of Differential Geometry, S.Kobayashi has investigated 
 the theory of `` Cartan connection ''[15]. This theory, in fact, 
 has ability to reinforce the above weak point. The brief 
 recapitulation is as follows. Let $E(B_n,F,G,P)$ be a fiber bundle
 (which we call Cartan-type bundle) associated with a principal 
 fiber bundle $P(B_n,G)$ where $B_n$ is a base manifold with  
 dimension `` $n$ '', $G$ is a structure group, $F$ is a fiber space 
 which is homogeneous and diffeomorphic with $G/G'$ where $G'$ 
 is a subgroup of $G$. Let $P'=P'(B_n,G')$ be a principal 
 fiber bundle, then $P'$ is a subbundle of $P$.
 Here let it be possible to decompose the Lie algebra ${\bf g}$ of 
 $G$ into the subalgebra ${\bf g}'$ of $G'$ and a vector space 
 ${\bf f}$ such as :   
 $$ {\bf g}={\bf g}'+{\bf f},\hspace{1cm}{\bf g}'\cap{\bf f}=0,      
 \eqno{(2.1)}$$                                            \label{2.1} 
 $$ [{\bf g'},{\bf g'}]\subset {\bf g'},\eqno{(2.2)}$$     \label{2.2}   
 $$ [{\bf g'},{\bf f}]\subset {\bf f},\eqno{(2.3)}$$       \label{2.3} 
 $$ [{\bf f},{\bf f}]\subset {\bf g'},\eqno{(2.4)}$$       \label{2.4} 
 where $dim{\bf f}=dimF=dimG-dimG'=dimB_n=n$. 
 The homogeneous space $F=G/G'$ is said to be `` weakly reductive '' 
 if there exists a vector space ${\bf f}$ satisfying (2.1) and (2.3).
 Further $F$ satisfying (2.4) is called `` symmetric space ''. 
 Let ${\bf \omega}$ denote the connection form of $P$ and 
 $\overline{\bf \omega}$ be the restriction of ${\bf \omega}$ to
 $P'$. Then $\overline{\bf \omega}$ is a ${\bf g}$-valued linear
 differential 1-form and we have : 
 $$  {\bf \omega}=g^{-1}\overline{\bf \omega}g+g^{-1}dg,
                  \eqno{(2.5)}$$                            \label{2.5} 
 where $g\in{G}$, $dg\in{T_g(G)}$. ${\bf \omega}$ is called the  
 form of `` Cartan connection '' in $P$. 
 \par
 Let the homogeneous space $F=G/G'$ be weakly reductive. The 
 tangent space $T_O(F)$ at $o\in{F}$ is isomorphic with ${\bf f}$
 and then $T_O(F)$ can be identified with ${\bf f}$ and also 
 there exists a linear ${\bf f}$-valued differential 
 1-form (denoted by ${\bf \theta}$) which we call the 
 `` form of soldering ''. Let ${\bf \omega}'$ 
 denote a ${\bf g}'$-valued 1-form in $P'$, we have : 
 $$  \overline{\bf \omega}={\bf \omega}'+{\bf \theta}.
                           \eqno{(2.6)}$$                    \label{2.6} 
 The dimension of vector space ${\bf f}$ and the dimension of  
 base manifold $B_n$ is the same `` $n$ '', and then ${\bf f}$  
 can be identified with the tangent space of $B_n$ at the same 
 point in $B_n$ and ${\bf \theta}$s work as $n$-bein fields. In this 
 case  ${\bf \omega}'$ and ${\bf \theta}$ unifyingly belong to 
 group $G$. Here let us call such a mechanism `` Soldering Mechanism ''.
 \par
 Drechsler has found out the useful aspects of this theory and 
 investigated a gravitational gauge theory based on 
 the concept of the Cartan-type bundle equipped with the 
 Soldering Mechanism[16]. He considered $F=SO(1,4)/SO(1,3)$ 
 model. Homogeneous space $F$ with $dim=4$ solders 4-dimensional 
 real space-time. The Lie algebra of $SO(1,4)$ corresponds to 
 ${\bf g}$ in (2.1), that of $SO(1,3)$ corresponds to ${\bf g}'$
 and ${\bf f}$ is 4-dimensional vector space. The 6-dimensional 
 spin connection fields are ${\bf g}'$-valued objects and vierbein 
 fields are ${\bf f}$-valued, both of which are unified into 
 the members of $SO(1,4)$ gauge group. We can make the metric 
 tensor ${\bf g}^{{\mu}{\nu}}$ as a bilinear function of 
 ${\bf f}$-valued vierbein fields.
 \par 
 Inheriting Drechsler's study, the author has investigated 
 the quantum theory of gravity which has already 
 appeared in Ref.[14]. The most important ingredient 
 of this investigation is that $F$ is a `` symmetric space '' 
 and then ${\bf f}$s are satisfied with (2.4). 
 Using this symmetric nature we can pursue making a 
 quantum gauge theory, that is, constructing ${\bf g}'$-valued 
 Faddeev-Popov ghost (denoted by ${\bf C}$), 
 anti-ghost (denoted by$\overline{\bf C}$)
 , gauge fixing (denoted by ${\bf B}$), 
 anti-gauge fixing (denoted by $\overline{\bf B}$),
 gaugeon (denoted by ${\bf G_{1}}$)
 and its pair field (denoted by ${\bf G_{2}}$) 
 as composite fusion fields of 
 ${\bf f}$-valued gauge fields `` ${\bf \theta}$ '' by use 
 of (2.4) and also naturally inducing {\bf BRS-invariance}  
 among them. In this way these six kinnds of fusion fields
 are made of ${\bf f}$-valued viebein fields. Here let us call
 these six fields together `` {\bf six-fields-set} '' : 
 $\{$ ${\bf C},\overline{\bf C},{\bf B},
 \overline{\bf B},{\bf G_{1}},{\bf G_{2}}$ $\}$
 \par
 Comparing with such a scheme of gravity, let us consider the
 Yang-Mills gauge theories. Usually when we make the Lagrangian 
 density ${\cal L}=tr({\cal F}\wedge{\cal F}^{\ast})$ 
 (${\cal F}$ is a field strength of the Yang-Mills fields), we must 
 borrow a metric tensor ${\bf g}^{{\mu}{\nu}}$ from gravity to 
 get ${\cal F}^{\ast}$ and also for Yang-Mills gauge fields to 
 propagate in the 4-dimensional real space-time. This fact seems to 
 mean that `` there is a hierarchy between gravity and other three 
 gauge fields (electromagnetic, strong, and weak) and gravity has the 
 outstanding position compared with others ''. But is it really the case ? 
 As an alternative thought let us think that all kinds of gauge fields are 
 `` equal ''. Then it would be natural for the question to arise :
 `` What kind of equality is that ? ''. In other words, it is the question 
 that `` What is the minimum structure of the gauge mechanism which 
 four kinds of forces are commonly equipped with ? ''. 
 For answering this question, let us begin from making an assumption :
 \newline $\;$ \newline
   \begin{em}\hspace{0.7cm}
  `` Gauge fields are Cartan connections equipped with Soldering Mechanism
   \end{em}. ''
 \newline  
 In this meaning four gauge fields are all equal. 
 In this scheme three gauge fields except gravity are also able to have 
 their own metric tensors `` ${\bf g}^{{\mu}{\nu}}_{a}$ `` ( where $a$
 means $electro magnetic$, $strong$ and $weak$.)  
 and to propagate in the real space-time without 
 the help of gravity. Such a model has already investigated in 
 Ref.[14]. 
 \par
 Let us discuss them briefly. It is found that there 
 are four types of sets of classical groups with small dimensions 
 which admit (2.1$\sim$4), that is, $F=SO(1,4)/SO(1,3)$, 
 $SU(3)/U(2)$, $SL(2,C)/GL(1,C)$ and $SO(5)/SO(4)$ with 
 $dimF=4$[17]. Note that the quality of 
 `` $dimension:\hspace{0.5mm}4$ ''
  is very important because it guarantees $F$ to solder 
 to 4-dimensional real space-time and all gauge fields 
 to work in it. The model of $F=SO(1,4)/SO(1,3)$ for gravity 
 is already mentioned. 
 Concerning other gauge fields, it seems to be appropriate  
 to assign $F=SU(3)/U(2)$ to QCD gauge fields, 
 $F=SL(2,C)/GL(1,C)$ to QED gauge fields and 
 $F=SO(5)/SO(4)$ to weak interacting gauge fields
 ( as is well known, $SO(4)$ is 
 locally isomorphic with $SU(2)\otimes{SU(2)}$, which 
 we set as $SU(2)_{L}\otimes{SU(2)_{R}}$).
 It is noted that four kinds of ${\bf g}'$-valued 
 gauge fields have each {\bf six-fields-set} of their own, 
 with the help of which and also with ${\bf g}^{{\mu}
 {\nu}}_{i}$ ($i=gravitational, 
 electro magnetic, strong, {\rm and}\, weak$) they can propagate 
 all over the universe. And also it is memorable that 
 our model expects that the {\bf six-fields-set} is not merely
 the mathematical tool for {\bf BRS}-invariance but {\bf really 
 exist at every point of the universe} 
 (speculatively in the cube of ${\rm (Plank\hspace{1mm}length)}^3$).
 Then ${\bf massless\; scalar\; fields}$ such as 
 $\{$ ${\bf C},\overline{\bf C},{\bf B},
 \overline{\bf B},{\bf G_{1}},{\bf G_{2}}$ $\}$
 cause the `` ${\bf repulsive\; forces}$ ''
 at every points of the universe. Especially fermionic scalars
 of ${\bf C}$ and $\overline{\bf C}$ may be thought to have 
 generated huge short distant repulsive forces at 
 the very early Universe by {\bf Pauli Exclusion Principle}.  
 Therefore they are possibly candidates for 
 `` ${\bf Dark\; Energy}$ ''. \par Concerning matter fields,         
 they  couple to ${\bf g}'$-valued gauge fields. As for   
 QCD, matter fields couple to the gauge fields of $U(2)$
 subgroup but $SU(3)$ contains, as is well known, three types 
 of $SU(2)$ subgroups and then after all they couple 
 to all members of $SU(3)$ gauge fields. In case of QED,  
 $GL(1,C)$ is locally isomorphic with $C^1\cong{U(1)}\otimes{R}$. 
 Then usual Abelian gauge fields are assigned to $U(1)$ 
 subgroup of $GL(1,C)$. Georgi and Glashow suggested that  
 the reason why the electric charge is quantized comes from 
 the fact that $U(1)$ electromagnetic gauge group is 
 a unfactorized subgroup of $SU(5)$[18]. Our model is 
 in the same situation because $GL(1,C)$ is an unfactorized 
 subgroup of $SL(2,C)$. For usual electromagnetic $U(1)$ 
 gauge group, the electric charge unit ``$e$''$(e>0)$ is for   
 $one\hspace{2mm} generator$ of $U(1)$ but in case of $SL(2,C)$ 
 which has $six\hspace{2mm} generators$, the minimal 
 unit of electric charge shared per one generator must be `` $e/6$ ''. 
 This suggests that quarks and leptons might have the substructure 
 simply because $e,\hspace{1mm}2e/3,\hspace{1mm}e/3>e/6$.
  Finally as for weak interactions we adopt 
 $F=SO(5)/SO(4)$. As is stated above, $SO(4)$ is 
 locally isomorphic with $SU(2)\otimes{SU(2)}$. Therefore 
 it is reasonable to think it the left-right symmetric 
 gauge group : $SU(2)_L\otimes{SU(2)}_R$. As two $SU(2)$s are 
 direct product, they are able to have coupling constants 
 (${\bf g}_L,{\bf g}_R$) independently. 
 This is convenient to explain the fact of the disappearance 
 of right-handed weak interactions in the low-energy region. 
 Possibility of composite structure of quarks and leptons 
 suggested by above mentioned $SL(2,C)$-QED would introduce 
 the thought that the usual left-handed weak interactions 
 are intermediated by massive composite vector bosons (usually
 denoted by {\bf W,Z}) same as ${\bf \rho}$-meson 
 in QCD and that they are residual interactions due to 
 substructure dynamics of quarks 
 and leptons. The elementary massless gauge fields , as
`` \begin{em}connection fields ''\end{em}, relate 
 intrinsically to the structure of the four dimensional 
 real space-time but on the other hand 
 the composite vector bosons have nothing to do with it. 
 Considering these discussions, we set the assumption : 
 \newline $\;$ \newline
 \begin{em}
     `` All kinds of gauge fields are elementary massless 
 fields, belonging to spontaneously unbroken 
 $SU(3)_C\otimes{SU(2)}^{h}_L\otimes{SU(2)}^{h}_R
 \otimes{U(1)}_{e.m}$ 
 gauge group and quarks and leptons and {\bf W, Z} are all 
 composite objects of the elementary matter fields\end{em}. ''

\section{Composite model}
 \hspace*{\parindent}
 As discussing in Section 2, the assumption : `` The minimal unit of 
 electric charge is $e/6$ `` leads us to think of compositeness 
 of quarks and leptons. However, other several phenomenological 
 facts tempt us to consider a composite model, e.g., repetition 
 of generations, quark-lepton parallelism of weak isospin doublet 
 structure, quark-flavor-mixings, etc..
 Especially Bjorken[3]'s and Hung and Sakurai[4]'s suggestion of an 
 alternative to usual unified electro-weak gauge theories have 
 invoked many studies of composite models including composite 
 weak bosons[5$\sim$11].  
 Our model stands on  the line of those studies. 
 There are two ways to make composite models, that is, 
 `` Preons are all fermions. '' or 
 `` Preons are both fermions and bosons, '' 
 which is denoted by FB-model. 
 The merit of the former is that it can avoid the problem of 
 a quadratically divergent self-mass of elementary scalar fields. 
 However, even in the latter case it is found that such a disease 
 is overcome if both fermions and bosons are the 
 supersymmetric pairs, both of which carry the same quantum 
 numbers except the nature of Lorentz transformation (
 spin-1/2 or spin-0)[19]. Pati and Salam have suggested    
 that the construction of a neutral fermionic composite object 
 (neutrino in practice) needs both kinds of preons : fermionic 
 and  bosonic, if they carry the same charge for the 
 Abelian gauge or belong to the same fundamental representation 
 for the non-Abelian gauge[20].  
 This is a very attractive idea for constructing the minimal 
 model. Further, according to the representation theory of  
 Poincar\'e group both integer and half-integer 
 spin angular momentum occur equally for massless particles[21], 
 and then equal existence of fermionic and bosonic elementary particle
 may be naturally acceptable. But on the contrary, if nature chooses 
 `` fermionic monism '', there must exist the additional special 
 reason to select it. Therefore in this point also, the thought 
 of the FB-model is minimal without any special conditions.
 Based on such considerations we propose a FB-model :
 \newline $\;$ \newline 
 \begin{em}`` Primodial elementary particles are 
 (spin 1/2)-fermion (denoted by $\Lambda$) and 
  (spin 0)-boson (denoted by $\Theta$) . ''
 \end{em}
 \newline $\;$ \newline
 (Preliminary version of this model has appeared in Ref.[14]). 
 Both have the same electric charge of `` $e/6$ ''  
 (Maki has first proposed the FB-model with the minimal 
 electric charge $e/6$.[22])
\renewcommand{\thefootnote}{\arabic{footnote}}
\footnote{The notations of $\Lambda$ and $\Theta$ are 
 inherited from those in Ref.[22]. After this we 
 call $\Lambda$ and $\Theta$ \par `` Primon '' named by Maki 
 which means `` primordial particle ''[22].}
 and admit the same transformation properties of the 
 fundamental representation ( 3, 2, 2) under the 
 spontaneously unbroken   
 gauge symmetry of $SU(3)_C\otimes{SU(2)_L^h}
 \otimes{SU(2)_R^h}$ (let us call $SU(2)_L^h\otimes{SU(2)_R^h}$ 
 `` hypercolor gauge symmetry ''). Then $\Lambda$ and $\Theta$ 
 come into the supersymmetric pair 
 which guarantees 'tHooft's naturalness condition[23]. 
 The $SU(3)_C$, $SU(2)_L^h$ and $SU(2)_R^h$ gauge fields 
 cause the confining forces with confining energy scales of 
 $\Lambda_c<< \Lambda_L<(or \cong) \Lambda_R$ 
 (Schrempp and Schrempp discussed this issue elaborately 
 in Ref.[11]). Here we call positive-charged 
 primons ($\Lambda$, $\Theta$) `` {\bf matter} '' 
 and negative-charged 
 primons ($\overline\Lambda$, $\overline\Theta$)
 `` {\bf antimatter} ''. Our final goal 
 is to build quarks, leptons 
 and ${\bf W, Z}$ from $\Lambda$ ($\overline\Lambda$)
  and $\Theta$ ($\overline\Theta$).   
 Let us discuss that scenario next.
 At the very early stage of the development of the universe, 
 the matter fields ($\Lambda$, $\Theta$) and their 
 antimatter fields ($\overline{\Lambda}$, $\overline{\Theta}$) 
 must have been created from the vacuum. After that they 
 would have combined with each other as the universe 
 was expanding. That would be the first step of the existence 
 of composite objects, which we call `` {\bf subquark} ''.
 There are ten types of them : 
\newline
 \hspace*{1.0cm}$spin\displaystyle{\frac{1}{2}}$
 \hspace{1.7cm}$spin0$
 \hspace{1.3cm}$e.m.charge$
 \hspace{1.4cm}$Y.M.representation$
{
 \setcounter{enumi}{\value{equation}}
 \addtocounter{enumi}{1}
 \setcounter{equation}{0} 
 \renewcommand{\theequation}{\theenumi\alph{equation}}
 \begin{eqnarray}
 \hspace{1.2cm}
     \Lambda\Theta\hspace{1.8cm}\Lambda\Lambda,
     \Theta\Theta\hspace{1.9cm}
     \frac{1}{3}e\hspace{2.0cm}(\overline{3},1,1)\hspace{2mm}
     (\overline{3},3,1)\hspace{2mm}(\overline{3},1,3),
     \hspace{0.8cm}(3.1)\nonumber\\
     \Lambda\overline\Theta,\overline\Lambda\Theta\hspace{1.4cm}
     \Lambda\overline\Lambda,\Theta\overline\Theta\hspace{2.0cm}
     0\hspace{2.2cm}(1,1,1)\hspace{2mm}(1,3,1)\hspace{2mm}(1,1,3),
     \hspace{0.8cm}(3.2)\nonumber\\
     \overline\Lambda\overline\Theta\hspace{1.8cm}\overline\Lambda
     \overline\Lambda,\overline\Theta\overline\Theta\hspace{1.45cm}
     -\frac{1}{3}e\hspace{2.0cm}(3,1,1)\hspace{2mm}(3,3,1)
     \hspace{2mm}(3,1,3)\hspace{1mm}.\hspace{0.75cm}(3.3)\nonumber     \label{3.3}
 \end{eqnarray}
 \setcounter{equation}{\value{enumi}}}
  In this step the confining forces are, in kind, in $SU(3)\otimes
 {SU(2)}_L^h\otimes{SU(2)}_R^h$ gauge symmetry but the 
 $SU(2)_L^h\otimes{SU(2)}_R^h$ confining forces must be main 
 because of the energy scale of $\Lambda_L,\Lambda_R>>\Lambda_c$
 and then the color gauge coupling $\alpha_s$ and e.m. coupling 
 constant $\alpha$ are negligible. 
 As is well known, the coupling constant of $SU(2)$ confining
 force are generally characterized 
 by $\varepsilon_i=\sum_a\sigma_p^a\sigma_q^a$,where 
 ${\sigma}s$ are $2\times2$ matrices of $SU(2)$, $a=1,2,3$, 
 $p,q=\Lambda,\overline\Lambda,\Theta,\overline
 \Theta$, $i=0$ for singlet and $i=3$ for triplet. 
 They are calculated as 
 $\varepsilon_0=-3/4$ which causes the attractive force and 
 and $\varepsilon_3=1/4$ causing the repulsive force. Next, 
 $SU(3)_C$ octet and sextet states are repulsive but singlet,
 triplet and antitriplet states are attractive and then the formers 
 are disregarded. Like this, two primons are confined into composite 
 objects in more than  one singlet state of any 
 $SU(3)_C,SU(2)_L,SU(2)_R$ as appeared in (3.1$\sim$3.3).
  Note that three primon systems 
 cannot make the singlet states of $SU(2)$ and we omit them. 
 In (3.2), the $(1,1,1)$-state is the `` most attractive channel ''. 
 Therefore $(\Lambda\overline\Theta)$, $(\overline\Lambda
 \Theta)$, $(\Lambda\overline\Lambda)$ and $(\Theta\overline
 \Theta)$ of $(1,1,1)$-states with neutral e.m. charge must 
 have been most abundant in the universe. 
 Further $(\overline{3},1,1)$- and 
 $(3,1,1)$-states in (3.1) and (3.3) are next attractive. 
 They presumably go into 
 $\{(\Lambda\Theta)(\overline\Lambda
 \overline\Theta)\}, \{(\Lambda\Lambda)(\overline\Lambda 
 \overline\Lambda)\}$, etc. of $(1,1,1)$-states 
 with neutral e.m. charge. Then
 these objects may be the candidates for the 
 `` {\bf Cold Dark Matter} '' if they have tiny masses. 
 Namely it may be said that `` {\bf Dark Matter is subquark}. '' 
 It is presumable that the ratio of the quantities between 
 the ordinary matters and the dark matters 
 firstly depends on the color and hypercolor charges 
 and the quantity of dark matter greatly surpasses that of the 
 ordinary matter(maybe the ratio is around $1/(2\times3)$).
\par 
 Finally the $(*,3,1)$-and $(*,1,3)$-states are remained 
 ($*$ is $1,3,\overline{3}$). 
 They are also stable
 because $|\varepsilon_0|>|\varepsilon_3|$. These subquarks 
 are, so to say, the ``intermediate clusters'' 
 towards constructing ordinary matters 
 (quarks, leptons and ${\bf W,Z}$)
\footnote{Such thoughts have been first proposed by Maki in Ref.[22].}
   and are denoted as follows :
\newline
\hspace*{5.0cm}$Y.M.representation$\hspace{1.5cm}$spin$
 \hspace{0.5cm}$e.m.charge$\hspace{2.0cm}
{
 \setcounter{enumi}{\value{equation}}
 \addtocounter{enumi}{1}
 \setcounter{equation}{0} 
 \renewcommand{\theequation}{\theenumi\alph{equation}}
 \begin{eqnarray}
    {\bf \alpha}&=&(\Lambda\Theta)\hspace{2.2cm}{\bf \alpha}_L:
    (\overline{3},3,1)\hspace{3mm}{\bf \alpha}_R:(\overline{3},1,3)
    \hspace{1.2cm}\frac{1}{2}\hspace{1.2cm}\frac{1}{3}e
    \hspace{2cm}(3.4)\nonumber\\
    {\bf \beta}&=&(\Lambda\overline\Theta)\hspace{2.2cm}{\bf \beta}_L: 
    (1,3,1)\hspace{3mm}{\bf \beta}_R:(1,1,3)\hspace{1.3cm}\frac{1}{2}
    \hspace{1.25cm}0\hspace{2.15cm}(3.5)\nonumber\\
    {\bf x}&=&(\Lambda\Lambda,\hspace{2mm}
    \Theta\Theta)\hspace{1.2cm}{\bf x}_L:
    (\overline{3},3,1)\hspace{3mm}{\bf x}_R:(\overline{3},1,3)    
    \hspace{1.3cm}0\hspace{1.25cm}\frac{1}{3}e
    \hspace{2cm}(3.6)\nonumber\\                               \label{3.7}
    {\bf y}&=&(\Lambda\overline\Lambda,\hspace{2mm}
    \Theta\overline\Theta)\hspace{1.2cm}{\bf y}_L:(1,3,1)
    \hspace{3mm}{\bf y}_R:(1,1,3)\hspace{1.3cm}0
    \hspace{1.3cm}0,\hspace{2.1cm}(3.7)\nonumber                             
 \end{eqnarray}
 \setcounter{equation}{\value{enumi}}}
and there are also their anti-subquarks[9].
\footnote{The notations of ${\bf \alpha}$,${\bf \beta}$,
 ${\bf x}$ and ${\bf y}$ are inherited from those in Ref.[9] 
 written by Fritzsch and Mandelbaum, because ours is, 
 in the subquark level, similar to theirs
 with two fermions and two bosons.
 R. Barbieri, R. Mohapatra and A. Masiero proposed 
 the similar model[9].}
\par
 Now we come to the step to build quarks, leptons and ${\bf W,Z}$. 
 The gauge symmetry of the confining forces in this step is also 
 $SU(2)_L^h\otimes{SU(2)}_R^h$ because the subquarks are in the
 triplet states of $SU(2)_{L,R}^h$ and then they are combined 
 into singlet states by the decomposition of 
 $3\otimes3=1\oplus{3}\oplus{5}$
 in $SU(2)$. We make the first generation of quarks and leptons 
 as follows :
\newline
 \hspace*{6cm}$e.m.charge$\hspace{1.2cm}$Y.M.representation$
 {
 \setcounter{enumi}{\value{equation}}
 \addtocounter{enumi}{1}
 \setcounter{equation}{0}
 \renewcommand{\theequation}{\theenumi\alph{equation}} 
 \begin{eqnarray}
     \hspace{0.5cm}
        <{\bf u}_l|&=&<{\bf \alpha}_l{\bf x}_l|\hspace{3.1cm}
                     \frac{2}{3}e\hspace{2.9cm}(3,1,1)
                     \hspace{3.1cm}(3.8)\nonumber\\
        <{\bf d}_l|&=&<\overline{\bf \alpha}_l\overline{\bf x}_l
                     {\bf x}_l|\hspace{2.2cm}-\frac{1}{3}e
                     \hspace{2.9cm}(3,1,1)                    
                     \hspace{3.1cm}(3.9)\nonumber\\
        <{\bf \nu}_l|&=&<{\bf \alpha}_l\overline{\bf x}_l|
                     \hspace{3.13cm}0\hspace{3.15cm}(1,1,1)
                     \hspace{3.0cm}(3.10)\nonumber\\
        <{\bf e}_l|&=&<\overline{\bf \alpha}_l\overline{\bf x}_l
                     \overline{\bf x}_l|\hspace{2.4cm}-e           \label{3.11}
                     \hspace{3.0cm}(1,1,1),\hspace{2.75cm}
                     (3.11)\nonumber                      
 \end{eqnarray}
 \setcounter{equation}{\value{enumi}}}
 where $l$ stands for $L$(left handed) or $R$(right handed).
\footnote{Subquark configurations in (3.8;9;10;11) are essentially the 
 same as those in Ref.[5] written by Kr\' olikowski, 
 who proposed the model of one fermion and one boson 
 with the same e.m. charge $e/3$.}
 Here we note that ${\bf \beta}$ and ${\bf y}$ do not appear.  
 In practice ($({\bf \beta}{\bf y}):(1,1,1)$)-particle 
 is a candidate for neutrino. But as Bjorken has pointed
 out[3], non-vanishing charge radius of neutrino is necessary
 for obtaining the correct low-energy effective weak interaction
 Lagrangian[11]. Therefore ${\bf \beta}$ is assumed not to 
 contribute to forming ordinary quarks and leptons.
 However $({\bf \beta}{\bf y})$-particle may be a candidate
 for `` sterile neutrino ''. 
 Presumably composite (${\bf \beta}$ ${\bf \beta}$)-;
 (${\bf \beta}\overline{\bf \beta}$)-;($\overline{\bf \beta}
 \overline{\bf \beta}$)-states may go into the dark matters. 
 It is also noticeable that in this model the leptons have 
 finite color charge radius and then $SU(3)$ gluons interact 
 directly with the leptons at energies of the order of, or 
 larger than $\Lambda_{L}$ or $\Lambda_{R}$[19]. \par
 Concerning the confinement of primon-level or subquark-level, 
 the confining forces of these levels are controled 
 by t he same spontaneously 
 `` unbroken '' $SU(2)_L^h\otimes{SU(2)}_R^h$ gauge symmetry.
 It is known that the running coupling constant 
 of the $SU(2)$ gauge theory satisfies the following equation : 
 {
 \setcounter{enumi}{\value{equation}}
 \addtocounter{enumi}{1}
 \setcounter{equation}{0}
 \renewcommand{\theequation}{\theenumi\alph{equation}}
 \begin{eqnarray}
       \hspace{1.5cm} 
       \frac{1}{\alpha^{a}_{W}(Q_{1}^{2})}&=&
                \frac{1}{\alpha^{a}_{W}(Q_{2}^{2})}
                +b_{a}\ln\left(\frac{Q_{1}^{2}}{Q_{2}^{2}}\right),
                \hspace{6.0cm}(3.12)\nonumber\\
       b_{a}&=&\frac{1}{4\pi}\left(\frac{22}{3}-\frac{2}{3}
       \cdot{N}_{f}-\frac{1}{12}\cdot{N}_{s}\right),
       \hspace{5.0cm}(3.13)\nonumber                 
 \end{eqnarray}
 \setcounter{equation}{\value{enumi}}}
 where $N_f$ and $N_s$ are the numbers of fermions and scalars   
 contributing to the vacuum polarizations,
 ($a=q$) for the confined subquarks in quark and ($a=sq$)
 for confined primons in subquark and 
 $Q_{1 or 2}^{2}$ is the effective 
 four momentum squuare of ${\bf g}_h$-exchange.
 We calculate $b_q=0.35$ which comes from that the number of 
 confined fermionic subquarks are $4$ (${\bf \alpha}_{i}, i=1,2,3$
 for color freedom, ${\bf \beta}$) and $4$ for bosons (${\bf x}_i,
 {\bf y}$) contributing to the vacuum polarization, and 
 $b_{sq}=0.41$ which  is calculated with three kinds of 
 $\Lambda$ and $\Theta$ owing to three color freedoms. 
 Experimentaly it is reported that $\Lambda_q>1.8$ TeV(CDF exp.)
 or $\Lambda_q>2.4$ TeV(D$\O$ exp.)[12].
 Extrapolations of $\alpha^{q}_{W}$ and $\alpha^{sq}_{W}$ to near 
 Plank scale are expected to converge to the same point and then
  tentatively, setting 
 $\Lambda_q=5$ TeV, $\alpha^{q}_{W}(\Lambda_q)=\alpha^{sq}_{W}
 (\Lambda_{sq})=\infty$, we get $\Lambda_{sq}=10^3\Lambda_q$,   
\par
 Next let us see the higher generations. Harari and Seiberg 
 have stated that the orbital and radial excitations 
 seem to have the wrong energy scale ( order of 
 $\Lambda_{L,R}$) and then the most likely type of 
 excitations is the addition of primon-antiprimon pairs[6,25].
 In our model the essence of generation is like `` $isotope$ "
 in case of atoms.
 Then using neutral ${\bf y}_{L,R}$ in (3.7) 
 we construct them as follows :
 {  
 \setcounter{enumi}{\value{equation}}
 \addtocounter{enumi}{1} 
 \setcounter{equation}{0}
 \renewcommand{\theequation}{\theenumi\alph{equation}}
 \begin{eqnarray}
 &&\left\{
 \begin{array}{lcl}
          <{\bf c}\hspace{0.5mm}|
                  &=&<{\bf \alpha}{\bf x}{\bf y}|\\
          <{\bf s}\hspace{0.5mm}|
                  &=&<\overline{\bf \alpha}\overline{\bf x}
                      {\bf x}{\bf y}|,
 \end{array} 
 \right.
 \hspace{6mm}
 \left\{
 \begin{array}{lcl}
          <{\bf \nu_\mu}\hspace{0.3mm}|
                       &=&<{\bf \alpha}\overline{\bf x}{\bf y}|\\          
          <{\bf \mu}\hspace{2mm}|&=&<\overline{\bf \alpha}
                    \overline{\bf x}\overline{\bf x}{\bf y}|,
 \end{array}
 \right.
 \hspace{0.6cm}\mbox{2nd generation}\hspace{1cm}
               \mbox{(3.14)}\nonumber\\
 &&\left\{
 \begin{array}{lcl}
          <{\bf t}\hspace{0.5mm}|
                   &=&<{\bf \alpha}{\bf x}{\bf y}{\bf y}|\\
          <{\bf b}|&=&<\overline{\bf \alpha}
                      \overline{\bf x}{\bf x}{\bf y}{\bf y}|,
 \end{array}
 \right.
 \hspace{0.3cm}
 \left\{
 \begin{array}{lcl}
          <{\bf \nu_\tau}\hspace{0.5mm}|
                &=&<{\bf \alpha}\overline{\bf x}
                              {\bf y}{\bf y}|\\
          <{\bf \tau}\hspace{2mm}|&=&<\overline{\bf \alpha}
          \overline{\bf x}\overline{\bf x}{\bf y}{\bf y}|,
 \end{array}
 \right.
 \hspace{5.0mm}\mbox{3rd generation}\hspace{1.06cm}
             \mbox{(3.15)}\nonumber                           \label{3.15}
 \end{eqnarray} 
 \setcounter{equation}{\value{enumi}}}
 where the suffix $L,R$s are omitted for brevity. 
 We can also make vector and scalar particles with (1,1,1) : 
 {
 \setcounter{enumi}{\value{equation}}
 \addtocounter{enumi}{1} 
 \setcounter{equation}{0}
 \renewcommand{\theequation}{\theenumi\alph{equation}}
 \begin{eqnarray}&&\left\{
 \begin{array}{lcl}
             <{\bf W}^+|&=&<{\bf \alpha}^\uparrow{\bf \alpha}^
                             \uparrow{\bf x}|\\
             <{\bf W}^-|&=&<\overline{\bf \alpha}^\uparrow
                           \overline{\bf \alpha}^\uparrow\overline{\bf x}|,
 \end{array}
 \right.\hspace{6mm}
 \left\{
 \begin{array}{lcl}
             <{\bf Z}_1^0|&=&<{\bf \alpha}^\uparrow\overline
                             {\bf \alpha}^\uparrow|\\
             <{\bf Z}_2^0|&=&<{\bf \alpha}^\uparrow\overline
                             {\bf \alpha}^\uparrow{\bf x}\overline{\bf x}|, 
 \end{array} 
 \right.\hspace{0.75cm}\mbox{Vector}\hspace{1.5cm}\mbox{(3.16)}\nonumber\\
 &&\left\{
 \begin{array}{lcl}
             <\hspace{2mm}{\bf S}^+\hspace{0.5mm}|
                  &=&<{\bf \alpha}^\uparrow{\bf \alpha}^
                           \downarrow{\bf x}|\\
             <\hspace{2mm}{\bf S}^-\hspace{0.5mm}|
                  &=&<\overline{\bf \alpha}^
                          \uparrow\overline{\bf \alpha}^\downarrow{\bf x}|,
 \end{array}
 \right.
 \hspace{6mm}
 \left\{
 \begin{array}{lcl}
            <{\bf S}_1^0|&=&<{\bf \alpha}^\uparrow\overline
                             {\bf \alpha}^\downarrow|\\
            <{\bf S}_2^0|&=&<{\bf \alpha}^\uparrow\overline
                            {\bf \alpha}^\downarrow{\bf x}
                            \overline{\bf x}|,
 \end{array}
 \right.
 \hspace{8mm}\mbox{Scalar}\hspace{1.5cm}\mbox{(3.17)}\nonumber           \label{3.17}
 \end{eqnarray}
 \setcounter{equation}{\value{enumi}}}
 where the suffix $L,R$s are omitted for brevity and $\uparrow, 
 \downarrow$ indicate $spin\hspace{1mm}up, spin\hspace{1mm}down$ states.
 They play the role of intermediate bosons same as ${\bf \pi}$, 
 ${\bf \rho}$ in the strong interactions. As (3.8$\sim$11) and (3.16;17) 
 contain only ${\bf \alpha}$ and ${\bf x}$ subquarks, we can 
 draw the `` $line\hspace{1mm}diagram$ '' of weak interactions as seen 
 in Fig (1). Equation (3.11) shows that the electron is constructed 
 from antimatters only. Therefore electrons are totally not matters 
 but antimatters. Actually we don`t know the exact reason why 
 `` electron is matter '' and this is merely the assumption.
 We know, phenomenologically, that this 
 universe is mainly made of protons, electrons, neutrinos,
 antineutrinos and unknown dark matters. It is said that  
 the universe contains  almost the same number of protons 
 and electrons.
  Our model show that one proton has the configuration
 of $({\bf u}{\bf u}{\bf d}): (2{\bf \alpha}, \overline{\bf \alpha}, 
 3{\bf x}, \overline{\bf x})$; electron: $(\overline{\bf \alpha}, 
 2\overline{\bf x})$; neutrino: $({\bf \alpha}, \overline{\bf x})$; 
 antineutrino: $(\overline{\bf \alpha}, {\bf x})$ and the dark 
 matters are constructed from the same amount of matters 
 and antimatters because of their neutral charges. Note that 
 proton is a mixture of matters and anti-matters and electrons
 is composed of anti-matters only. These ideas may lead the thought 
 that `` {\bf The universe is the matter-antimatter-even object.} '' 
 And then there exists a conception-leap between 
 `` proton-electron abundance '' and `` matter abundance '' 
 if our composite scenario is admitted 
 (as for the possible way to realize the proton-electron 
 excess universe, see Ref.[14]).
 This idea is different from the current thought that
 the universe is made of matters only. 
 Then in our model the problem 
 about CP violation in the early universe does not occur.
\par
 Our composite model contains two steps, namely the first is 
 `` subquarks made of primons '' and the second is `` quarks and 
 leptons made of subquarks ''. Here let us discuss about 
 the mass generation mechanism  of quarks and leptons as 
 composite objects. Our model has only one  kind of fermion 
 : $\Lambda$  and boson : $\Theta$. The first  step of 
 `` subquarks made of primons '' seems to have nothing 
 to do with 'tHooft's anomaly matching condition[23] 
 because there is no global symmetry with 
 $\Lambda$ and $\Theta$. Therefore from this line of thought 
 it is  impossible to say anything about that ${\bf \alpha}$, 
 ${\bf \beta}$,  ${\bf x}$ and ${\bf y}$ are massless or massive. 
 However, if it is the case that the neutral  
 (1,1,1)-states of primon-antiprimon composites (as is stated above) 
 construct the dark matters, the masses of them are presumably less 
 than the order of MeV from the phenomenological aspects of 
 astrophysics. 
 Then we may assume that these subquarks are 
 massless or almost massless compared with $\Lambda_{L,R}$ 
 in practice, that is, utmost a few MeV. 
 In the second step, the arguments of 'tHooft's anomaly 
 matching condition are meaningful. The confining of subquarks 
 must occur at the energy scale of $\Lambda_{L,R}>>\Lambda_c$ 
 and then it is natural that $\alpha_s, \alpha \rightarrow0$ 
 and that the gauge symmetry group is the spontaneously 
 unbroken $SU(2)_L\otimes{SU(2)}_R$ gauge group. 
 Seeing (3.8$\, \sim$11), we find quarks and leptons are composed of 
 the mixtures of subquarks and antisubquarks. 
 Therefore it is proper to 
 regard subquarks and antisubquarks as different kinds of 
 particles. From (3.4$\,$;5) we find eight kinds of fermionic 
 subquarks ( 3 for ${\bf \alpha}$, $\overline{\bf \alpha}$ and 
 1 for ${\bf \beta}$, $\overline{\bf \beta}$). So the global 
 symmetry concerned is $SU(8)_L\otimes{SU(8)}_R$.
\par 
 Then we  arrange : 
       $$ ({\bf \beta},\overline{\bf \beta},
          {\bf \alpha}_i,\overline{\bf \alpha}_
          i\hspace{3mm}i=1,2,3\hspace{1mm})_{L,R}
        \hspace{0.5cm}in\hspace{0.5cm}(SU(8)_L
        \otimes{SU(8)}_R)_{global},\hspace{2cm}
        \eqno{(3.18)}$$                         \label{3.18}
 where $i$ is color freedom.
 Next, the fermions in (3.18) are confined into the singlet 
 states of the local $SU(2)_L\otimes{SU(2)}_R$ gauge symmetry 
 and make up quarks and leptons as seen in (3.8$\sim${11}) 
 (eight fermions). \par 
 Then we arrange :
         $$({\bf \nu_e},{\bf e},{\bf u}_i,{\bf d}_i\hspace{3mm}i
           =1,2,3\hspace{1mm})_{L,R}\hspace{0.5cm}in
            \hspace{0.5cm}(SU(8)_L\otimes{SU(8)}_R)_{global},
            \hspace{2cm}\eqno{(3.19)}$$                         \label{3.19}
  where $i$s are color freedoms. From(3.18) and (3.19) 
 the anomalies of the subquark level and the quark-lepton level 
 are matched and then all composite quarks and leptons (in the 1st 
 generation) are remained massless or almost massless. 
 Note again that presumably, 
 ${\bf \beta}$ and $\overline{\bf \beta}$ in(3.18) are composed 
 into ``bosonic'' (${\bf \beta}$${\bf \beta}$), 
 (${\bf \beta}$$\overline{\bf \beta}$) and 
 ($\overline{\bf \beta}$$\overline{\bf \beta}$), 
 which vapour out to the dark matters. Schrempp and Schrempp have 
 discussed about a confining $SU(2)_L\otimes{SU(2)}_R$ gauge 
 model with three fermionic preons and stated that it is 
 possible that not only the left-handed quarks and leptons   
 are composite but also the right-handed ones are so on the condition  
 that $\Lambda_R/\Lambda_L$$\sim{O}$$(10^3)$[11].
  As seen in (3.16) the existence of composite 
 ${\bf W}_R$, ${\bf Z}_R$ is predicted. As concerning, 
 the fact that they are not observed yet means  that the masses of 
 ${\bf W}_R$, ${\bf Z}_R$ are larger than those of 
 ${\bf W}_L$, ${\bf Z}_L$ because of $\Lambda_R>\Lambda_L$. 
 Owing to 'tHooft's 
 anomaly matching condition the small mass nature of the 1st 
 generation comparing to $\Lambda_L$ is guaranteed but 
 the evidence that the quark masses of the 2nd and the 3rd 
 generations become larger as the generation numbers increase 
 seems to have nothing to do 
 with the anomaly matching mechanism in our model, because, as 
 seen in (3.11;12), these generations are obtained by just 
 adding neutral scalar ${\bf y}$-particles. 
  This is different from Abott and Farhi's model 
 in which all fermions of three generations are equally
  embedded in $SU(12)$ global symmetry group and all 
  members take part in the anomaly matching mechanism[8,26]. 
 Equation(3.16;17) shows that the difference in $Z^{0}$ 
 and $S^{0}$ essentially originates from the combination of 
 two spins(up-spin and down-spin) of ${\bf \alpha}$- and
 $\overline{\bf \alpha}$- subquark. $S^{0}$ has the 
 combination of up- and down-spin and $Z^{0}$ has that of 
 up- and up-spin.This situation is similar 
 to hadronic mesons. They are the composite objects of 
 a quark($q$) and a anti-quark($\overline{q}$). namely, 
 ${\rho}$-${\pi}$, $K^{*}$-$K$, $D^{*}$-$D$, $B^{*}$-$B$.
 Each vector meson mass(denoted by $M(V)$) is larger than 
 the mass(denoted by $M(Ps)$) of its  pseudo-scalar partner. 
 The mass differences between $M(V)$ and $M(Ps)$ are 
 qualitatively explained by the hyperfine spin-spin 
 interaction in Breit-Fermi Hamiltonian[28].
 As the model of the hadronic mass spectra by the 
 Breit-Fermi Hamiltonian is described by use of the 
 semi-relativistical approach, it has some defects 
 in the quantitative estimations, especially in the 
 small mass mesons (such as ${\rho}$-${\pi}$ and 
 $K^{*}$-$K$) but qualitatively it is not so bad, 
 namely the explanation of the fact that : $M(V)>
 M(Ps)$(and else $M(J=3/2\hspace{1.5mm}
 \rm{baryon})>M(J=1/2\hspace{1.5mm} \rm{baryon})$).
 The hyperfine interaction Hamiltonian(denoted 
 by $H_{q{\overline q}}^{l}$) causing mass split between 
 $M(V)$ and $M(Ps)$ is described as :
  $$H_{q{\overline{q}}}^{l=0}=
         -\frac{8{\pi}}{3m_{q}m_{\overline{q}}}
         \overrightarrow{S}_{q}\overrightarrow{S}_
         {\overline{q}}\delta(|\overrightarrow{r}|),
         \hspace{5cm}\eqno{(3.20)}$$                 \label{3.20}  
 where $\overrightarrow{S}_{q(\overline{q})}$ is a 
 operator of $q(\overline{q})$'s spin with its 
 eigenvalue of 1/2 or -1/2, $m_{{q}(\overline{q})}$ is
 quark (anti-quark) mass, $l$ is the orbital angular
 momentum between $q$ and $\overline{q}$ and 
 $|\overrightarrow{r}|=|\overrightarrow{r}_{q}
 -\overrightarrow{r}_
 {\overline{q}}|$[28]. 
 \par In QCD theory eight gluons are intermediate gauge 
 bosons belonging to {\bf 8} representation which 
 is real adjoint representation. Quarks(anti-quarks)
 belong to ${\bf 3}({\bf \overline{3}})$ representation
 which is complex fundamental representation. Therefore  
 gluons can discriminate between quarks and anti-quarks 
 and couple to them in the '' $opposite\hspace{2mm}sign$ ''.
  The strength of their couplings to different color 
 quarks and anti-quarks is described as :
\begin{eqnarray}
      \hspace{2cm}
    +g\frac{\lambda_{ij}^{a}}{2}&:&\hspace{2.8cm}
    \rm{for\hspace{5mm} quark}\nonumber\\
    -g\frac{\lambda_{ij}^{a}}{2}&:&\hspace{2.7cm}
   \rm{for\hspace{5mm}anti-quark},
   \hspace{4.5cm}(3.21)\nonumber                            \label{3.21}
\end{eqnarray}
 where $a(=1\sim8)$ : gluon indices; $i,j(=1,2,3)$ : 
 quark indices; ${\lambda}$'s : SU(3) matrices
 and $g$ : the coupling constant of gluons to quarks 
 and anti-quarks(See Fig.(3)). 
 The wave function of a color singlet 
 $q\overline{q}$(meson) system is ${\delta}_{ij}/
 \sqrt{3}$, corresponding to :
   $$|q\overline{q}>=(1/\sqrt{3})\displaystyle
   {\sum_{i=1}^{3}}|q_{i}\overline{q_{i}}>.\hspace{5.0cm}
   \eqno{(3.22)}$$                                           \label{3.22} 
 By use of (3.22) the effective coupling for the 
 $q{\overline{q}}$ system(denoted by ${\alpha}_{s}$)
 is given by : 
  \begin{eqnarray}
            \hspace{2cm}
     {\alpha}_{s}&=& {\displaystyle{\sum_{a,b}}{\sum_{i,j,k,l}}}
                   {\frac{1}{\sqrt{3}}}{\delta}_{ij}
                   \left({\frac{g}{2}}{\lambda}_{ik}^{a}\right)
                   \left(-{\frac{g}{2}}{\lambda}_{lj}^{b}\right)
                   {\frac{1}{\sqrt{3}}}{\delta}_{kl}
                  =-{\frac{g^{2}}{12}}{\displaystyle{\sum_{a,b}}{\sum_{j,l}}}
                   {\lambda}_{jl}^{a}{\lambda}_{lj}^{b}\nonumber\\
                 &=&-{\frac{g^{2}}{12}}\displaystyle{\sum_{a,b}}
                   \rm{Tr}\left({\lambda}^{a}{\lambda}^{b}\right)
                  =-{\frac{g^{2}}{6}}{\displaystyle
                   {\sum_{ab}}}{\delta}_{ab}\nonumber\\
                 &=&-{\frac{4}{3}}g^{2}.\hspace{9.8cm}(3.23)\nonumber
                                                                      \label{3.23}
  \end{eqnarray} 
 Making use of (3.22) and (3.23) let us write the quasi-static 
 Hamiltonian for a bound state of a quark and a anti-quark is
 given as :  
     $$ \hspace{0.5cm}H
           =H_{0}+{\alpha}_{s}H_{q{\overline{q}}}^{l=0}.
            \hspace{6.7cm}\eqno{(3.24)}$$                        \label{3.24}  
 Calculating the eigenvalue of $H$ in (3.24) we have :
      $$ M(V{\rm{or}}S)=M_{0}+{\xi}_{q}
      <\overrightarrow{S}_{q}\overrightarrow{S}_
      {\overline{q}}>,\hspace{4.5cm}\eqno{(3.25)}$$               \label{3.25}  
  where ${\xi}_{q}$ is a positive constant which incldes 
  the calculation of ${\alpha}_{s}$.
  In (3.25) it is found that  
  $<\overrightarrow{S}_{q}\overrightarrow{S}_{\overline{q}}>=-3/4$
  for pseudoscalar mesons and 
  $<\overrightarrow{S}_{q}\overrightarrow{S}_{\overline{q}}>=1/4$
  for vector mesons and then we have :
  \begin{eqnarray}
       \hspace{3cm}
    M(Ps)&=&M_{0}-{\frac{3}{4}}{\xi}_{q}\nonumber\\
    M(\hspace{1mm}V\hspace{1mm})\hspace{0.7mm}
    &=&M_{0}+{\frac{1}{4}}{\xi}_{q}.\hspace{7.2cm}(3.26)\nonumber     \label{3.26}
  \end{eqnarray}
  By (3.26) it is resulted that :   
     $$ M(V)>M(Ps).\hspace{6cm}\eqno{(3.27)}$$                                                    
  Here let us turn discussons to `` intermediate weak bosons ''.
  As seen in (3.16;17) $Z^{0}$ weak boson has its scalar 
  partner $S^{0}$ and both of them contain `` $fermionic$ ''
  ${\alpha}_{L}$ and $\overline{\alpha}_{L}$ as subquark elements.  
  Referring(3.4) we find that both of ${\alpha}_{L}$ and 
  $\overline{\alpha}_{L}$ belong to `` adjoint ${\bf 3}$ '' state 
  of $SU(2)_{L}$(which is the real representation) 
  and then $SU(2)_{L}$-hypercolor gluons cannot distingush    
  ${\alpha}_{L}$ from $\overline{\alpha}_{L}$. Therefore 
  the hypercolor gluons couple to ${\alpha}_{L}$ and
  $\overline{\alpha}_{L}$ in the `` $same\hspace{2mm} sign$ ''.
  This point is distinguishably dfferent from hadronic mesons
  (Refer (3.21)).
  The wave function of a hypercolor singlet 
  (${\alpha}{\overline{\alpha}}$)-system is 
  ${\delta}_{ij}/{\sqrt{3}}$, corresponding to 
   $|{\bf \alpha}_{i}\overline{\bf \alpha}_{i}>
  =(1/{\sqrt{3}}){\displaystyle|{\sum_{i=1}^{3}}}|
  {\bf \alpha}_{i}\overline{\bf \alpha}_{i}>$
  where $i=1,2,3$ are different three states of the triplet 
  of $SU(2)_{L}$.
   The strength of their couplings to different hypercolor 
  subquarks and anti-subquarks is described as :
\begin{eqnarray}
     \hspace{1.5cm}
    +g_{h}\frac{{\tau}_{ij}^{a}}{2}
              &:&\hspace{2.5cm}
              \rm{for\hspace{6mm}subquark}\nonumber\\
    +g_{h}\frac{{\tau}_{ij}^{a}}{2}
              &:&\hspace{2.5cm}
              \rm{for\hspace{2mm}anti-subquark},
              \hspace{4.5cm}(3.28)\nonumber                           \label{3.28}
  \end{eqnarray}   
  where $a(=1,2,3)$ : hypercolor gluon indices; $i,j(=1,2,3)$ :
  subquark and anti-subquark indices and ${\tau}$ : $SU(2)$ 
  matrices and $g_{h}$ : the coupling constant of hypergluons 
 to the subquarks and anti-subquarks(See Fig.(3)). 
  By use of (3.28) the effective coupling 
  (denoted by ${\alpha_{W}}$) is given by : 
   \begin{eqnarray}
     {\alpha}_{W}&=& {\displaystyle{\sum_{a,b}}{\sum_{i,j,k,l}}}
             {\frac{1}{\sqrt{3}}}{\delta}_{ij}
             \left({\frac{g_{h}}{2}}{\tau}_{ik}^{a}\right)
             \left({\frac{g_{h}}{2}}{\tau}_{lj}^{b}\right)
             {\frac{1}{\sqrt{3}}}{\delta}_{kl}
                  ={\frac{g^{2}_{h}}{12}}{\displaystyle
             {\sum_{a,b}}{\sum_{j,l}}}
             {\tau}_{jl}^{a}{\tau}_{lj}^{b}\nonumber\\
                 &=&{\frac{g^{2}_{h}}{12}}\displaystyle{\sum_{a,b}}
             \rm{Tr}\left({\tau}^{a}{\tau}^{b}\right)
                  ={\frac{g^{2}_{h}}{6}}{\displaystyle
             {\sum_{ab}}}{\delta}_{ab}\nonumber\\
                 &=&{\frac{1}{2}}g^{2}_{h},\hspace{11.8cm}
                    (3.29)\nonumber                                   \label{3.29}
  \end{eqnarray}
  where $a,b=1,2,3$; $i,j,k,l=1,2,3$. Note that ${\alpha}_{s}$ 
  (in (3.23)) is `` $negative$ '' but ${\alpha}_{W}$(in (3.29)) 
  `` $positive$ ''. Through the same procedure as  hadronic mesons 
  the masses of $Z^{0}$ and $S^{0}$ are described as :  
     $$ M(Z^{0}\hspace{0.6mm}\rm{or}\hspace{0.6mm}S^{0})
        =M_{0}-{\xi}_{sq}<\overrightarrow{S}_{\alpha}
        \overrightarrow{S}_{\overline{\alpha}}>,\hspace
        {5.5cm}\eqno{(3.30)}$$                                   \label{3.30}
    where ${\xi}_{sq}$ is a positive constant which includes 
  the calculation of ${\alpha}_{W}$ and 
  $\overrightarrow{S}_{\alpha(\overline{\alpha})}$ is 
  the spin operator of ${\alpha}(\overline{\alpha})$.          
   In (3.30) it is calculated that 
  $<\overrightarrow{S}_{\alpha}\overrightarrow{S}_
  {\overline{\alpha}}>=-3/4$
  for scalar : $S^{0}$ and 
  $<\overrightarrow{S}_{\alpha}\overrightarrow{S}_
  {\overline{\alpha}}>=1/4$
  for vector : $Z^{0}$ and then we get :
   \begin{eqnarray}
       \hspace{2cm}
    M(S^{0})&=&M_{0}+{\frac{3}{4}}{\xi}_{sq}\nonumber\\
    M(Z^{0})&=&M_{0}-{\frac{1}{4}}{\xi}_{sq}.\hspace{8.3cm}
              (3.31)\nonumber                                        \label{3.31}
   \end{eqnarray}
  From this it follows that : 
       $$M(S^{0})>M(Z^{0}).\hspace{7cm}\eqno{(3.32)}$$             \label{3.32}   
  Here let us define :
   \begin{eqnarray}
   \hspace{2.2cm}\tilde{M}
      &=&\frac{1}{2}\left(M(S^{0})+M(Z^{0})\right),\nonumber\\
   {\Delta}
      &=&M(S^{0})-M(Z^{0}),\nonumber\\
     R&=&\frac{\Delta}{\tilde{M}}.\hspace{10cm}(3.33)\nonumber        \label{3.33} 
  \end{eqnarray}
  Experimentally it is reported : $M(Z^{0})=91$GeV[24], 
  with which by use of (3.33) we obtain :
   \begin{eqnarray}
    \hspace{2.1cm}R&=&
      0.2\hspace{1cm}M(S^{0})\approx{110}
      \hspace{2mm}\rm{GeV},\nonumber\\
    R&=&
      0.3\hspace{1cm}M(S^{0})\approx{120}
      \hspace{2mm}\rm{GeV}.\hspace{6cm}(3.34)\nonumber
   \end{eqnarray}                                                      \label{3.34}
  Therefore if the existence of the scalar particle 
  whose mass is a little above $Z^{0}$'s mass
  is confirmed in future it may be a scalar partner of 
  $Z^{0}$ and that might suggest the possibility of the 
  subquark structure.
  \par One of the experimental evidences inspiring the SM is 
 the `` universality '' of the coupling strength among the 
 weak interactions. Of course if the intermediate bosons are 
 gauge fields, they couple to the matter fields universally. 
 But the inverse of this statement is not always true, namely 
 the quantitative equality of the coupling strength of the 
 interactions does not necessarily imply that the intermediate 
 bosons are elementary gauge bosons. In practice the interactions 
 of ${\bf \rho}$ and ${\bf \omega}$ are regarded as indirect 
 manifestations of QCD. In case of chiral 
 $SU(2)\otimes{SU(2)}$ the pole dominance works very well
 and the predictions of current algebra and PCAC seem  
 to be fulfilled within about $5$\%[19]. 
 Fritzsch and Mandelbaum[9][19] and Gounaris, K\"ogerler 
 and Schildknecht[10][27] have elaborately discussed about 
 universality of weak interactions appearing as a 
 consequence of current algebra and ${\bf W}$-pole 
 dominance of the weak spectral functions from the 
 stand point of the composite model. 
 Extracting the essential points from their  
 arguments we mention our case as follows. 
 In the first generation let the weak charged currents be 
 written in terms of the subquark fields as :
    $$\hspace{2cm}{\bf J}_{\mu}^{+}=\overline{U}h_{\mu}
          D,\hspace{2cm}{\bf J}_{\mu}^{-}=\overline{D}
          h_{\mu}U,\hspace{4.5cm}\eqno{(3.35)}$$                      \label{3.35}
  where $U=({\bf \alpha}{\bf x})$, $D=(\overline{\bf \alpha}
 \overline{\bf x}{\bf x})$ and $h_{\mu}=\gamma_{\mu}
 (1-\gamma_5)$.
 Reasonableness of (3.35) may given by the fact that
 $M_W<<\Lambda_{L,R}$ (where $M_W$ is ${\bf W}$-boson mass).
 Further, let $U$ and $D$ belong to the doublet of 
 the global weak isospin $SU(2)$ group and ${\bf W}^+$, 
 ${\bf W}^-$, $(1/\sqrt{2})({\bf Z}_1^0-{\bf Z}_2^0)$ 
 be in the triplet and $(1/\sqrt{2})({\bf Z}_1^0+
 {\bf Z}_2^0)$ be in the singlet of $SU(2)$. 
 These descriptions seem to be natural 
 if we refer the diagrams in Fig.(1). 
 The universality of the weak interactions are inherited 
 from the universal coupling strength of the algebra 
 of the global weak isospin $SU(2)$ group with the 
 assumption of ${\bf W}$-, ${\bf Z}$-pole dominance. 
 The universality including the 2nd and the 3rd 
 generations are investigated in the next section 
 based on the above assumptions  
 and in terms of the flavor-mixings.

\section{\hspace{0.5cm}Flavor-mixing matrix element by subquark 
\newline \hspace{0.3cm} dynamics}
\hspace*{\parindent}

 The quark-flavor-mixings in the weak interactions are 
 usually expressed by Cabbibo-Kobayashi-Maskawa (CKM) matrix 
 based on the SM. Its nine matrix elements 
 (in case of three generations) are " free '' parameters 
 (in practice four parameters with the unitarity) 
 and this point is said to be one of the drawbacks 
 of the SM along with the origins of the quark-lepton mass 
 spectrum and generations. 
 In the SM, the quarks and leptons  are elementary and then 
 we are able to investigate, 
 at the utmost, the external relationship among them. 
 On the other hand if quarks are the composites of 
 substructure constituents, 
 the quark-flavor-mixing phenomena must be understood 
 by the substructure dynamics and the values of CKM
 matrix elements become materials for studying these. 
 Terazawa and Akama have investigated quark-flavor-mixings 
 in a three spinor subquark model with higher generations 
 of radially excited state of the up (down) quark 
 and stated that a quark-flavor-mixing matrix element 
 is given by an overlapping integral of two radial 
 wave functions of the subquarks which depends on 
 the momentum transfer between quarks[13][31].
 \par In our model we set the assumption :
\newline $\;$ \newline 
 \begin{em}\hspace{0.6cm}The quark-flavor-mixings occur 
           by creation$\,$(or annihilation) 
           of ${\bf y}$-particles \par from(or into) vacuum  
           inside quarks.
 \end{em}
\newline $\;$ \newline
 The ${\bf y}$-particle is a neutral scalar   
 subquark in the {\bf3}-state of $SU(2)_L$ group (as seen in(3.7)).
  and then couples to two hypercolor gluons 
 (denoted by ${\bf g}_h$) (see Fig.(2)). 
 This is analogous to $\pi^0 \to 2\gamma$.
\par Here we propose the annother important assumption :
 \newline $\;$ \newline
 \begin{em}
     \hspace{0.6cm}
     The (${\bf y}\rightarrow{2}{\bf g}_h$)-process is 
     factorized from the net ${\bf W}^{\pm}$ exchange 
     interactions.
 \end{em}
 \newline $\;$ \newline
 This assumption is plausible because the effective 
 energy of this process may be in a few TeV energy 
 region comparing to a hundred GeV energy region of 
 {\bf W}-exchange processes.
 Let us write the contribution of 
 (${\bf y}\rightarrow{2}{\bf g}_h$)-process to charged 
 weak interactions as : 
   $$\hspace{0.5cm}A_{i}=\alpha^{q}_{W}(Q_{i}^{2})^{2}\cdot{B}
     \hspace{2cm}\mbox{$i$\hspace{0.5mm}
     =\hspace{2mm}{\bf s},{\bf c},{\bf b},{\bf t}},
     \hspace{1.0cm}\eqno{(4.1)}$$                                 \label{4.1} 
 where $\alpha_{W}$ is a running coupling constant 
 of the hypercolor gauge theory appearing in (3.12)
 , $Q_{i}$ is the effective four momentum of 
 ${\bf g}_h$-exchange among subquarks inside 
 the $i$-quark and $B$ is a dimensionless 
 `` $complex$ '' free parameter originated from 
 the unknown primon dynamics and may depend on 
 $<0|f(\overline\Lambda\cal{O}$${\Lambda}$,{\rm and/or},
 $\overline\Theta\cal{O}$${\Theta})|{\bf y}>$
 ($\cal{O}$ is some operator).
 \par
  The weak charged currents of quarks are 
 taken as the matrix elements of subquark currents 
 between quarks which are not the eigenstates of 
 the weak isospin[13]. 
 Using (3.14;15), (3.18) and (4.1) with the above 
 assumption we have :
{
\setcounter{enumi}{\value{equation}}
\addtocounter{enumi}{1}
\setcounter{equation}{0}
\renewcommand{\theequation}{\theenumi\alph{equation}}
\begin{eqnarray}
    \hspace{1cm}
    V_{ud}\overline{\bf u}h_{\mu}{\bf d}
     &=&<{\bf u}|\overline{U}h_{\mu}D|{\bf d}>,                     \label{4.2}
       \hspace{8cm}(4.2)\nonumber\\
    V_{us}\overline{\bf u}h_{\mu}{\bf s}
     &=&<{\bf u}|\overline{U}h_{\mu}(D{\bf y})|{\bf s}>\cong
       <{\bf u}|\overline{U}h_{\mu}D|{\bf s}>
       \cdot{A}_s,\hspace{3.8cm}(4.3)\nonumber\\
    V_{ub}\overline{\bf u}h_{\mu}{\bf b}
     &=&<{\bf u}|\overline{U}h_{\mu}(D{\bf y}{\bf y})|
       {\bf b}>\cong<{\bf u}|\overline{U}h_{\mu}D|
       {\bf b}>{\bf \cdot}{2}A_{b}^{2},\hspace{3.2cm}(4.4)\nonumber\\
    V_{cd}\overline{\bf c}h_{\mu}{\bf d}
     &=&<{\bf c}|(\overline{U}{\bf y})h_{\mu}D|{\bf d}>\cong
       <{\bf c}|\overline{U}h_{\mu}D|{\bf d}>
       \cdot{A}_c,\hspace{3.8cm}(4.5)\nonumber\\
    V_{cs}\overline{\bf c}h_{\mu}{\bf s}
     &=&<{\bf c}|(\overline{U}{\bf y})h_{\mu}(D{\bf y})|
       {\bf s}>,\hspace{7cm}(4.6)\nonumber\\
    V_{cb}\overline{\bf c}h_{\mu}{\bf b}
     &=&<{\bf c}|(\overline{U}{\bf y})h_{\mu}(D{\bf y}{\bf y})
       |{\bf b}>\cong<{\bf c}|(\overline{U}{\bf y})
       h_{\mu}(D{\bf y})|{\bf b}> \cdot{A}_b,\hspace{1.8cm}
       (4.7)\nonumber\\
    V_{td}\overline{\bf t}h_{\mu}{\bf d}
     &=&<{\bf t}|(\overline{U}{\bf y}{\bf y})h_{\mu}D|{\bf d}>
       \cong<{\bf t}|\overline{U}h_{\mu}D|{\bf d}>
       \cdot{2}A_{t}^{2},\hspace{3.4cm}(4.8)\nonumber\\
    V_{ts}\overline{\bf t}h_{\mu}{\bf s}
     &=&<{\bf t}|(\overline{U}{\bf y}{\bf y})h_{\mu}(D{\bf y})
       |{\bf s}>\cong<{\bf t}|(\overline{U}{\bf y})h_{\mu}
       (D{\bf y})|{\bf s}>\cdot{A}_t,\hspace{2.1cm}(4.9)\nonumber\\
    V_{tb}\overline{\bf t}h_{\mu}{\bf b}
     &=&<{\bf t}|(\overline{U}{\bf y}{\bf y})h_{\mu}
       (D{\bf y}{\bf y})|{\bf b}>,\hspace{6.4cm}(4.10)\nonumber     \label{4.10}
\end{eqnarray}
\setcounter{equation}{\value{enumi}}}
 where $V_{ij}$s are flavor-mixing coupling constants 
 (complex number in general)
 ,which correspond to CKM-matrices in the SM and \{ ${\bf u}$, 
 ${\bf d}$, ${\bf s}$, etc.\} in the left sides of 
 the equations are quark-mass eigenstates. 
 Here we need some explanations. 
 In transitions from the 3rd to the 1st 
 generation in (4.4;8) there are two types :  
 One is that two (${\bf y}\rightarrow{2}{\bf g}_h$)-processes 
 occur simultaneously and the other is that 
 ${\bf y}$ annihilates into $2{\bf g}_{h}$ in a cascade 
 way . Then let us describe the case of (4.4) as :
\begin{eqnarray}
   <{\bf u }|\overline{U}h_{\mu}(D{\bf y}{\bf y})|{\bf b}>
       &\cong&<{\bf u}|\overline{U}h_{\mu}D|{\bf b}>
       \cdot{A}_{b}^{2}+<{\bf u}|\overline{U}h_{\mu}(D{\bf y})
       |{\bf b}>\cdot{A}_{b}\nonumber\\
       &\cong&<{\bf u}|\overline{U}h_{\mu}D|{\bf b}>
              \cdot{A}_{b}^{2}+<{\bf u}|\overline{U}h_{\mu}D
              |{\bf b}>\cdot{A}_{b}^{2}\nonumber\\
       &=&<{\bf u}|\overline{U}h_{\mu}D|{\bf b}>
          \cdot{2}A_{b}^{2}.\hspace{5.9cm}(4.11)\nonumber                     \label{4.11}
\end{eqnarray}
 The case of (4.8) is also same as (4.11) . 
 If we admit the assumption of factorizability of (${\bf y}
 \rightarrow{2}{\bf g}_{h}$)-process, it is natural that the 
 universality of the net weak interactions among three generations 
 are realized. The net weak interactions are essentially same 
 as $({\bf u}\rightarrow{\bf d})$-transitions(Fig.(1)). 
 Then we may think that : 
{
\setcounter{enumi}{\value{equation}}
\addtocounter{enumi}{1}
\setcounter{equation}{0}
\renewcommand{\theequation}{\theenumi\alph{equation}}
\begin{eqnarray}
      \hspace{1cm}|<{\bf u}|\overline{U}h_{\mu}D|{\bf d}>| &\cong&
      |<{\bf u}|\overline{U}h_{\mu}D|{\bf s}>| \cong
      |<{\bf u}|\overline{U}h_{\mu}D|{\bf b}>|\hspace{2.6cm}\nonumber\\                            
                                               &\cong&
      |<{\bf c}|\overline{U}h_{\mu}D|{\bf d}>|\cong|
      <{\bf t}|\overline{U}h_{\mu}D|{\bf d}>|,\hspace{2cm}(4.12)\nonumber\\
      |<{\bf c}|(\overline{U}{\bf y})h_{\mu}(D{\bf y})|{\bf s}>|
                                              &\cong&
      |<{\bf c}|(\overline{U}{\bf y})h_{\mu}(D{\bf y})|{\bf b}>|\nonumber\\
                                              &\cong&
      |<{\bf t}|(\overline{U}{\bf y})h_{\mu}(D{\bf y})|{\bf s}>|,
      \hspace{4.7cm}(4.13)\nonumber                         
\end{eqnarray}
\setcounter{equation}{\value{enumi}}}     \label{4.13}

 and additionally we  may assume :
 {
\setcounter{enumi}{\value{equation}}
\addtocounter{enumi}{1}
\setcounter{equation}{0}
\renewcommand{\theequation}{\theenumi\alph{equation}}
\begin{eqnarray}
    \hspace{1cm}|<{\bf u}|\overline{U}h_{\mu}D|{\bf d}>|
                    &\cong&
      |<{\bf c}|(\overline{U}{\bf y})h_{\mu}(D{\bf y})|
                                {\bf s}>|\nonumber\\
      \hspace{1.5cm}&\cong&
      |<{\bf t}|(\overline{U}{\bf y}{\bf y})h_{\mu}
                   (D{\bf y}{\bf y})
      |{\bf b}>|.\hspace{4.2cm}(4.14)\nonumber 
 \end{eqnarray}
\setcounter{equation}{\value{enumi}}}
                                                                                  \label{4.14}
 In (4.13) and (4.14) ${\bf y}$-particles are the 
 `` $spectators$ '' for the weak interactions.
 \par
 Concerning the left sides of (4.2$\sim$4.10), 
 \{ $\overline{\bf u}h_{\mu}
 {\bf d}$, $\overline {\bf u}h_{\mu}{\bf s}$, etc.\} operate as 
 the current operators coupling to the  
 $W$-boson current when only weak interactions switch on. 
 In our subquark model we think that weak interactions 
 occur as the residual 
 ones among subquarks inside any kinds of quarks. 
 Therefore in this scenario, \{ $\overline{\bf u}h_{\mu}
 {\bf d}$, $\overline {\bf u}h_{\mu}{\bf s}$, etc.\} 
 act identically in the weak interactions.
  Then it seems natural to assume : 
      $$\hspace{1cm}\overline{\bf u}h_{\mu}{\bf d}
              =\overline{\bf u}h_{\mu}
       {\bf s}=\overline{\bf u}h_{\mu}{\bf b}
              =\overline{\bf c}
        h_{\mu}{\bf d}=\cdots.\hspace{3.0cm}\eqno{(4.15)}$$                   \label{4.15}
Namely they make equal operations as current operators 
because ${\bf y}$-particles work as spectators.
\par
 Using (4.1$\sim$4.10) and (4.12$\sim$4.15) we find :
{
\setcounter{enumi}{\value{equation}}
\addtocounter{enumi}{1}
\setcounter{equation}{0}
\renewcommand{\theequation}{\theenumi\alph{equation}}
\begin{eqnarray}
       \hspace{3cm}
      \frac{|V_{us}|}{|V_{ud}|}&=&
             |A_{s}|=\alpha^{q}_{W}(Q_{s}^{2})^{2}\cdot|B|,
             \hspace{5.3cm}(4.16)\nonumber\\                          \label{4.16}
      \frac{|V_{cd}|}{|V_{ud}|}&=&
             |A_{c}|=\alpha^{q}_{W}(Q_{c}^{2})^{2}\cdot|B|,
             \hspace{5.3cm}(4.17)\nonumber\\
      \frac{|V_{cb}|}{|V_{cs}|}&=&
             |A_{b}|=\alpha^{q}_{W}(Q_{b}^{2})^{2}\cdot|B|,
             \hspace{5.3cm}(4.18)\nonumber\\
      \frac{|V_{ts}|}{|V_{cs}|}&=&
             |A_{t}|=\alpha^{q}_{W}(Q_{t}^{2})^{2}\cdot|B|,
             \hspace{5.3cm}(4.19)\nonumber\\
      \frac{|V_{ub}|}{|V_{ud}|}&=&
             2|A_{b}|^{2}=2\{\alpha^{q}_{W}
             (Q_{b}^{2})^{2}\cdot|B|\}^2,\hspace{4.15cm}(4.20)
                                 \nonumber\\
      \frac{|V_{td}|}{|V_{ud}|}&=&
             2|A_{t}|^{2}=2\{\alpha^{q}_{W}
             (Q_{t}^{2})^{2}\cdot|B|\}^2.\hspace{4.2cm}(4.21)
                                 \nonumber                            \label{4.21}
\end{eqnarray}
\setcounter{equation}{\value{enumi}}}.
 Here let us discuss how the substructure dynamics 
 inside quarks generate quark masses. 
 In our composite model quarks are composed of 
 ${\bf \alpha}$, ${\bf x}$, ${\bf y}$. As seen in(3.14;15) 
  ${\bf c}$-quark is composed of 
 three subquarks; ${\bf t}$-quark : four subquarks; 
 ${\bf s}$-quarks : four subquarks; ${\bf b}$-quark : 
 five subquarks. As discussing in Section 3, 
 subquark masses are expected to be almost massless 
 and then it may be thought that
 quark masses are proportional to the sum of 
 the average kinetic energies of the subquarks 
 (denoted by $<T_{i}>$,\hspace{1mm}$i={\bf s}, 
 {\bf c},{\bf b},{\bf t}$). 
 The proportional constants (denoted by 
 $K_{s}\hspace{1mm}(s=up,down)$,
 which may depend on details of subquark dynamics)
 are assumed common in the up (down)-quark sector 
 and different between the up- and the down-quark sector 
 from the hierarchical pattern of quark masses. 
 According to `` Uncertainty Principle ''
 the kinetic energies of the constituent particles moving 
 inside the composite particle are, in general, inversely
 proportional to the radius of that composite particle. 
 The radii of quarks may be around $1/\Lambda_{L,R}$. 
 So the kinetic energies of subquarks may be more than 
 hundreds GeV. Therefore the masses of quarks 
 may essentially depend on two parts, namely the kinetic 
 energies of subquarks and such a large binding energies 
 as counterbalane them. The $<T_{i}>$ may be considered 
 in inverse proportion to the average interaction 
 length among subquarks (denoted by $<r_{i}>$). 
 Further, it is presumable that $\sqrt{Q^2_{i}}$($Q_{i}$ 
 is the effective four momentum of ${\bf g}_h$-exchange 
 among subquarks inside the $i$-quark seen in (4.1)) is 
 inversely proportional to $<r_{i}>$. 
\par
 Then we have :
{
\setcounter{enumi}{\value{equation}}
\addtocounter{enumi}{1}
\setcounter{equation}{0}
\renewcommand{\theequation}{\theenumi\alph{equation}}
\begin{eqnarray}
      \hspace{2cm}
      \frac{m_{b}}{m_{s}}
      &=&\frac{5K_{down}<T_{b}>}{4K_{down}<T_{s}>}
      =\frac{5}{4}\cdot\frac{<r_{s}>}{<r_{b}>}\nonumber\\
      &=&\frac{5}{4}\cdot\sqrt{\frac{Q_{b}^2}{Q_{s}^2}},
      \hspace{9cm}(4.22)\nonumber\\
      \frac{m_{t}}{m_{c}}
      &=&\frac{4K_{up}<T_{t}>}{3K_{up}<T_{c}>}
      =\frac{4}{3}\cdot\frac{<r_{c}>}{<r_{t}>}\nonumber\\
      &=&\frac{4}{3}\cdot\sqrt\frac{Q_{t}^2}{Q_{c}^2},
      \hspace{9cm}(4.23)\nonumber                                        \label{4.23}
\end{eqnarray}
\setcounter{equation}{\value{enumi}}}
 where $m_{i}$ is the mass of $i$-quark.
 In the Review of Particle Physics[29] we find :
 $m_{s}=0.095$ GeV; $m_{b}=4.2$ GeV; $m_{c}=1.25$ GeV; 
 $m_{t}=174.2$ GeV, by which we obtain $m_{b}/m_{s}=44.2$
 and $m_{t}/m_{c}=139.4$. Using them we get by (4.22;23) : 
{
\setcounter{enumi}{\value{equation}}
\addtocounter{enumi}{1}
\setcounter{equation}{0}
\renewcommand{\theequation}{\theenumi\alph{equation}}
\begin{eqnarray}
      \hspace{3.5cm}\frac{Q_{b}^2}{Q_{s}^2}&\cong&(35.4)^2,
      \hspace{7.9cm}(4.24)\nonumber\\
      \hspace{2cm}\frac{Q_{t}^2}{Q_{c}^2}&\cong&(104.5)^2.
      \hspace{7.8cm}(4.25)\nonumber                                \label{4.24;25}
\end{eqnarray}
\setcounter{equation}{\value{enumi}}}
 Note again  that it seems to be meaningless to estimate 
 $Q_{s}^2/Q_{t}^2$ or $Q_{c}^2/Q_{b}^2$ 
 because the up-quark sector and the down-quark sector 
 possibly have the different aspects of 
 substructure dynamics (that is $K_{up} \neq K_{down}$). 
\par
 Review of Particle Physics [29] has reported that : 
\begin{eqnarray}
         \hspace{1.5cm}
         |V_{ud}|&=&0.9735\pm0.0008,\hspace{1cm}|V_{us}|=0.2196
                    \pm0.0023,\nonumber\\
         |V_{cd}|\hspace{0.5mm}&=&0.224\pm0.016,\hspace{1.4cm}|
                    V_{cb}|=(41.6\pm0.6)
                   \times10^{-3},\hspace{2.3cm}(4.26)\nonumber\\
         |V_{cs}|\hspace{0.5mm}&=&1.04\pm0.16,\hspace{1.8cm}|V_{ub}|
                    =(3.84^{+0.67}_{-0.49})\times10^{-3},
                    \nonumber
                                                                           \label{4.26}
\end{eqnarray}
which are `` experimental '' results without unitarity assumption. 
\par
 Relating these data to the scheme of our composite model, 
 let us investigate the quark-flavor-mixing phenomena 
 in terms of the subquark dynamics. 
 Using (4.16;18) and $|V_{us}|$, 
 $|V_{cb}|$ in (4.26) we get :

 $$\hspace{3cm}\frac{\alpha^{q}_{W}(Q_{s}^{2})}{\alpha^{q}_{W}
 (Q_{b}^{2})}=2.30,\hspace{8.0cm}\eqno{(4.27)}$$         \label{4.27}
 where we assume $|V_{ud}|=|V_{cs}|$.
 Applying $N_{f}=N_{s}=4$ (as is stated in Section 3) 
 to (3.13) 
 we have :
  $$\hspace{2.2cm}b_{q}=0.35.\hspace{9.3cm}\eqno{(4.28)}$$          \label{4.28}
 Here we rewrite (3.12) as :
 $$ \hspace{2.3cm}\alpha^{q}_{W}(Q_{1}^{2})
  =\frac{1-\displaystyle{\frac{\alpha^{q}_{W}
 (Q_{1}^{2})}{\alpha^{q}_{W}(Q_{2}^{2})}}}
 {b_{q}\ln\left(\displaystyle{\frac{Q_{1}^{2}}
 {Q_{2}^{2}}}\right)}.
 \hspace{6.7cm}\eqno{(4.29)}$$                                               \label{4.29}
 Inserting the values of (4.24;27;28), into (4.29)
 we have :
     $$\hspace{2.3cm}\alpha^{q}_{W}(Q_{s}^{2})=0.520,
       \hspace{7.9cm}\eqno{(4.30)}$$                                     \label{4.30}
  where $Q_{s}$,$(Q_{b})$ corresponds to 
 $Q_{1}$,$(Q_{2})$ in (4.29). 
 Combining $|V_{ud}|$, $|V_{us}|$ in (4.26) and (4.30) 
 with (4.16) we obtain :
     $$\hspace{2.4cm}|B|=0.835,\hspace{8.6cm}\eqno{(4.31)}$$             \label{4.31}
 and using (4.30) to (4.27) we get :
          $$\hspace{2.3cm}\alpha^{q}_{W}(Q_{b}^{2})
            =0.226.\hspace{8cm}\eqno{(4.32)}$$                                \label{4.32}
 By use of $|V_{ud}|$, $|V_{cd}|$ in (4.26) and (4.31) 
 to (4.17) we have :
          $$\hspace{2.3cm}\alpha^{q}_{W}(Q_{c}^{2})
            =0.525.\hspace{7.9cm}\eqno{(4.33)}$$                              \label{4.33}
  Using (4.25; 28; 33) and setting $t$ ($c$) to $1$ ($2$) 
 in (4.29) we obtain :
     $$\hspace{2.3cm}\alpha^{q}_{W}(Q_{t}^{2})
     =0.197.\hspace{8cm}\eqno{(4.34)}$$                                \label{4.34}
 Inserting (4.31;32), to the right side of (4.20) we have :
     $$\hspace{2.2cm}|V_{ub}|
     =3.54\times10^{-3}.\hspace{7.6cm}\eqno{(4.35)}$$                     \label{4.35} 
 Comparing this with the experimental value 
 of $|V_{ub}|$ in (4.26) the consistency between them seems good. 
 \par Finally using (4.31;34) to (4.19;21) we predict :
     $$\hspace{1cm}|V_{ts}|=3.06\times10^{-2},\hspace{1.5cm}
     |V_{td}|=1.92\times10^{-3},\hspace{3.0cm}\eqno{(4.36)}$$            \label{4.36} 
 where we adopt $|V_{ud}|=|V_{cs}|=0.974$ from (4.26).
 \par Comparing the values of (4.36) with 
 $|V_{ts}|=0.039\pm0.004$ and  
 $|V_{td}|=0.0085\pm0.0045$ [29] obtained by assuming 
 unitarity with three generations, 
 we find that our results are rather smaller than them. 
 The origin of these results is presumably in the fact that 
 `` the top-quark mass is heavy. ''
 We wish the direct measurements of ($t\rightarrow{d},s$) 
 transitions in leptonic 
 and/or semileptonic decays of top-quark mesons .
 \par
 So far we have discussed absolute values of $V_{qq^{'}}$ 
 but they are `` $complex$ '' in principle 
 because $B$ (in (4.1)) is originated from ${\bf y}$-subquark 
 annihilation(creation) to(from) vacuum. 
 Therefore let us make the definition : 
 $$\hspace{2.5cm}<0|f({\bf \Lambda}{\cal O}
 \overline{\bf \Lambda}, \rm{and/or,} 
 {\bf \Theta}{\cal O}{\overline{\bf \Theta}})|{\bf y}>
 \equiv|B|e^{i{\theta}},\hspace{3.0cm}\eqno{(4.37)}$$                    \label{4.37}
 where ${\cal O}$ is some operator.
  Then preparing for discussions in following sections, 
 let the generation-changing 
 flavor-mixing-coupling-constants 
 of $V_{qq^{'}}$ in (4.3$\sim$5) and (4.7$\sim$9)
 be parametrized as :
{  
\setcounter{enumi}{\value{equation}}
\addtocounter{enumi}{1}
\setcounter{equation}{0}
\renewcommand{\theequation}{\theenumi\alph{equation}}

\begin{eqnarray}
      \hspace{1.7cm}V_{us}&=&{\lambda}e^{i\delta}\hspace{1.3cm}
      V_{cb}={\lambda}^{2}e^{i\delta}
      \hspace{1.6cm}V_{ub}={\lambda}^{3}e^{i{\delta}^{'}},\nonumber\\
      V_{cd}&=&{\lambda}e^{-i{\delta}}\hspace{1.1cm}
      V_{ts}={\lambda}^{2}e^{-i{\delta}}
      \hspace{1.4cm}V_{td}={\lambda}^{3}e^{-i{\delta}^{'}},
      \hspace{3.15cm}(4.38)\nonumber                                           \label{4.38}
\end{eqnarray}

\setcounter{equation}{\value{enumi}}}
 here $\delta(\delta^{'})$corresponds to one(two) ${\bf y}$-
 subquark(s) creation from vacuum and $-\delta(-\delta^{'})$ 
 one(two) ${\bf y}$-subquark(s) annihilation to vacuum 
 and we use $\lambda=0.22$ from Wolfenstein's 
 parametrization[70]. In case of (4.2;6;10)
 ${\bf y}$-subquark is a `` spectator '' 
 and then $V_{qq^{'}}$ are real, which we set for simplicity :               
         $$\hspace{3.5cm}V_{ud}=V_{cs}=V_{tb}=1.
         \hspace{7.9cm}\eqno{(4.39)}$$                              \label{4.39}
 It is important to note that $V_{qq^{'}}$s are different 
 from the CKM matrix elements. They do not necessarily 
 satisfy the strict unitarity-condition with three 
 generations because in our composite model the 
 generating-mechanism of `` generation '' originates from 
 subquark dynamics (adding ${\bf y}$-subquarks). 
 The seeming disappearance of the forth generation may 
 be caused by imbalance of kinetic and binding energies 
 of internal subquark system inside quarks. 
\section{Mass difference ${\Delta}M_P$
 by $P^{0}-\overline{P^{0}}$ mixing }
\subsection
{\bf Historical summary}

\par
 Mass difference ${\Delta}M_P$ originates from the mixing 
 between a neutral pseudo scalar meson ($P^0$) and its 
 antimeson ($\overline{P^{0}}$). There are six types of mixing 
 , e.g., $K^{0}-\overline{K^{0}}$, 
 $D^{0}-\overline{D^{0}}$, $B^{0}_{d}-\overline{B^{0}_{d}}$, 
 $B^{0}_{s}-\overline{B^{0}_{s}}$, $T^{0}_{u}-
 \overline{T^{0}_{u}}$ and $T^{0}_{c}-\overline{T^{0}_{c}}$ 
 mixings. 
 Theoretically they have been considered to be one of the most 
 sensitive probes of higher-order effects of the weak 
 interactions in the SM. 
 The basic tool to investigate them is the `` box diagram ''. 
 By using this diagram to the $K_{L}$-$K_{S}$ mass 
 difference, Gaillard and Lee predicted the mass of the 
 charm quark[37]. Later, Wolfenstein suggested that the 
 contribution of the box diagram which is called the 
 short-distance (SD) contribution cannot supply the whole 
 of the mass difference ${\Delta}M_K$ and there are 
 significant contributions arising from the long-distance 
 (LD) contributions associated with low-energy 
 intermediate hadronic states[38]. 
 As concerns, the LD-phenomena occur in the energy range 
 of few hundred MeV and the SD-phenomena around $100$ GeV 
 region.
  Historically there are various investigations for 
 $P^{0}$-$\overline{P^{0}}$ mixing problems[36][39$\sim$48] 
 and many authors have examined them by use of LD- and 
 SD-contributions. 
 \par
 In summary, the comparison between the theoretical results 
 and the experiments about ${\Delta}M_P$ 
 ($P=K,D$ and $B_d$) are as follows :
  
{\setcounter{enumi}{\value{equation}}
\addtocounter{enumi}{1}
\setcounter{equation}{0}
\renewcommand{\theequation}{\theenumi\alph{equation}}
\begin{eqnarray}
 \hspace{2cm}{\Delta}M^{LD}_{K}&\approx&{\Delta}M^{SD}_{K}\approx
 {\Delta}M^{exp}_{K},\hspace{7cm}(5.1)\nonumber\\
 {\Delta}M^{SD}_{D}&\ll&{\Delta}M^{LD}_{D}
 (\ll{\Delta}M^{exp}_{D},\rm{upper}\hspace{2mm}\rm{bound}),
 \hspace{4.15cm}(5.2)\nonumber\\
 {\Delta}M^{LD}_{B_d}&\ll&{\Delta}M^{SD}_{B_d}\simeq
 {\Delta}M^{exp}_{B_d}.\hspace{7.1cm}(5.3)\nonumber                  \label{5.1;2;3}
\end{eqnarray}
\setcounter{equation}{\value{enumi}}}
 Concerning (5.1) it is explain that ${\Delta}M_{K}=
 {\Delta}M^{SD}_{K}+m{\Delta}M^{LD}_{K}$ where `` $m$ '' is 
 a numerical value of order $O(1)$. As for (5.3), 
 they found that ${\Delta}M^{LD}_{B_d}\approx10^{-16}$ GeV 
 and ${\Delta}M^{SD}_{B_d}\approx10^{-13}$ GeV, 
 then the box diagram is the most important 
 for $B^{0}_{d}$-$\overline{B^{0}_{d}}$ mixing. 
 Computations of ${\Delta}M^{SD}_{B_d}$ and 
 ${\Delta}M^{SD}_{B_s}$ from the box diagrams 
 in the SM give : 
  $$\frac{{\Delta}M^{SD}_{B_s}}{{\Delta}M^{SD}_{B_d}}\simeq
 \left|\frac{{\cal{V}}_{ts}}{{\cal{V}}_{td}}\right|^2
 \frac{B_{B_s}{f}^2_{B_s}}
 {B_{B_d}{f}^2_{B_d}} \frac{M_{B_s}}{M_{B_d}}\zeta,
 \hspace{6cm}\eqno{(5.4)}$$  \label   {5.4}                                                                                             where ${\cal{V}}_{ij}$s stand for CKM matrix elements; $M_P$ :
 P-meson mass; $\zeta$ : a QCD correction of order $O(1)$; 
 $B_B$ : Bag factor of B-meson and $f_B$ : decay constant 
 of B-meson.
 Measurements of ${\Delta}M^{exp}_{B_d}$ and 
 ${\Delta}M^{exp}_{B_s}$ are, therefore, said to be useful 
 to determine $|{\cal{V}}_{ts}/{\cal{V}}_{td}|$[49][50]. 
 Concerning (5.2), they found that ${\Delta}M^{LD}_{D}
 \approx10^{-15}$ GeV and ${\Delta}M^{SD}_{D}
 \approx10^{-17}$ GeV[36][44] but the experimental 
 measurement is ${\Delta}M^{exp}_{D}<4.6\times10^{-14}$ 
 GeV with CL=$90\verb+%+$ [29]. 
 Further there is also a study that ${\Delta}M^{LD}_D$ is 
 smaller than $10^{-15}$ GeV by using the heavy quark 
 effective theory[45]. 
 Then many people state that it would be a signal of new 
 physics beyond the SM if the future experiments confirm 
 that ${\Delta}M^{exp}_D\simeq10^{-14}\sim10^{-13}$ 
 GeV[39$\sim$45][60].
 \par
 On the other hand some researchers have studied these 
 phenomena in the context of the theory explained by 
  the single dynamical origin. 
 Cheng and Sher[68], Liu and Wolfenstein[47], and G\'erard 
 and Nakada[48] have thought that all 
 $P^{0}$-$\overline{P^0}$ mixings occur only by the 
 dynamics of the TeV energy region which is essentially 
 the same as the idea of Super-Weak (denoted by SW) 
 originated by Wolfenstein[35]. 
 They extended the original SW-theory (which explains indirect 
 CP violation in the $K$-meson system) to other flavors by 
 setting the assumption that some neutral $spin\hspace{2mm}0$ 
 particle with a few TeV mass 
 (denoted by $H$) contributes to the `` real part '' of 
 $M_{ij}$ which determines ${\Delta}M_P$ and also the 
 `` imaginary part '' of $M_{ij}$ which causes the indirect 
 CP violation. 
 The ways of extensions are that $H$-particles couple to 
 quarks by the coupling proportional to $\sqrt{{m}_i
 {m}_j}$[47][68], $({m}_i/{m}_j)^n\hspace
 {2mm}n=0,1,2$[47] and $({m}_i+{m}_j)$[48] 
 where $i,j$ are flavors of quarks coupling to $H$. 
 It is suggestive that the SW-couplings depend on quark 
 masses (this idea of `` mass-dependence ''is adopted in 
 our model discussed below). 
 Cheng and Sher[68] and Liu and Wolfenstein[47] 
 obtained that ${\Delta}M_D=({m}_c/{m}_s)
 {\Delta}M^{exp}_{K}\approx10^{-14}$ GeV with the 
 assumption that $H$-exchange mechanism saturates the 
 ${\Delta}M^{exp}_K$ bound, which is comparable to 
 ${\Delta}M^{exp}_{D}<4.6\times10^{-14}$ GeV[29].
 \par
 Concerning $B$-meson systems they found that 
 ${\Delta}M_{B_s}/{\Delta}M_{B_d}={m}_s/{m}_d
 \simeq20$. But from the experimental data we have
 $({\Delta}M_{B_s}/{\Delta}M_{B_d})_{exp}=36.8$ [29][62]. 
  Further using their scheme it is calculated that : 
$$\hspace{1.5cm}
 \frac{{\Delta}M_{B_d}}{{\Delta}M_K}=
 \frac{B_{B_d}{f}^2_{B_d}}{B_K{f}^{2}_{K}}
 \frac{M_{B_d}}{M_{K}}
 \frac{m_b}{m_s}
 \simeq300,\hspace{5.6cm}\eqno{(5.5)} $$                      \label{5.5}
 where we use $m_b=4.3$ GeV, $m_s=0.2$ GeV, 
 $M_{B_d}=5.279$ GeV, $M_K=0.498$ GeV, 
 $B_{B_d}{f}^2_{B_d}=(0.22{\rm GeV})^2$, 
 $B_K{f}^2_K=(0.17{\rm GeV})^2$. 
 This is larger than 
 $({\Delta}M_{B_d}/{\Delta}M_K)_{exp}=89$ [29] and 
 is caused by large b-quark mass value.

\subsection
   {\bf $P^{0}-\overline{P^{0}}$ mixing by subquark dynamics}
 \par
 Various ideas discussed in former subsection seem to be 
 hard to explain all mass defferences as a whole. 
 So in order to overcome this difficulty
 let us discuss $P^0$-$\overline{P^0}$ mixings by 
 using our subquark model. The discussions start from the 
 assumption : Mass mixing matrix $M_{ij}(P)$ $(i(j)=
 1(2)$ denotes $P^0(\overline{P^0}))$ is described only by the 
 `` {\bf y}-exchange '' interactions causing 
 $P^{0}-\overline{P^{0}}$ transitions. 
 We calculate ${\Delta}M_P$ as :
{  
\setcounter{enumi}{\value{equation}}
\addtocounter{enumi}{1}
\setcounter{equation}{0}
\renewcommand{\theequation}{\theenumi\alph{equation}}
\begin{eqnarray}
     \hspace{2.8cm}M_{12}(P)&=&
     <\overline{P^0}|{\cal H}^{\bf y}|P^0>,
     \hspace{6.3cm}(5.6)\nonumber\\
     \hspace{2cm}{\Delta}M_P\hspace{2mm}&=&M_H-M_L
     \simeq2|M_{12}(P)|,
     \hspace{4.77cm}(5.7)\nonumber                             \label{5.6;7}
\end{eqnarray}
\setcounter{equation}{\value{enumi}}}
 where ${\cal H}^{\bf y}$ is Hamiltonian for 
 $P^0$-$\overline{P^0}$ transition interaction 
 by ${\bf y}$-exchange and we assume 
 $ImM_{12}\ll{ReM_{12}}$ which is 
 experimentally acceptable[36][52] and $M_{H(L)}$ stands 
 for heavier (lighter) $P^0(\overline{P^{0}})$-meson mass.\par
 Applying the vacuum-insertion calculation to the hadronic 
 matrix element as : $<\overline{P^0}|[\overline{{q}_{i}}
 \gamma_{\mu}(1-\gamma_5){q}_j]^2|P^0>\sim
 B_P{f}^2_{P}M^2_P$ [36] we get : 
 $$ \hspace{2cm}
    M_{12}(P)=\frac{1}{12\pi^2}B_P{f}^2_{P}
         M_{P}{\cal M}_P,\hspace{6.3cm}\eqno{(5.8)}$$           \label{5.8}
  here  ${\cal M}_P$ is a matrix element contributed by ${\bf y}$-
 exchange diagram (which is seen in Fig.(3)).  
 $P^0$-$\overline{P^0}$ mixings occur due to 
 `` {\bf y}-exchange '' between two quarks inside the present 
 $P^0(\overline{P^{0}})$-meson. 
 This is a kind of the realizations of Wolfenstein's 
 SW-idea [35]. The schematic illustration is as follows : 
 two particles ( that is, quarks) with radius order of 
 $1/{\Lambda}_q$ (maybe a few ${\rm TeV}^{-1}$) are moving  
 inside a sphere ( that is, meson) with radius order of 
 ${\rm GeV}^{-1}$. 
 The ${\bf y}$-exchange interactions would occur when 
 two quarks inside $P^0(\overline{P^{0}})$-meson 
 interact in contact with 
  each other because ${\bf y}$-particles are confined 
 inside quarks. 
 As seen in Fig.(3), the contributions of 
 ${\bf y}$-exchanges are common among various 
 $P^0(\overline{P^{0}})$-mesons. \par 
 Then we set the assumption :
 \newline $\;$ \newline 
 \begin{em} \hspace{2cm}
      Universality of the ${\bf y}$-exchange interactions,
 \end{em}
 \newline $\;$ \newline 
 which means that these interactions
 are independent of a variety of quarks.
 \par
 Further let us describe ${\cal M}_P$ as :
 $$\hspace{2cm}
        {\cal M}_P={n}_P{\eta}(P)
        {\tilde {\cal M}}_{l}(P),\hspace{7cm}\eqno{(5.9)}$$   \label{5,9}
  where $n_P=1$ for $P=K, D, B_d, T_u$; $n_P=2$ for $P=B_s, 
 T_c$, $l=1$ for $K, D, B_s, T_u$; $l=2$ for $B_d, T_c$ and
 ${\tilde {\cal M}}_{l}(P)$ is the `` net '' matrix element of
 ${\bf y}$-exchange interaction.
 \par
 `` Universality '' means explicitly that :
\begin{eqnarray}
 \hspace{2cm}
 {\tilde {\cal M}}_{1}(K)&=&{\tilde {\cal M}}_{1}(D)=
 {\tilde {\cal M}}_{1}(B_s)={\tilde {\cal M}}_{1}(T_c),
 \nonumber\\
 &=&{\tilde {\cal M}}_{2}(B_d)
 ={\tilde {\cal M}}_{2}(T_u).\hspace{6.3cm}(5.10)
 \nonumber                                                 \label{5.10}
\end{eqnarray}
 The explanation of ${n}_P$ is such that $K$ and $D$ 
 have one ${\bf y}$-particle and one ${\bf y}$-particle 
 exchanges; $B_d$ and $T_u$ have two ${\bf y}$-particles 
 and both of them exchange simultaneously, so 
 we set ${n}_P=1$ for them. On the other hand 
$B_s$ and $T_c$ have two 
 ${\bf y}$-particles but one of them exchanges, 
 so they have ${n}_P=2$ because the probability becomes 
 double. 
 The `` ${l}$ '' means the number of exchanging 
 ${\bf y}$-particles in the present diagram.
\par 
 Concerning ${\eta}(P)$, we explain as follows : 
 In our FB-model $P^0$-$\overline{P^0}$ mixing occurs by 
 the `` contact interaction '' of two quarks colliding 
 inside $P^0(\overline{P^0})$-meson. 
 Therefore the probability of this interaction may be 
 considered inverse proportional to the volume of the 
 present $P^0(\overline{P^{0}})$-meson, e.g., 
 the larger radius $K$-meson 
 gains the less-valued probability of the colliding than 
 the smaller radius $D$- (or $B_s$-) meson. 
 The various aspects of hadron dynamics seem to be 
 successfully illustrated by the semi-relativistic 
 picture with `` Breit-Fermi Hamiltonian ''[28]. 
 Assuming the power-law potential 
 $V(r)\sim{r}^{\nu}$($\nu$ is a real number), the radius 
 of $P^0(\overline{P^{0}})$-meson (denoted by ${\bf r}_P)$ is  
 proportional to ${\mu}_P^{-1/(2+\nu)}$, where 
 ${\mu}_P$ is the reduced mass of two quark-masses inside 
 $P^0(\overline{P^{0}})$-meson [28]. 
 Then the volume of $P^0(\overline{P^{0}})$-meson is 
 proportional to ${\bf r}_P^{3}\sim{\mu}_P^{-3/(2+\nu)}$. 
 After all we could assume for ${\eta}(P)$ in (5.9) as :
{  
\setcounter{enumi}{\value{equation}}
\addtocounter{enumi}{1}
\setcounter{equation}{0}
\renewcommand{\theequation}{\theenumi\alph{equation}}
\begin{eqnarray}
 \hspace{1cm}{\eta}(P)&=&{\xi}(\frac{{\mu}_P}
 {{\mu}_K})^{1.0}\hspace{2cm}{\rm for\hspace{3mm}
 linear-potential},\hspace{4.0cm}(5.11)\nonumber\\
 &=&{\xi}(\frac{{\mu}_P}{{\mu}_K})^{1.5}
 \hspace{2cm}{\rm for\hspace{8mm}
 log-potential},\hspace{4.0cm}(5.12)\nonumber                 \label{5.11;12}
\end{eqnarray}
\setcounter{equation}{\value{enumi}}}
 where $\xi$ is a dimensionless numerical factor 
 depending on the details of the dynamics 
 of the quark-level. The  
 ${\eta}(P)$ is normalized by ${\mu}_K$ (reduced mass 
 of $s$- and $d$-quark in $K$ meson) for convenience. 
\par 
 The present experimental results of $\Delta M_P$ are as 
 follows [29][51][62] : 
{  
\setcounter{enumi}{\value{equation}}
\addtocounter{enumi}{1}
\setcounter{equation}{0}
\renewcommand{\theequation}{\theenumi\alph{equation}}
\begin{eqnarray}
 \hspace{1.9cm}{\Delta}M_{K}^{exp}&=&(3.489\pm0.008) \times 10^{-15}
                 \hspace{2cm}{\rm GeV},
                 \hspace{2.93cm}(5.13)\nonumber\\
 {\Delta}M_{D}^{exp}&<&4.6 \times 10^{-14}
                 \hspace{4.2cm}{\rm GeV},
                 \hspace{2.93cm}(5.14)\nonumber\\
 {\Delta}M_{B_d}^{exp}&=&(3.12 \pm 0.11) \times 10^{-13}
                 \hspace{2.4cm}{\rm GeV},
                 \hspace{3cm}(5.15)\nonumber\\
{\Delta}M_{B_s}^{exp}&=&11.47 \times 10^{-12}
                 \hspace{3.85cm}{\rm GeV}.
                 \hspace{3cm}(5.16)\nonumber         \label{5.13;14;15;16}
\end{eqnarray}
\setcounter{equation}{\value{enumi}}}
 Using (5.7;8) and (5.13$\sim$16), we have :
{  
\setcounter{enumi}{\value{equation}}
\addtocounter{enumi}{1}
\setcounter{equation}{0}
\renewcommand{\theequation}{\theenumi\alph{equation}}
\begin{eqnarray}
   \hspace{1.8cm}|{\cal M}_D|&<& \hspace{2mm}1.4|{\cal M}_K|,
                  \hspace{8.6cm}(5.17)\nonumber\\
 |{\cal M}_{B_d}| &=&   4.92 |{\cal M}_K|,
                  \hspace{8.6cm}(5.18)\nonumber\\
 |{\cal M}_{B_s}| &=&   142 |{\cal M}_K|.
                  \hspace{8.8cm}(5.19)\nonumber       \label{5.17;18;19}
\end{eqnarray}
\setcounter{equation}{\value{enumi}}}
\par
 At the level of ${\cal M}_P$, it seems that : 
 $$ \frac{|{\cal M}_P|}{|{\cal M}_K|}
      \hspace{2.5mm}\simeq\hspace{2.5mm}
    O(1) \sim O(100),\hspace{7.7cm}\eqno{(5.20)}$$               \label{5.20}
 where $P=D, B_d, B_s$. 
\par 
Here adopting `` ${\tilde {\cal M}}_{l}(P)$ '' instead 
of ${\cal M}_P$, let us make following discussions.
By use of (5.9$\sim$12) and (5.18) we obtain :
{  
\setcounter{enumi}{\value{equation}}
\addtocounter{enumi}{1}
\setcounter{equation}{0}
\renewcommand{\theequation}{\theenumi\alph{equation}}
\begin{eqnarray}
 \hspace{0.8cm}
 {\mu}_{B_d}  &=&  4.91{\mu}_K
                \hspace{3cm}{\rm for\hspace{3mm}
 linear-potential},\hspace{3.6cm}(5.21)\nonumber\\
              &=&  2.88{\mu}_K
                \hspace{3cm}{\rm for\hspace{8mm}
 log-potential},\hspace{3.6cm}(5.22)\nonumber                     \label{5.21;22}           
\end{eqnarray}
\setcounter{equation}{\value{enumi}}}
 where $B_{B_d}{f}^2_{B_d}=(0.22{\rm GeV})^2$, 
 $B_K{f}^2_K=(0.17{\rm GeV})^2$ are used. 
 This result does not seem `` $unnatural$ ''.
 Comparing with the case of (5.5), 
 we can evade the large enhancement by $b$-quark mass 
 effect. 
 This is because the quark mass dependence is  
 introduced through the `` $reduced\hspace{2mm} mass$ ''
  (in which the 
 effect of heavier mass decreases). 
 Some discussions are as follows : 
 If we adopt the pure non-relativistic picture 
 it may be that 
 ${\mu}_K \simeq {\mu}_{B_d} \simeq {m}_d
 \simeq ({\mu}_D \simeq {\mu}_{T_u})$ but from the 
 semi-relativistic standpoint it seems preferable that 
 ${\mu}_K (< {\mu}_{D}) < 
  {\mu}_{B_d}(<{\mu}_{T_u})$ because the effective 
 mass value of `` $d$-quark '' in $B_d$-meson is considered 
 larger than that in $K$-meson, which  
 may be caused by that the kinetic energy of 
 `` $d$-quark '' in $B_d$-meson is larger than that in 
 $K$-meson (Refer to the discussion in Section 3).
 Then we can expect the plausibility of (5.21;22). 
 \par
 Next, let us discuss ${\Delta}M_D$. 
 Here we write ${\Delta}M^{\bf y}_{P}$ as the mass difference 
 of $P^{0}$and $\overline{P^{0}}$ by ${\bf y}$-exchange 
 interaction.
 \par
 Using (5.7$\sim$10) we obtain :
   $$\hspace{2cm}{\Delta}M_{D}^{\bf y}= 
      \frac{B_{D}{f}^2_{D}}{B_K{f}^2_K}\frac{M_{D}}{M_K}
      \frac{{\eta}(B_s)}{{\eta}(K)}
      {\Delta}M_K^{\bf y},\hspace{5.5cm}\eqno{(5.23)}$$      \label{5.23}     
 If we set ${\mu}_D={\mu}_K$ tentatively 
 in (5.11;12) we obtain :
 $$\hspace{2.1cm}{\Delta}M_D^{\bf y}=4.67 \times 
       {\Delta}M_{K}^{exp}= 1.6  \times 10^{-14}
       \hspace{1cm} {\rm GeV},\hspace{2.7cm}\eqno{(5.24)}$$   \label{5.24}
 where we assume ${\Delta}M_{K}={\Delta}M_{K}^{exp}$ 
 in (5.13) and use $B_D{f}_D^2=(0.19 {\rm GeV})^2$.
 In the same way, assuming 
 ${\mu}_D=1.5 \times {\mu}_K$ for example we have :
 $$ \hspace{2.3cm}{\Delta}M_D^{\bf y}
  =(2.9 \sim 5.4) \times 10^{-14}\hspace{2.6cm}
   {\rm GeV},\hspace{2.65cm}\eqno{(5.25)}$$                     \label{5.25}
where the parenthesis means that (linear-potential $\sim$ 
log-potential).
 This result is consistent and comparable with (5.14).
 These values are similar to the results 
 by Cheng and Sher [68] and Liu and Wolfenstein [47].
 Concerning compilation of various studies about 
 ${\Delta}M_D$, see Ref.[53]. 
\par 
 The study of ${\Delta}M_{B_s}$ is as follows. 
 Both $s$- and $b$-quark in $B_s$-meson are rather massive 
 and then supposing availability of the non-relativistic 
 scheme we have :
 $$\hspace{3cm}{\mu}_{B_s}
    =\frac{{m}_s {m}_b}{{m}_s+{m}_b}
    =0.19 \hspace{3.0cm}{\rm GeV},
    \hspace{2.5cm}\eqno{(5.26)}$$                               \label{5.26}
 where ${m}_s=0.2$ GeV and ${m}_b=4.3$ GeV
 are used.
 If we adopt ${\mu}_K=0.01$ GeV($\simeq{m}_d$) 
 for example we obtain :
 {  
\setcounter{enumi}{\value{equation}}
\addtocounter{enumi}{1}
\setcounter{equation}{0}
\renewcommand{\theequation}{\theenumi\alph{equation}}
\begin{eqnarray}
      \hspace{2.8cm}{\eta}(B_s) &=& 19.0 {\xi}
\hspace{2cm}{\rm for\hspace{3mm}linear-potential},
             \hspace{2.5cm}(5.27)\nonumber\\
             &=& 82.8 {\xi}
\hspace{2cm}{\rm for\hspace{8mm}log-potential},
             \hspace{2.5cm}(5.28)\nonumber                        \label{5.27;28}
\end{eqnarray}
\setcounter{equation}{\value{enumi}}}
   By using (5.7$\sim$10) we have :
   $$\hspace{2cm}{\Delta}M_{B_s}^{\bf y}= 
 \frac{2B_{B_s}{f}^2_{B_s}}{B_K{f}^2_K}
 \frac{M_{B_s}}{M_K}
 \frac{{\eta}(B_s)}{{\eta}(K)}
 {\Delta}M_K^{\bf y},\hspace{5.3cm}\eqno{(5.29)}$$                       \label{5.29}
 where factor 2 comes from $n_{B_s}=2$ in (5.9).
 Assuming that ${\Delta}M_K^{\bf y}={\Delta}M_K^{exp}$ 
 in (5.13) and using (5.26;27) we obtain :
   $$\hspace{2.2cm} {\Delta}M_{B_s}^{\bf y}=
       (3.1\sim14)\times10^{-12}\hspace{3.0cm}{\rm GeV},
       \hspace{2.7cm}\eqno{(5.30)}$$                               \label{5.30}
 where we use $B_{B_s}{f}^2_{B_s}=(0.25{\rm GeV})^2$[49]
 (the parenthesis means the same as (5.25)).
 \par
 This estimation is consistent with (5.16) and note that 
 it is obtained by inputting the information 
 of ${\Delta}M_K^{exp}$.
  \par  
 Finally let us estimate ${\Delta}M_{T_u}^{\bf y}$ and 
 ${\Delta}M_{T_c}^{\bf y}$. 
 Setting ${\mu}_{T_u}={\mu}_{B_d}$
 (though ${\mu}_{T_u}>{\mu}_{B_d}$ in practice)  
 and using (5.7$\sim$10) we find : 
      $$ \hspace{2.2cm}{\Delta}M_{T_u}^{\bf y}= 
 \frac{B_{T_u}{f}^2_{T_u}}{B_{B_d}{f}^2_{B_d}}
 \frac{M_{T_u}}{M_{B_d}}{\Delta}M_{B_d}^{\bf y}
      =7.3 \times 10^{-10}\hspace{0.6cm}{\rm GeV},
      \hspace{2cm}\eqno{(5.31)}$$                                    \label{5.31} 
 where we use $B_{T_u}{f}^2_{T_u}=(1.9{\rm GeV})^2$ [36], 
 $M_{B_d}=5.279$ GeV, $M_{T_u}=171$ GeV and set 
 ${\Delta}M_{B_d}^{\bf y}={\Delta}M_{B_d}^{exp}$ in (5,15).
 \par 
 For evaluating ${\Delta}M_{T_c}$, we calculate :
 $$ \hspace{2.9cm}{\mu}_{T_c}
    =\frac{{m}_c {m}_t}{{m}_c+{m}_t}
    =1.34 \hspace{3.8cm}{\rm GeV},\hspace{1.1cm}
    \eqno{(5.32)}$$                                                   \label{5.32}
 where ${m}_c=1.35$ GeV and ${m}_t=170$ GeV are used.
 \par
 Then from (5.11;12) we get :
 {  
\setcounter{enumi}{\value{equation}}
\addtocounter{enumi}{1}
\setcounter{equation}{0}
\renewcommand{\theequation}{\theenumi\alph{equation}}
\begin{eqnarray}
 \hspace{2.4cm}{\eta}(T_c) &=& 134 {\xi}
\hspace{1.8cm}{\rm for\hspace{3mm}linear-potential},
       \hspace{3.2cm}(5.33)\nonumber \\
          &=& 1551 {\xi}
\hspace{1.6cm}{\rm for\hspace{7.5mm}log-potential},
       \hspace{3.2cm}(5.34)\nonumber                                 \label{5.33; 34}
\end{eqnarray}
\setcounter{equation}{\value{enumi}}}
 where we set ${\mu}_K=0.01$ GeV.
 \par
 With (5.7$\sim$10) and (5.31;32) we obtain :
    $$\hspace{1.5cm}{\Delta}M_{T_c}^{\bf y}= 
 \frac{2B_{T_c}{f}^2_{T_c}}{B_K{f}^2_K}
 \frac{M_{T_c}}{M_{K}}
 \frac{{\eta}(T_c)}{{\eta}(K)}
      {\Delta}M_K^{\bf y}
      =(4\sim47) \times 10^{-8}\hspace{2mm}{\rm GeV},
      \hspace{0.9cm}\eqno{(5.35)}$$                                        \label{5.35} 
 where we adopt $n_{T_c}=2$, $B_{T_c}{f}^2_{T_c}=
 (1.9{\rm GeV})^2$[36], $M_{T_u}=171$ GeV and 
 ${\Delta}M_K^{\bf y}={\Delta}M_K^{exp}$ in (5.13)
 and the parenthesis means the same as (5.24).

\section
  { Indirect CP violation in $P^0$-$\overline{P^0}$ mixing}
 Here we discuss indirect CP violation by mass-mixings 
 which is assumed 
 to be saturated by the `` ${\bf y}$-exchange interactions ''. 
 In the CP-conserving limit in the $P^0
 (\overline{P^{0}})$-meson systems, 
 $M_{12}(P)$s are supposed to be real positive. 
 Note that $CP|P_H>=-|P_H>$ and 
 $CP|P_L>=|P_L>$ where $H$ $(L)$ means heavy (light). 
 If the CP-violating ${\bf y}$-exchange interactions are 
 switched on, $M_{12}(P)$ becomes complex.
\par 
 Following G\'erard and Nakada's notation [48][52],
  we write as : 
  $$ \hspace{2.5cm}M_{12}=|M_{12}|\exp(i{\theta}_P),
       \hspace{7cm}\eqno{(6.1)}$$                                \label{6.1}
 where ${\theta}_P$ is defined by :
  $$ \hspace{2.5cm}\tan{\theta}_P
     =\frac{{\rm{Im}}M_{12}(P)}{{\rm{Re}}M_{12}(P)}.
      \hspace{7.2cm}\eqno{(6.2)}$$                                  \label{6.2}
 As we assume that the ${\bf y}$-exchange interaction 
 saturates indirect CP violation, we can write :
    $$ \hspace{2.5cm}{\rm{Im}}
       <\overline{P^0}|{\cal H}^{\bf y}|P>
          ={\rm{Im}}M_{12}(P).\hspace{5.4cm}\eqno{(6.3)}$$        \label{6.3}
 From (5.6;8;9) we obtain :  
    $$ \hspace{2.6cm}{\rm{Im}}M_{12}(P)={\cal C} 
       \cdot {\rm{Im}}{\tilde {\cal M}}_{l}(P),
       \hspace{6.2cm}\eqno{(6.4)}$$                                 \label{6.4}
 where ${\cal C}=
   (1/12\pi^2)B_P{f}^2_PM_P{\eta}(P)$. 
 Therefore the origin of indirect CP violation of 
 $P^0(\overline{P^{0}})$-meson system is only 
 in ${\tilde {\cal M}_{l}(P)}$. 
 The Factor `` ${\cal C}$ '' in (6.4) is common also in 
 ${\rm{Re}}M_{12}(P)$ and then we have :
   $$  \hspace{2.6cm}\frac{{\rm{Im}}M_{12}(P)}
     {{\rm{Re}}M_{12}(P)}=
     \frac{{\rm{Im}}{\tilde {\cal M}}_{l}(P)}
     {{\rm{Re}}{\tilde {\cal M}}_{l}(P)}.
     \hspace{7cm}\eqno{(6.5)}$$                                     \label{6.5}
 If the universality of (5.10) is admitted , we obtain :
     $$\hspace{2.5cm}{\theta}_K
          ={\theta}_D={\theta}_{B_d}={\theta}_{B_s}
          = {\theta}_{T_c}={\theta}_{T_u}.
          \hspace{5.3cm}\eqno{(6.6)}$$                                \label{6.6}
 \par         
 These are the predictions about indirect CP violation from 
 the stand point of our subquark-model.  
 As for the experimental result it is reported that[47] :
   $$ \hspace{2.6cm}{\theta}_K
         =(6.5\pm0.2) \times 10^{-3}.
         \hspace{7cm}\eqno{(6.7)}$$                  \label{6.7}
  
\section{\hspace{0.5cm}\bf  Direct  and mixing-induced CP 
         Violation by \newline \hspace{0.5cm}Subquark Dynamics}
\hspace*{\parindent}         
In our model direct and mixing-induced CP violations occur 
by subquark dynamics same as indirect CP violations discussed         
in Section 6 and essentially originate from the phases :         
`` $\delta$ '' and `` $\delta^{'}$ '' which appeared in          
(4.38). Recently experimental measuements of various CP-         
asymmetries are already available, which are $B_d$-meson decays           
at KEK and SLAC and $B_s$-meson decays at Fermilab. By use          
of these experiments we can get the informations about          
$\delta$ and $\delta^{'}$.         
\subsection{Preliminaries}
\hspace*{\parindent}  
 First we denote the amplitude of $P^{0}(\overline{P^{0}})$
-meson decaying into some final state (denoted by $\it{f}$) 
as $A(P^0\to\it{f})$ and $A(\overline{P^o}\to\it{f})$.
In order to calculate direct CP violation
(denoted by ${\cal A}^{dir}_{cp}(P^{0}\to\it{f})$) and 
mixing-induced CP violation
(denoted by ${\cal A}^{mix}_{cp}(P^{0}\to\it{f})$) we 
introduce `` $\xi(P^{0}\to\it{f})$ '', which is defined
as : 
 $$\xi(P^{0}\to\it{f})\equiv{\pm}e^{-i\theta_{P}}
 \frac{A(\overline{P^o}\to\it{f})}{A(P^0\to\it{f})},
 \hspace{3cm}\eqno{(7.1)}$$                               \label{7.1}
where $e^{-i\theta_{P}}=\sqrt{M^{*}_{12}/M_{12}}$
($M_{12}$ and $\theta_{P}$ appeared in (6.1) ); (-)-sign
for $\it{f}=(P_{S},P_{S})$ CP-even final state and 
(+)-sign for $\it{f}=(P_{S},V)$ CP-odd final state.  
\par
By use of (7.1) we obtain :
 {  
\setcounter{enumi}{\value{equation}}
\addtocounter{enumi}{1}
\setcounter{equation}{0}
\renewcommand{\theequation}{\theenumi\alph{equation}}
\begin{eqnarray}
 \hspace{2cm}
{\cal A}^{dir}_{cp}(P^{0}\to\it{f})
  =\frac{1-|\xi(P^{0}\to\it{f})|^2}
        {1+|\xi(P^{0}\to\it{f})|^2},
        \hspace{6.2cm}(7.2)\nonumber\\
{\cal A}^{mix}_{cp}(P^{0}\to\it{f})
  =\frac{2{\rm{Im}}\xi(P^{0}\to\it{f})}
        {1+|\xi(P^{0}\to\it{f})|^2}.
        \hspace{6.2cm}(7.3)\nonumber                                              
\end{eqnarray}
\setcounter{equation}{\value{enumi}}}                       \label{7.2; 3}

\subsection{CP violation through $B_{d}\to{J}/{\psi}K_{s}$}   \label{SS 2}
\hspace*{\parindent}
 We write ($B_{d}\to{J}/{\psi}K_{s}$)-decay amplitude as :
\begin{eqnarray} 
    A(B_{d}\to{J}/{\psi}K_{s})
      \hspace{0.2cm}&=&V_{cb}^{*}V_{cs}A_{T}
      +V_{ub}^{*}V_{us}A_{P}^{u}
      +V_{cb}^{*}V_{cs}A_{P}^{c}
      +V_{tb}^{*}V_{ts}A_{P}^{t}\nonumber\\
                    &=&
      {\lambda}^{2}e^{-i{\delta}}A_{T}
      +{\lambda}^{2}e^{-i{\delta}}A_{P}^{c}
      +e^{-i{\delta}}A_{P}^{t}
      +{\lambda}^{2}e^{i({\delta-\delta^{'}})}A_{P}^{u},
      \hspace{1.8cm}(7.4)\nonumber                            \label{7.4}
 \end{eqnarray}
 where $A_{i}^{q}$ ($i=T :\rm{tree}$ ; $P :\rm{penguin}$ ;
$q$ : quark name) is the amplitude of strong interaction  
and the equations of (4.38) and (4.39) are used.
\par
Here let us abbreviate (7.4) and obtain :
  $$A(B_{d}\to{J}/{\psi}K_{s})\propto
         e^{-i{\delta}}\{1+\gamma_{1}\lambda^{2}
         e^{i({2\delta-\delta^{'}})}\cdot
         e^{i\theta_{1}}\},\hspace{2cm}\eqno{(7.5)}$$       \label{7.5}
where $\gamma_{1}e^{i\theta_{1}}\equiv{A^{u}_{P}}/
      (A_{T}+{A^{c}_{P}}+{A^{t}_{P}})$ ; $\theta_{1}$ is 
CP invariant strong phase.
\par
For ($\overline{B}_{d}\to{J}/{\psi}K_{s}$)-process
we have :
  $$\overline{A}(\overline{B}_{d}\to{J}/{\psi}K_{s})
               \propto
      e^{i{\delta}}\{1+\gamma_{1}\lambda^{2}
      e^{-i({2\delta-\delta^{'}})}\cdot
      e^{i\theta_{1}}\}.\hspace{2cm}\eqno{(7.6)}$$          \label{7.6} 
The application of (7.5) and (7.6) to (7.1) yields :
  $$\xi(B_{d}\to{J}/{\psi}K_{s})
               =+e^{-i({2\delta+\theta_{B_{d}}})}\cdot
               \frac 
 {1+{\gamma_{1}}{\lambda^{2}}e^{-i({2\delta-\delta^{'}})}
               \cdot{e^{i\theta_{1}}}}
 {1+\gamma_{1}{\lambda^{2}}e^{i({2\delta-\delta^{'}})}
              \cdot{e^{i\theta_{1}}}}.
               \hspace{0.9cm}\eqno{(7.7)}$$                  \label{7.7}
Putting (7.7) into (7.2;3) we obtain :
 $${\cal A}^{dir}_{cp}(B_{d}\to{J}/{\psi}K_{s})
           =\frac
{-2\gamma_{1}\lambda^{2}\rm{sin}
           ({2\delta-\delta^{'}})\rm{sin}\theta_{1}}
{1+2\gamma_{1}\lambda^{2}\rm{cos}
           ({2\delta-\delta^{'}})\rm{cos}\theta_{1}},
        \hspace{1.9cm}\eqno{(7.8)}$$                       \label{7.8}
and
  $$\hspace{2.2cm}{\cal A}^{mix}_{cp}(B_{d}\to{J}/{\psi}K_{s})
           =\frac
{\rm{sin}({2\delta-\theta_{B_{d}})+
     2\gamma_{1}\lambda^{2}\rm{sin}
     ({\delta^{'}}-\theta_{B_{d}})\rm{cos}\theta_{1}}}
{1+2\gamma_{1}\lambda^{2}\rm{cos}
     ({2\delta-\delta^{'}})\rm{cos}\theta_{1}}.
     \hspace{0.5cm}\eqno{(7.9)}$$                       \label{7.9}
Taking $\gamma_{1}\lambda^{2}\sim{\rm{O}(10^{-2})}\ll{1}$ 
($\gamma_{1}\sim0.3$ : QCD calculation) into account
for (7.8) and (7.9) we have :
 $${\cal A}^{dir}_{cp}(B_{d}\to{J}/{\psi}K_{s})
      \cong0+\rm{O}(\gamma_{1}\lambda^{2}),
      \hspace{4.5cm}\eqno{(7.10)}$$                       \label{7.10}
and
 $${\cal A}^{mix}_{cp}(B_{d}\to{J}/{\psi}K_{s})
             \cong
      {\rm{sin}({2\delta-\theta_{B_{d}})
      +{\rm{O}(\gamma_{1}\lambda^{2})}}}.
      \hspace{2.5cm}\eqno{(7.11)}$$                       \label{7.11}
      
 The Belle and Babar recentry reported that :
{  
 \setcounter{enumi}{\value{equation}}
 \addtocounter{enumi}{1} 
 \setcounter{equation}{0}
 \renewcommand{\theequation}{\theenumi\alph{equation}}
 \begin{eqnarray}
 \mbox{${\cal A}^{mix}_{cp}(B_{d}
  \to{J}/{\psi}K_{s})=$ }
 &&\hspace{-0.8cm}\left\{
 \begin{array}{lcl}
    0.642\pm0.030(stat.)\pm0.017(syst.)
       \hspace{0.2cm}(\rm{Belle}\hspace{0.4cm} [32]\,)\\      \label{7.12}
          \hspace{9.8cm}(7.12)\\
    0.715\pm0.034(stat.)\pm0.019(syst.)
          \hspace{0.2cm}(\rm{BaBar}\hspace{0.1cm} [33]\,).
          \nonumber      
 \end{array} 
 \right.
\end{eqnarray}
 \setcounter{equation}{\value{enumi}}}
 Putting the numerical values of (7.12) into the left side
 of (7.11)
  we obtain : 
 {  
 \setcounter{enumi}{\value{equation}}
 \addtocounter{enumi}{1} 
 \setcounter{equation}{0}
 \renewcommand{\theequation}{\theenumi\alph{equation}}
 \begin{eqnarray}
 \hspace{1.3cm}\mbox{$\delta\cong$ }
 &&\hspace{-0.7cm}\left\{
 \begin{array}{lcl}
    (20\pm2)^{\circ}\hspace{0.3cm}\rm{or}
       \hspace{0.3cm}(70\pm2)^{\circ}
       \hspace{1cm}(\rm{Belle})\\      
          \hspace{11.8cm}(7.13)\\                         \label{7.13}        
    (23\pm2)^{\circ}\hspace{0.3cm}\rm{or}
       \hspace{0.3cm}(67\pm2)^{\circ}
          \hspace{1.0cm}(\rm{BaBar}),
          \nonumber      
 \end{array} 
 \right.
\end{eqnarray}
 \setcounter{equation}{\value{enumi}}}
where $\theta_{B_{d}}\ll{1^{\circ}}$ is neglected.
\par
We see a two-fold ambiguity in (7.13). The $\rm{Babar}$ 
Collaboration has showed the interesting information 
about the value of `` ${\rm{cos}}2\delta$ ''
 in $B_{d}\to{D^{*0}}{\it{h}}^{0}$ 
(where ${\it{h}}^{0}$ is ${\it{\pi}}^{0}, {\it{\eta}},
{\it{\eta}}^{'}$ or $\it{\omega}$)
and claimed that $\delta=(23\pm2)^{\circ}$ is prefererable 
to $\delta=(67\pm2)^{\circ}$$\:$[33]. 
\subsection {CP violation through $B_{d}\to{\it{\pi}}^{+}
{\it{\pi}}^{-}$}                                              \label{SS 3}
 \hspace*{\parindent}With the help of (4.39$\sim${41})
 we obtain :
\begin{eqnarray} 
    A(B_{d}\to{\it{\pi}}^{+}{\it{\pi}}^{-})                                  
      \hspace{0.2cm}&=&V_{ub}^{*}V_{ud}A_{T}
      +V_{ub}^{*}V_{ud}A_{P}^{u}
      +V_{cb}^{*}V_{cd}A_{P}^{c}
      +V_{tb}^{*}V_{td}A_{P}^{t}\nonumber\\
                    &\propto&
      e^{-i{\delta}^{'}}\{1+\gamma_{2}
         e^{-i({2\delta-\delta^{'}})}\cdot
         e^{i\theta_{2}}\},\hspace{2cm}\hspace{3.3cm}(7.14)              
      \nonumber                                              \label{7.14}
 \end{eqnarray} 
where $\gamma_{2}e^{i\theta_{2}}\equiv{A^{c}_{P}}/
      (A_{T}+{A^{u}_{P}}+{A^{t}_{P}})$ ; $\theta_{2}$ is 
CP invariant strong phase. And concerning 
$A({\overline{B}}_{d}\to{\it{\pi}}^{+}{\it{\pi}}^{-})$  
we obtain :
$$\hspace{2cm}A(\overline{B}_{d}\to{\it{\pi}}^{+}{\it{\pi}}^{-})
              \propto
e^{i{\delta}^{'}}\{1+\gamma_{2}
         e^{i({2\delta-\delta^{'}})}\cdot
         e^{i\theta_{2}}\},\hspace{2cm}\hspace{2.5cm}(7.15)              
      \nonumber$$                                             \label{7.15}
using (7.14;15) to  (7.1) we get :
 $$\xi(B_{d}\to{\it{\pi}}^{+}{\it{\pi}}^{-})
               =-e^{-i({2\delta^{'}+\theta_{B_{d}}})}\cdot
               \frac 
 {1+{\gamma_{2}}e^{i({2\delta-\delta^{'}})}
               \cdot{e^{i\theta_{2}}}}
 {1+\gamma_{2}e^{-i({2\delta-\delta^{'}})}
              \cdot{e^{i\theta_{2}}}}.
               \hspace{1.6cm}\eqno{(7.16)}$$                  \label{7.16}
The applcation of (7.16) to (7.2;3) yields :
 $${\cal A}^{dir}_{cp}(B_{d}\to{\it{\pi}}^{+}{\it{\pi}}^{-})
           =\frac
{2\gamma_{2}\rm{sin}
           ({2\delta-\delta^{'}})\rm{sin}\theta_{2}}
{1+2\gamma_{2}\rm{cos}
           ({2\delta-\delta^{'}})\rm{cos}\theta_{2}},
        \hspace{2.9cm}\eqno{(7.17)}$$                       \label{7.17}
and
  $$\hspace{1.0cm}{\cal A}^{mix}_{cp}(B_{d}
     \to{\it{\pi}}^{+}{\it{\pi}}^{-})
     =-\frac
     {\rm{sin}({2\delta^{'}-\theta_{B_{d}})+
     2\gamma_{2}\rm{sin}
     (3{\delta^{'}}-2\delta-\theta_{B_{d}})
     \rm{cos}\theta_{2}}}
     {1+2\gamma_{2}\rm{cos}
     ({2\delta-\delta^{'}})\rm{cos}\theta_{2}}.
     \hspace{0.5cm}\eqno{(7.18)}$$                       \label{7.18}
\par Carrying out further approximation of O($\gamma_{2}$) 
we have :

$${\cal A}^{dir}_{cp}(B_{d}\to{\it{\pi}}^{+}{\it{\pi}}^{-})
           \cong
     {2\gamma_{2}\rm{sin}({2\delta-\delta^{'}})
     \rm{sin}\theta_{2}},\hspace{4cm}\eqno{(7.19)}$$        \label{7.19}
and

 $$\hspace{0.3cm}{\cal A}^{mix}_{cp}
  (B_{d}\to{\it{\pi}}^{+}{\it{\pi}}^{-})
         \cong
     -{\rm{sin}({2\delta^{'}-\theta_{B_{d}})\nonumber
           +2\gamma_{2}\rm{sin}
     (2{\delta}-\delta^{'})\rm{cos}(2\delta^{'}-\theta_{B_{d}})
     \rm{cos}\theta_{2}}}.\eqno{(7.20)}$$                               \label{7.20}
The update experimental informations are as follows : 
{  
 \setcounter{enumi}{\value{equation}}
 \addtocounter{enumi}{1} 
 \setcounter{equation}{0}
 \renewcommand{\theequation}{\theenumi\alph{equation}}
 \begin{eqnarray}
 \hspace{1.5cm}\mbox{${\cal A}^{dir}_{cp}(B_{d}
  \to{\it{\pi}}^{+}{\it{\pi}}^{-})=$ }
 &&\hspace{-0.8cm}\left\{
 \begin{array}{lcl}
    +0.55\pm0.08\pm0.05
       \hspace{0.2cm}(\rm{Belle}\hspace{0.4cm} [50]\,)\\      \label{7.21}
          \hspace{8.8cm}(7.21)\\
    +0.16\pm0.11\pm0.03
          \hspace{0.2cm}(\rm{BaBar}\hspace{0.1cm} [51]\,),
          \nonumber      
 \end{array} 
 \right.
\end{eqnarray}
\setcounter{equation}{\value{enumi}}}
and 
{  
 \setcounter{enumi}{\value{equation}}
 \addtocounter{enumi}{1} 
 \setcounter{equation}{0}
 \renewcommand{\theequation}{\theenumi\alph{equation}}
 \begin{eqnarray}
 \hspace{1.4cm}\mbox{${\cal A}^{mix}_{cp}(B_{d}
  \to{\it{\pi}}^{+}{\it{\pi}}^{-})=$ }
 &&\hspace{-0.8cm}\left\{
 \begin{array}{lcl}
    -0.61\pm0.10\pm0.04
       \hspace{0.2cm}(\rm{Belle}\hspace{0.4cm} [50]\,)\\      \label{7.22}
          \hspace{8.8cm}(7.22)\\
    -0.53\pm0.14\pm0.02
          \hspace{0.2cm}(\rm{BaBar}\hspace{0.1cm} [51]\,),
          \nonumber      
 \end{array} 
 \right.
\end{eqnarray}
\par
Both experiments comparatively coincides in 
${\cal A}^{mix}_{cp}(B_{d}
  \to{\it{\pi}}^{+}{\it{\pi}}^{-})$ but contradict in
${\cal A}^{dir}_{cp}(B_{d}
  \to{\it{\pi}}^{+}{\it{\pi}}^{-})$. Therefore let us 
investigate ${\cal A}^{mix}_{cp}(B_{d}
  \to{\it{\pi}}^{+}{\it{\pi}}^{-})$. Assuming that 
$\gamma_{2}\sim0.3$ in (7.20), the second term is estimated 
to be a few 10\% contribution and there approximately
exists (10$\sim$30)\% error in (7.21). So at present 
 stage in order to 
get the information about `` $\delta^{'}$ '' it may well be 
adopted that :
$${\cal A}^{mix}_{cp}(B_{d}\to{\it{\pi}}^{+}{\it{\pi}}^{-})
        \cong
 {\rm{sin}}({2\delta^{'}}-\theta_{B_{d}}).
        \hspace{5cm}\eqno{(7.23)}$$                           \label{7.23}
Using (7.22) to (7.23) we obtain :        
 {  
 \setcounter{enumi}{\value{equation}}
 \addtocounter{enumi}{1} 
 \setcounter{equation}{0}
 \renewcommand{\theequation}{\theenumi\alph{equation}}
 \begin{eqnarray}
 \hspace{1.5cm}\mbox{$\delta^{'}\cong$ }
 &&\hspace{-0.7cm}\left\{
 \begin{array}{lcl}
    (19\pm5)^{\circ}\hspace{0.3cm}\rm{or}
       \hspace{0.3cm}(71\pm5)^{\circ}
       \hspace{1cm}(\rm{Belle})\\      
          \hspace{11.5cm}(7.24)\\                         \label{7.24}        
    (16\pm5)^{\circ}\hspace{0.3cm}\rm{or}
       \hspace{0.3cm}(74\pm5)^{\circ}
          \hspace{1.0cm}(\rm{BaBar}),
          \nonumber      
 \end{array} 
 \right.
\end{eqnarray}
 \setcounter{equation}{\value{enumi}}}
where $\theta_{B_{d}}(\ll{1^{\circ}})$ is neglected 
same as (7.13). From (7.19) we estimate :
$$|{\cal A}^{dir}_{cp}(B_{d}\to{\it{\pi}}^{+}
     {\it{\pi}}^{-})|\le0.3,\hspace{7cm}\eqno{(7.25)}$$      \label{7.25}
where we use $\delta=21.5^{\circ}$ 
and $\delta^{'}=(17.5\;{\rm{or}}\;72.5)^{\circ}$ 
by averaging Belle and BaBar in (7.13) and (7.24) ;
$|\rm{sin}\theta_{2}|\le{1}$ ; $\gamma_{2}=0.3$.   
Then the estimation of (7.25) suggests that BaBar 
is preferable to Belle in (7.21).
\subsection{CP violation through $B_{d}\to{\it{\phi}
            }K_{s}$}                                          \label{SS 4}
\hspace*{\parindent}
The ($B_{d}\to{\it{\phi}}K_{s}$)-decay has a CP-odd 
final state and receives contribution from only 
penguin topologies with $b\to\overline{s}s\overline{s}$              
quark level processes. 
\par With the help of (4.39$\sim$41),
the decay amplitude is described as :
\begin{eqnarray} 
    A(B_{d}\to{\it{\phi}}K_{s})                                  
      \hspace{0.2cm}&=&
      +V_{ub}^{*}V_{us}A_{P}^{u}
      +V_{cb}^{*}V_{cs}A_{P}^{c}
      +V_{tb}^{*}V_{ts}A_{P}^{t}\nonumber\\
                    &\propto&
      e^{-i{\delta}}\{1+\gamma_{3}\lambda^{2}
         e^{i({2\delta-\delta^{'}})}\cdot
         e^{i\theta_{3}}\},\hspace{2cm}
         \hspace{3.5cm}(7.26)\nonumber                      \label{7.26}
 \end{eqnarray} 
where $\gamma_{3}e^{i\theta_{3}}\equiv{A^{u}_{P}}/
      ({A^{c}_{P}}+{A^{t}_{P}})$ and 
$\gamma_{3}\sim\rm{O}(1)$ is expected. 
\par For ($\overline{B}_{d}\to{\it{\phi}}K_{s}$)-process 
 we have :   
$$\hspace{1.7cm}
    \overline{A}(\overline{B}_{d}\to{\it{\phi}}K_{s})
              \propto
e^{i{\delta}^{'}}\{1+\gamma_{3}\lambda^{2}
         e^{-i({2\delta-\delta^{'}})}\cdot
         e^{i\theta_{3}}\}.
         \hspace{2cm}\hspace{2.5cm}(7.27)\nonumber$$        \label{7.27}
Putting (7.26) and (7.27) into (7.1) and using (7.2;3)
we obtain :
 $$\hspace{1.6cm}
{\cal A}^{dir}_{cp}(B_{d}\to{\it{\phi}}K_{s})
           =-\frac
{2\gamma_{3}\lambda^{2}\rm{sin}
           ({2\delta-\delta^{'}})\rm{sin}\theta_{3}}
{1+2\gamma_{3}\lambda^{2}\rm{cos}
           ({2\delta-\delta^{'}})\rm{cos}\theta_{3}},
        \hspace{2.9cm}\eqno{(7.28)}$$                       \label{7.28}
and
  $$\hspace{1.6cm}
    {\cal A}^{mix}_{cp}(B_{d}\to{\it{\phi}}K_{s})
           =\frac
{\rm{sin}({2\delta-\theta_{B_{d}})+
     2\gamma_{3}\lambda^{2}\rm{sin}
     (2{\delta^{'}}-\theta_{B_{d}})\rm{cos}\theta_{3}}}
{1+2\gamma_{3}\lambda^{2}\rm{cos}
     ({2\delta-\delta^{'}})\rm{cos}\theta_{3}}.
     \hspace{0.5cm}\eqno{(7.29)}$$                       \label{7.29}
\par Taking $\gamma_{3}\lambda^{2}\sim$
$O(\lambda^{2})\ll{1}$ in to account in (7.28;29)
we have :
 $$\hspace{1.6cm}
{\cal A}^{dir}_{cp}(B_{d}\to{\it{\phi}}K_{s})
\cong0+\rm{O}(\lambda^{2}),\hspace{5cm}\eqno{(7.30)}$$     \label{7.30}
and
$$\hspace{1.6cm}
    {\cal A}^{mix}_{cp}(B_{d}\to{\it{\phi}}K_{s})
  \cong
    \rm{sin}(2{\delta}-\theta_{B_{d}})+
    \rm{O}(\lambda^{2}).\hspace{5cm}\eqno{(7.31)}$$     \label{7.31}
From (7.11) and (7.31) we obtain :
$${\cal A}^{mix}_{cp}(B_{d} \to{J}/{\psi}K_{s}) 
         \cong 
{\cal A}^{mix}_{cp}(B_{d}\to{\it{\phi}}K_{s}).
         \hspace{4cm}\eqno{(7.32)} $$                    \label{7.32}
\par The experimental situation is as follows :
$$\hspace{1.5cm}{\cal A}^{mix}_{cp}(B_{d} \to{\phi}K_{s}) 
     =0.50\pm0.21\pm0.06.\;
      [32]\hspace{3.3cm}\eqno{(7.33)}$$           \label{7.33)}       
Comparing (7.33) with (7.12) consistency between 
them can be observed. 
\subsection{CP violation through $B_{d}\to{K}^{+}
{\it{\pi}}^{-}$ and $B_{u}^{+}\to{K}^{+}{\it{\pi}}^{0}$}   \label{ss 5}
The ($B\to{K}{\it{\pi}}$)-decay amplitudes are given as: 
$$ A(B_{d}\to{K}^{+}{\it{\pi}}^{-})                                  
      \hspace{0.2cm}=V_{ub}^{*}V_{us}A_{T}^{\rm{ex}}
      +V_{ub}^{*}V_{us}A_{P}^{u}
      +V_{cb}^{*}V_{cs}A_{P}^{c}
      +V_{tb}^{*}V_{ts}A_{P}^{t},
      \hspace{1cm}\eqno{(7.34)}$$                          \label{7.34}
$$ A(B_{u}^{+}\to{K}^{+}{\it{\pi}}^{0})                                  
      \hspace{0.2cm}=V_{ub}^{*}V_{us}A_{T}^{\rm{in}}
      +V_{ub}^{*}V_{us}A_{P}^{u}
      +V_{cb}^{*}V_{cs}A_{P}^{c}
      +V_{tb}^{*}V_{ts}A_{P}^{t}.
      \hspace{1cm}\eqno{(7.35)}$$                          \label{7.35}
Both amplitudes are contributed by tree and 
penguin modes but it is noticeable that 
the tree diagram of the former is `` external 
tree '' (denoted by $A_{T}^{\rm{ex}}$) and that
of latter is `` internal tree '' (denoted by 
$A_{T}^{\rm{in}}$). Therefore they possibly
differ in absolute values and/or phases.
By use of (4.39$\sim$41) equations of
 (7.34) and (7.35) are rewritten as :
   $$A(B_{d}\to{K}^{+}{\it{\pi}}^{-})                                  
      \hspace{0.2cm}\propto
            e^{-i{\delta}}\{1+\frac{\lambda^{2}}
            {\gamma_{\rm{ex}}}
         e^{i({2\delta-\delta^{'}})}\cdot
         e^{i\theta_{\rm{ex}}}\},
         \hspace{3.5cm}\eqno{(7.36)}$$                      \label{7.36}
where $(1/\gamma_{\rm{ex}})e^{i\theta_{\rm{ex}}}
      \equiv(A_{T}^{\rm{ex}}+{A^{u}_{P}})/
      ({A^{c}_{P}}+{A^{t}_{P}})$, and
 $$A(B_{u}^{+}\to{K}^{+}{\it{\pi}}^{0})                                  
      \hspace{0.2cm}\propto
            e^{-i{\delta}}\{1+\frac{\lambda^{2}}
            {\gamma_{\rm{in}}}
         e^{i({2\delta-\delta^{'}})}\cdot
         e^{i\theta_{\rm{in}}}\},
         \hspace{3.5cm}\eqno{(7.37)}$$                      \label{7.37}
where $(1/\gamma_{\rm{in}})e^{i\theta_{\rm{in}}}
      \equiv(A_{T}^{\rm{in}}+{A^{u}_{P}})/
      ({A^{c}_{P}}+{A^{t}_{P}})$.
Passing through the same procedure as previous 
sections we have :      
  $${\cal A}^{dir}_{cp}(B_{d}\to{K}^{+}{\it{\pi}}^{-})
           =-\frac
{2\frac{\lambda^{2}}{\gamma_{\rm{ex}}}
   \rm{sin}(2\delta-\delta^{'})\rm{sin}\theta_{\rm{ex}}}
{1+2\frac{\lambda^{2}}{\gamma_{\rm{ex}}}
   \rm{cos}({2\delta-\delta^{'}})\rm{cos}\theta_{\rm{ex}}},
        \hspace{2.9cm}\eqno{(7.38)}$$                       \label{7.38}
  $${\cal A}^{mix}_{cp}(B_{d}\to{K}^{+}{\it{\pi}}^{-})
           =-\frac
{\rm{sin}(2\delta-\theta_{B_{d}})-
     2\frac{\lambda^{2}}{\gamma_{\rm{ex}}}\rm{sin}
     (2\delta+\theta_{\rm{ex}}-\theta_{B_{d}})
\rm{cos}(2\delta-\delta^{'})}
{1-2\frac{\lambda^{2}}{\gamma_{\rm{ex}}}\rm{cos}
     ({2\delta-\delta^{'}})\rm{cos}\theta_{\rm{ex}}},
     \hspace{0.0cm}\eqno{(7.39)}$$                       \label{7.39}
and
  $${\cal A}^{dir}_{cp}(B_{u}\to{K}^{+}{\it{\pi}}^{0})
           =-\frac
{2\frac{\lambda^{2}}{\gamma_{\rm{in}}}
   \rm{sin}(2\delta-\delta^{'})\rm{sin}\theta_{\rm{in}}}
{1+2\frac{\lambda^{2}}{\gamma_{\rm{in}}}
   \rm{cos}({2\delta-\delta^{'}})\rm{cos}\theta_{\rm{in}}},
        \hspace{2.9cm}\eqno{(7.40)}$$                       \label{7.40}
From (7.38;40) we obtain :
  $${\cal A}^{dir}_{cp}(B_{d}\to{K}^{+}{\it{\pi}}^{-})
           =-2\frac{\lambda^{2}}{\gamma_{\rm{ex}}}
            \rm{sin}(2\delta-\delta^{'})
            \rm{sin}\theta_{\rm{ex}}
            +\rm{O}(\lambda^{4}/\gamma_{\rm{ex}}^{2}),
            \hspace{1.6cm}\eqno{(7.41)}$$                      \label{7.41}
and
  $${\cal A}^{dir}_{cp}(B_{u}\to{K}^{+}{\it{\pi}}^{0})
           =-2\frac{\lambda^{2}}{\gamma_{\rm{in}}}
            \rm{sin}(2\delta-\delta^{'})
            \rm{sin}\theta_{\rm{in}}
            +\rm{O}(\lambda^{4}/\gamma_{\rm{in}}^{2}).
            \hspace{2cm}\eqno{(7.42)}$$                      \label{7.42}
Experimental situations are as follows :
\begin{eqnarray}
\hspace{1.3cm}{\cal A}^{dir}_{cp}(B_{d}\to{K}^{+}
         {\it{\pi}}^{-})
       &=&-0.115\pm0.018,\;{[63]}
        \hspace{5,1cm}(7.43)\nonumber\\                         \label{7.43}    
\hspace{0.7cm}{\cal A}^{dir}_{cp}(B_{u}\to{K}^{+}
         {\it{\pi}}^{0})
       &=&0.016\pm0.041\pm0,012.\;[71]
        \hspace{4.0cm}(7.44)\nonumber
\end{eqnarray}                                                \label{7.44}
Omitting the second term of (7.41) we nave :
$$\rm{sin}\theta_{\rm{ex}}
    \cong-
 \frac{{\cal A}^{dir}_{cp}(B_{d}\to{K}^{+}{\it{\pi}}^{-})}           
   {\rm{sin}(2\delta-\delta^{'})}.
   \hspace{6.5cm}\eqno{(7.45)}$$                                    \label{7.45}
 Tentatively substituting \{ $\lambda=0.22:
\:{\gamma_{\rm{ex}}}=0.3\:;
\:\delta=21.5^{\circ}\:;\: 
\delta^{'}=(17.5\;\rm{or}\:72.5)^{\circ}$ (average values of
Belle and BaBar from (7.13;24) ) and 
${\cal A}^{dir}_{cp}(B_{d}\to{K}^{+}{\it{\pi}}^{-}) =-0.115$
from (7.43) \} into (7.45) we get :   
$$\theta_{\rm{ex}}\cong-56^{\circ}(\:\delta^{'}=17.5^{\circ}\:)
\;{\rm{or}}\; 46^{\circ}(\:\delta^{'}=72.5^{\circ}\:),
\hspace{4cm}\eqno{(7.46)}$$                                      \label{7.46}                        
(As seen in (7.54) of next section 
$\theta_{\rm{ex}}\cong46^{\circ}$ is favorable.)
On the other hand if we assume :
${\cal A}^{dir}_{cp}(B_{u}\to{K}^{+}{\it{\pi}}^{0})\cong0$
in (7.44) we  could estimate from (7.42) that :
$$\rm{sin}\theta_{\rm{in}}\cong0\;(\rm{that\; is,\;
\theta_{in}\cong0^{\circ}\;\rm{or}\;180^{\circ}}).
\hspace{5.6cm}\eqno{(7.47)}$$                                        \label{7.47}
From (7.46) and (7.47) we observe : 
$\theta_{\rm{ex}}\ne{\theta_{\rm{in}}}$, which causes :
${\cal A}^{dir}_{cp}(B_{d}\to{K}^{+}{\it{\pi}}^{-})
\ne{\cal A}^{dir}_{cp}(B_{u}\to{K}^{+}{\it{\pi}}^{0})$.
 \subsection{CP violation through $B_{s}\to{J}/{\psi}\phi$}   \label{SS 6}
\hspace*{\parindent}
 The$(B_{s}\to{J}/{\psi}\phi)$-decay mode is
the counterpart of the $(B_{d}\to{J}/{\psi}K_{s})$-decay, 
where the down quark of $B_{d}$ meson is replaced 
by the strange quark. The final state of 
$(B_{s}\to{J}/{\psi}\phi)$-decay is an admixture of 
different CP eigenstates in comparison 
to  $(B_{d}\to{J}/{\psi}K_{s})$-decay and then its 
study for CP violation is complex but the angular analysis
of the ${J}/{\psi}[\to{l^{+}l^{-}}]$$\phi[\to{K^{+}K^{-}}]$
decay products can solve that complexity${\:}$[73].
As the strange quark in $B_{s}$ meson is the spectator,
the decay amplitude of $(B_{s}\to{J}/{\psi}\phi)$-decay is 
completely analogous to that of 
$(B_{s}\to{J}/{\psi}\phi)$-decay mode$\:$[73]. 
Then we have :
\begin{eqnarray} 
   A(B_{s}\to{J}/{\psi}{\phi})                                  
      \hspace{0.2cm}&=&
      +V_{bc}^{*}V_{cs}A_{T}
      +V_{ub}^{*}V_{us}A_{P}^{u}
      +V_{cb}^{*}V_{cs}A_{P}^{c}
      +V_{tb}^{*}V_{ts}A_{P}^{t}\nonumber\\
                    &\propto&
      e^{-i{\delta}}\{1+\gamma_{4}\lambda^{2}
         e^{i({2\delta-\delta^{'}})}\cdot
         e^{i\theta_{4}}\},\hspace{2cm}
         \hspace{3.5cm}(7.48)\nonumber                      \label{7.48}
 \end{eqnarray} 
where $\gamma_{4}e^{i\theta_{4}}\equiv{A^{u}_{P}}/
      (A_{T}+{A^{c}_{P}}+{A^{t}_{P}}).$
Passing through the same procedure as the
$(B_{d}\to{J}/{\psi}K_{s})$-decay we have :
\begin{eqnarray} 
    \hspace{0.3cm}{\cal A}^{dir}_{cp}(B_{s}\to{J}/{\psi}{\phi})                                  
      \hspace{0.1cm}&=&
                 \frac
{-2\gamma_{4}\lambda^{2}\rm{sin}
           ({2\delta-\delta^{'}})\rm{sin}\theta_{1}}
{1+2\gamma_{4}\lambda^{2}\rm{cos}
           ({2\delta-\delta^{'}})\rm{cos}\theta_{1}},
        \hspace{1.9cm}\nonumber\\
                          &\cong&
0+\rm{O}(\:{\gamma_{4}\lambda^{2}}),                    
               \hspace{7,5cm}(7.49)\nonumber                      \label{7.49}
 \end{eqnarray} 
 and
 \begin{eqnarray}
  \hspace{0.8cm}{\cal A}^{mix}_{cp}(B_{s}\to{J}/{\psi}{\phi})
           &=&\frac
{\rm{sin}({2\delta-\theta_{B_{s}})+
     2\gamma_{4}\lambda^{2}\rm{sin}
     ({\delta^{'}}-\theta_{B_{d}})\rm{cos}\theta_{4}}}
{1+2\gamma_{4}\lambda^{2}\rm{cos}
     ({2\delta-\delta^{'}})\rm{cos}\theta_{4}}.
     \nonumber\\
           &\cong&
          \rm{sin}(2\delta)+
          \rm{O}(\:{\gamma_{4}\lambda^{2}}),
         \hspace{5.5cm}(7.50)\nonumber                  \label{7.50}
\end{eqnarray}
where $\theta_{B_{s}}$ is neglected.
Recently D$\O$ Collaboration at Fermilab has reported 
 CP violating phase of $(B_{s}\to{J}/{\psi}{\phi})$-
decay process as :
\begin{eqnarray}
\hspace{0,7cm}\phi_{s}=|0.79\pm0.56|_{\rm{rad}}
         \;&=&\:(45\pm32)^{\circ},
         \nonumber\\
  \rm{or}\hspace{0.5cm}|2.35\pm0.56|_{\rm{rad}}\:
           &=&\;(135\pm32)^{\circ}.\hspace{0.5cm}[74]
           \hspace{5.5cm}(7.51)\nonumber                \label{7.51}
\end{eqnarray}
The `` $\phi_{s}$ '' corresponds to `` $2\delta$ '' and
then it can be said that `` $2\delta\cong43^{\circ}$ ''
from (7.13) is in good agreement with the D$\O$
experiment : $\phi_{s}\;=\:(45\pm32)^{\circ}.$
\subsection {CP violation through $B_{d}
 (\overline{B_{d}})\to{D^{(*)\pm}}
{\it{\pi}}^{\pm}$}                                              \label{SS 7}
 \hspace*{\parindent} These processes are very 
interesting because the only one `` tree diagram ''
contributes to them. The $B_{d}\to{D^{(*)-}}
{\it{\pi}}^{+}$ and $\overline{B_{d}}\to{D^{(*)+}}
{\it{\pi}}^{-}$ are Cabibbo-favoured decay process 
(CFD) and $B_{d}\to{D^{(*)+}}{\it{\pi}}^{-}$ 
and $\overline{B_{d}}\to{D^{(*)-}}{\it{\pi}}^{+}$ 
are doubly-Cabibbo-suppressed decay process (DCSD).                      
\par Here let us study the mixing induced CP violation
between  $B_{d}\to{D^{-}}{\it{\pi}}^{+}$ and $\overline
{B_{d}}\to{D^{-}}{\it{\pi}}^{*}$ decay . For these 
decay amplitudes we obtain : 
$$\hspace{1cm}A(B_{d}\to{D^{-}}{\it{\pi}}^{+})                                  
      \hspace{0.2cm}=V_{cb}^{*}V_{ud}A_{T}
                    =\lambda^{2}e^{-i\delta}A_{T}
      \hspace{1cm}(\rm{CFD-MODE}),\hspace{0.5cm}\eqno{(7.52)}$$       \label{7.52}
 and  
$$\hspace{1cm}\overline{A}(\overline{B_{d}}
      \to{D^{-}}{\it{\pi}}^{+})                                  
      \hspace{0.2cm}=V_{cd}^{*}V_{bu}A_{T}^{'}
                    =\lambda^{4}e^{i(\delta-\delta^{'})}
                     A_{T}^{'}
      \hspace{0.5cm}(\rm{DCSD-MODE}).
               \hspace{0.9cm}\eqno{(7.53)}$$                         \label{7.53}
Using (7.52) and (7.53) we have :
\begin{eqnarray}
      \hspace{1cm}{\xi} (B_{d}\to{D^{-}}{\it{\pi}}^{+}) 
      &=&e^{-i\theta_{B_{d}}}\frac{\overline{A}}{A}
       =-e^{-i\theta_{B_{d}}}
 \cdot{\frac
{\lambda^{4}e^{i(\delta-\delta^{'})}} 
{\lambda^{2}e^{-i\delta}}}
\cdot{\frac{A_{T}^{'}}{A_{T}}}\nonumber\\
       &=&
-\lambda^{2}e^{-i(2\delta-\delta^{'}-\theta_{B_{d}}
-\theta_{5})},       
 \hspace{5.8cm}(7.54)\nonumber                               \label{7.54}
\end{eqnarray}
where $e^{i\theta_{5}}\equiv{A_{T}^{'}/A_{T}}$ by
assuming : $|A_{T}^{'}|=|A_{T}|.$
From (7.54) we obtain :
\begin{eqnarray}
  \hspace{0.8cm}{\cal A}^{mix}_{cp}(B_{d}\to
                {D^{-}}{\it{\pi}}^{+})
           &=&\frac
{-2\lambda^{2}\rm{sin}
   (2\delta-\delta^{'}-\theta_{B_{d}}-\theta_{5})}
   {1+\lambda^{4}}\nonumber\\
           &\cong&
-2\lambda^{2}\rm{sin}(2\delta-\delta^{'}-\theta_{5}).
         \hspace{5.0cm}(7.55)\nonumber
\end{eqnarray}                                                \label{7.55}
There are theoretical arguments : the still-unmeasured
values of $\theta_{5}$ for both ${D^{*}}{\it{\pi}}$
and $D{\it{\pi}}$ are small$\:$[75][76], so we set 
$\theta_{5}=0$. 
By use of $\delta=(21.5\pm2.0)^{\circ}$ ; 
$\delta^{'}=(17.5\pm5.0)^{\circ}$ or 
 $\delta^{'}=(72.5\pm5.0)^{\circ}$ (average of
 Belle and BaBar) we get :
 {  
 \setcounter{enumi}{\value{equation}}
 \addtocounter{enumi}{1} 
 \setcounter{equation}{0}
 \renewcommand{\theequation}{\theenumi\alph{equation}}
 \begin{eqnarray}
 \hspace{0.5cm}\mbox{${\cal A}^{mix}_{cp}(B_{d}\to
                {D^{-}}{\it{\pi}}^{+})=$}
 &&\hspace{-0.7cm}\left\{
 \begin{array}{lcl}
    -0.059\pm0.032\;
   \{\delta^{'}=(17.5\pm5.0)^{\circ}\}
     \nonumber\\ 
          \hspace{9.6cm}(7.56)\\                         \label{7.56}        
    +0.047\pm0.013\;
\{\delta^{'}=(72.5\pm5.0)^{\circ}\}    
          \nonumber      
 \end{array} 
 \right.
\end{eqnarray}
 \setcounter{equation}{\value{enumi}}}
Recently Belle Collaboration published 
measurements of CP violation in 
$B_{d}\to{D^{(*)-}}{\it{\pi}}^{+}$ decays$\:$[75]
and they informed that :
$${\cal A}^{mix}_{cp}(B_{d}\to{D^{-}}{\it{\pi}}^{+})
=0.068\pm0.029\pm0,012\hspace{5.3cm}\eqno{(7.57)}$$                 \label{7.57}
Comparing (7.57) with (7.56), the case of 
$\{\delta^{'}=(72.5\pm5.0)^{\circ}\}$ seems 
to be favorable. Refering above discussions combined 
with the result of (7.13) and [33] we may well
conclude :
$$\delta=(21.5\pm2.0)^{\circ}\hspace{1cm}\rm{and}\hspace{1cm}
\delta^{'}=(72.5\pm5.0)^{\circ}.\hspace{4cm}\eqno{(7.58)}$$         \label{7.58}   
 \subsection{CP violation through $D^{0}$ meson decay}              \label{SS 8}
\hspace*{\parindent}
\subsubsection{
  a. $(D^{\circ}\to{K^{\pm}}{\it{\pi}}^{\mp})$-decay mode}
 
\par These processes occur only through the `` tree ''
topologies and we have :
$$\hspace{1cm}A(D^{0}\to{K^{+}}{\it{\pi}}^{-})                                  
  \hspace{0.2cm}=V_{cd}^{*}V_{su}A_{T}
                =\lambda^{2}e^{-i\delta}e^{i\delta}A_{T}
                =A_{T},\hspace{4.5cm}\eqno{(7.59)}$$       \label{7.59}
 and  
$$\hspace{1cm}\overline{A}(\overline{D^{0}}
     \to{K^{+}}{\it{\pi}}^{-})                                  
      \hspace{0.2cm}=V_{cs}^{*}V_{ud}A_{T}^{'}
                    =1{\cdot}1{\cdot}A_{T}^{'}                    
                    =A_{T}^{'}.
                 \hspace{5.4cm}\eqno{(7.60)}$$               \label{7.60}
By use of them we obtain :
$$\xi(D^{0}\to{K^{+}}{\it{\pi}}^{-})
    \hspace{0.2cm}=\gamma_{6}e^{-i(\theta_{D}-\theta_{6})},
             \hspace{6.0cm}\eqno{(7.61)}$$                       \label{7.61}
where $\gamma_{6}e^{-i\theta_{6}}\equiv{A_{T}^{'}}/A_{T}$,
which leads :
 $$\hspace{1.2cm}
{\cal A}^{dir}_{cp}(D^{0}\to{K^{+}}{\it{\pi}}^{-})
    =\frac{1-\gamma_{6}^{2}}{1+\gamma_{6}^{2}}
         \hspace{9cm}\eqno{(7.62)}$$                            \label{7.62}
and
$$\hspace{1.2cm}
    {\cal A}^{mix}_{cp}(D^{0}\to{K^{+}}{\it{\pi}}^{-})
    =\frac{2\gamma_{6}^{2}}{1+\gamma_{6}^{2}}
     \rm{sin}(\theta_{6}-\theta_{D}).
     \hspace{3.8cm}\eqno{(7.63)}$$                                \label{7.63}
Here we may expect : $\gamma_{6}\cong1$ and get 
the same result about $(D^{0}\to{K^{-}}
{\it{\pi}}^{+})$-decay process. Then we have :
$${\cal A}^{dir}_{cp}(D^{0}\to{K^{\pm}}
{\it{\pi}}^{\mp})\cong0
      \hspace{0.5cm}\rm{and}\hspace{0.5cm}
{\cal A}^{mix}_{cp}(D^{0}\to{K^{\pm}}{\it{\pi}}^{\mp})
                 \cong\rm{sin}\theta_{6}.
       \hspace{1.7cm}\eqno{(7.64)}$$                                  \label{7.64} 
\hspace*{\parindent}
\subsubsection{
  b. $(D^{\circ}\to{K^{\pm}}{K^{\mp}})$-decay mode}
\par 
Only `` penguin '' type amplitudes contribute 
to these decays.
Then we describe them as : 
\begin{eqnarray} 
    \hspace{0.9cm}A(D^{0}\to{K}^{+}{K}^{-})                                  
                 &=&
       V_{cd}^{*}V_{ud}A_{P}^{d}
      +V_{cs}^{*}V_{us}A_{P}^{s}
      +V_{cb}^{*}V_{ub}A_{P}^{b}\nonumber\\
                    &\propto&
      e^{-i{\delta}}\{1+\gamma_{7}\lambda^{3}
      e^{i({2\delta-\delta^{'}})}\cdot
      e^{i\theta_{7}}\},\hspace{4.4cm}(7.65)\nonumber                  \label{7.65}
 \end{eqnarray} 
where $\gamma_{7}e^{i\theta^{7}}\equiv{A_{P}^{b}}/(A_{P}^{d}
+A_{P}^{s})$. And also we have :
$$\hspace{0.8cm}\overline{A}
(\overline{D^{0}}\to{K}^{+}{K}^{-}) 
       \hspace{0.3cm}\propto\hspace{0.3cm}
 e^{i{\delta}}\{1+\gamma_{7}\lambda^{3}
      e^{-i(2\delta-\delta^{'})}\cdot
      e^{i\theta_{7}}\}.\hspace{5.6cm}\eqno{(7.66)}$$                  \label{7.66}
 Using (7.65;66) we have :
 $${\cal A}^{dir}_{cp}(D^{0}\to{K}^{+}{K}^{-})
           =\frac
{2\gamma_{7}\lambda^{3}\rm{sin}
           ({2\delta-\delta^{'}})\rm{sin}\theta_{7}}
{1+2\gamma_{7}\lambda^{3}\rm{cos}
           ({2\delta-\delta^{'}})\rm{cos}\theta_{7}},
        \hspace{3.7cm}\eqno{(7.67)}$$                       \label{7.67}
and
  $$\hspace{0.8cm}{\cal A}^{mix}_{cp}
            (D^{0}\to{K}^{+}{K}^{-})
           =-\frac
{\rm{sin}({2\delta-\theta_{D})+
     2\gamma_{7}\lambda^{3}\rm{sin}
     ({\delta^{'}}-\theta_{D})\rm{cos}\theta_{7}}}
{1+2\gamma_{7}\lambda^{3}\rm{cos}                  
     ({2\delta-\delta^{'}})\rm{cos}\theta_{7}}.
     \hspace{0.0cm}\eqno{(7.68)}$$                      \label{7.68}
 From (7.67;68) we obtain :    
 $$\hspace{0.9cm}
{\cal A}^{dir}_{cp}(D^{0}\to{K}^{+}{K}^{-})
\cong0+\rm{O}(\gamma_{7}\lambda^{3}),
\hspace{5cm}\eqno{(7.69)}$$                               \label{7.69}
and
$$\hspace{0.9cm}
    {\cal A}^{mix}_{cp}(D^{0}\to{K}^{+}{K}^{-})
  \cong
    -\rm{sin}(2{\delta}-\theta_{D})+
    \rm{O}(\gamma_{7}\lambda^{3})
  =-0.68,  
    \hspace{2cm}\eqno{(7.70)}$$                           \label{7.70}
where we use $\delta=21.5^{\circ}$ 
and neglect $\theta_{D}$.
\subsubsection{
  c. $(D^{\circ}\to{\it{\pi}}^{+}
    {\it{\pi}}^{-})$-decay mode}
\par This process have both tree and penguin modes.
Its amplitude is described as :
\begin{eqnarray} 
\hspace{0.5cm}
    A(D^{0}\to{\it{\pi}}^{+}{\it{\pi}}^{-})                                  
      \hspace{0.2cm}&=&V_{cd}^{*}V_{ud}A_{T}
      +V_{cd}^{*}V_{ud}A_{P}^{u}
      +V_{cs}^{*}V_{us}A_{P}^{s}
      +V_{cb}^{*}V_{ub}A_{P}^{b}\nonumber\\
                    &\propto&
      e^{-i{\delta}}\{1+\gamma_{8}\lambda^{4}
         e^{-i({2\delta-\delta^{'}})}\cdot
         e^{i\theta_{8}}\},
        \hspace{4.6cm}(7.71)\nonumber             \label{7.71}
 \end{eqnarray} 
where $\gamma_{8}e^{i\theta_{8}}\equiv{A^{}_{P}}/
      (A_{T}+{A^{u}_{P}}+{A^{s}_{P}})$. 
From (7.71) we obtain :
 $$\hspace{0.5cm}
{\cal A}^{dir}_{cp}(D^{0}\to{\it{\pi}}^{+}
    {\it{\pi}}^{-})
           =\frac
{2\gamma_{8}\lambda^{4}\rm{sin}
           ({2\delta-\delta^{'}})\rm{sin}\theta_{8}}
{1+2\gamma_{8}\lambda^{4}\rm{cos}
           ({2\delta-\delta^{'}})\rm{cos}\theta_{8}}
           \hspace{6cm}\nonumber$$
$$\hspace{0.5cm}\cong0+\rm{O}(\gamma_{8}\lambda^{4}),
        \hspace{5.4cm}\eqno{(7.72)}$$
and                       \label{7.72}
  $$\hspace{0.5cm}{\cal A}^{mix}_{cp}
            (D^{0}\to{\it{\pi}}^{+}{\it{\pi}}^{-})
           =-\frac
{\rm{sin}({2\delta-\theta_{D})+
     2\gamma_{8}\lambda^{4}\rm{sin}
     ({\delta^{'}}-\theta_{D})\rm{cos}\theta_{8}}}
{1+2\gamma_{8}\lambda^{4}\rm{cos}                  
     ({2\delta-\delta^{'}})\rm{cos}\theta_{8}}
     \hspace{4.5cm}\nonumber$$ 
$$\hspace{1.7cm}\cong-\rm{sin}(2{\delta}-\theta_{D})+
    \rm{O}(\gamma_{8}\lambda^{4})
                \cong-0.68,  
    \hspace{2.6cm}\eqno{(7.73)}$$                           \label{7.73}
where $\delta=21.5^{\circ}$ is used and 
$\theta_{D}$ is neglected. In this connection 
CLEO Collaboration reported $\;$[79] :                     
$${\cal A}^{dir}_{cp}(D^{0}\to{K^{+}}{K^{-}}\;{,}\;
{\it{\pi}}^{+}{\it{\pi}}^{-})
         \cong0.\hspace{7cm}\eqno{(7.74)}$$                 \label{7.74}
 \subsection{CP violation through 
$K\to{\it{\pi}}{\it{\pi}}$ }                                    \label{SS 9}
Decay amplitudes for $K^{0}(\overline{K^{0}})
\to{\pi}{\pi}$ decays can be described by two parts, 
which are in isospin $I=0$ and $I=2$ states due to 
Bose statistics of S wave pions. 
On the other hand in the description of quark 
levels there are two types, namely : the `` tree ''- and 
the `` penguin ''-type graph. The former has both of 
$I=0$ and $I=2$ contributions but the latter has only 
$I=0$ contribution. Then let us write decay 
amplitudes$\;$(denoted by $A_{I}, I=0,2$ for isospin) as :
{  
\setcounter{enumi}{\value{equation}}
\addtocounter{enumi}{1}
\setcounter{equation}{0}
\renewcommand{\theequation}{\theenumi\alph{equation}}
\begin{eqnarray}
   \hspace{0.5cm}
  A_{0}&=&V^{*}_{us}V_{ud}A_{0T}
         +V^{*}_{us}V_{ud}A^{u}_{0P}
         +V^{*}_{cs}V_{cd}A^{c}_{0P}
         +V^{*}_{ts}V_{td}A^{t}_{0P}
          \nonumber\\ 
       &\propto&
      \lambda\left(e^{-i{\delta}}+\gamma_{9}\lambda^{4}
         e^{-i(\delta-\delta^{'})}\cdot
         e^{i\theta_{9}}\right),
        \hspace{7.1cm}(7.75)\nonumber\\                       \label{7.75}
    A_{2}&=&V^{*}_{us}V_{ud}A_{2T}
          ={\lambda}e^{-i{\delta}}|A_{2T}|,
          \hspace{8cm}(7.76)\nonumber                         \label{7.76}
\end{eqnarray}
\setcounter{equation}{\value{enumi}}}
where $\gamma_{9}e^{i\theta_{9}}\equiv
A_{0P}^{t}/(A_{0T}+A^{u}_{0P}+A_{0P}^{c})$
and for the phase convention $A_{2T}$ is set
real. As is well known$\,$[36][80],
CP violation in K meson system is analysed by 
two parameters, namely, $\varepsilon$ for indirect
CP violation and ${\varepsilon}^{'}$ for direct
CP violation. Concerning $\varepsilon$, its value
has been experimentally confirmed as :
$$\hspace{2.5cm}\varepsilon
    =(2.280\pm0.013)\times{10^{-3}}
     \cdot{e^{i\frac{\pi}{4}}}
     \hspace{6.1cm}\eqno{(7.77)}$$                             \label{7.77}
On the other hand $\:$${\varepsilon}^{'}$ is
described as :
$$   {\varepsilon}^{'}=\frac{i}{\sqrt{2}}
   {\omega}(t_{2}-t_{0})
   e^{i({\theta}_{2}-{\theta}_{0})},
   \hspace{6cm}\eqno{(7.78)}\nonumber$$                        \label{7.78}
 where  $\theta_{I}$ is the phaseshifts of strong 
 interactions ; $t_{I}=$Im$A_{I}/$Re$A_{I}$
 $\;$$(I=0, 2)$ and ${\omega}=e^{i({\theta}_{2}-
 {\theta}_{0})}$$\cdot$(Re$A_{2}/$Re$A_{0}$).
 \par
  In CPT symmetry limit phases of ${\varepsilon}$ 
 and ${\varepsilon}^{'}$ are accidentally almost 
 equal and then usually we discuss 
 by using the equation :
 $$ \rm{Re}\left(\frac{{\varepsilon}^{'}}
    {\varepsilon}\right)\approx\left|
    \frac{{\varepsilon}^{'}}{\varepsilon}\right|.
    \hspace{7.5cm}\eqno{(7.79)}$$                               \label{7.79}
 \par
 From (7.78) with $\;$ ${\varepsilon}$  
 we have :
 {  
 \setcounter{enumi}{\value{equation}}
 \addtocounter{enumi}{1}
 \setcounter{equation}{0}
 \renewcommand{\theequation}{\theenumi\alph{equation}}
 \begin{eqnarray}
   \hspace{2.5cm}
   \left|\frac{{\varepsilon}^{'}}{\varepsilon}
   \right|&=&\frac{|\omega|}{\sqrt{2}|{\varepsilon}|}
  {\kappa},\hspace{8.8cm}(7.80)\nonumber\\                       \label{7.80}
  {\kappa}\hspace{0.15cm}&=&\left|\frac{{\rm{Im}}A_{2}}
  {{\rm{Re}}A_{2}}-\frac{{\rm{Im}}A_{0}}
  {{\rm{Re}}A_{0}}\right|.
  \hspace{7.3cm}(7.81)\nonumber                                  \label{7.81}         
 \end{eqnarray}
 \setcounter{equation}{\value{enumi}}}
 \par
By using (7.75;76) to (7.81) we obtain :
$$ \hspace{2.9cm}{\kappa}\hspace{0.15cm}
          =2\gamma_{9}\lambda^{4}
     |\tan{\delta}|\left|\displaystyle
     {\frac{\sin(\Delta+\theta_{9})}{\sin{\delta}}}
     +\displaystyle{\frac{\cos(\Delta+\theta_{9})}
     {\cos{\delta}}}\right|,
     \hspace{2.6cm}\eqno{(7.82)}\nonumber$$                      \label{7.82}
 where $\Delta\equiv{\delta-\delta^{'}}$ and 
 $(\gamma_{9}\lambda^{4})^{2}$-terms are neglected.
 \par
 From here let us extract numerical informations of 
 above quantities from experimental results. 
 In $K$ meson system it is reported that$\;$[78] :
 $$\hspace{3cm}{\omega}=\frac{1}{29.5},
     \hspace{7cm}\eqno{(7.83)}
     \nonumber$$                                                    \label{7.83}
 which includes the isospin breaking effect 
 and also reported that$\;$[73] :
   $$  \hspace{3cm}
     \rm{Re}\left(\frac{{\varepsilon}^{'}}
     {\varepsilon}\right)
       =
     (16.6\pm{1.6})\times{10}^{-4}
     \hspace{5cm}\eqno{(7.84)}\nonumber$$                         \label{7.84}
 By use of (7.77;78) and (7.82$\;\sim\;$84) 
 we obtain :
         $$\hspace{3cm}{\theta_{9}}
                 =
          (46\pm11)^{\circ},
          \hspace{7cm}\eqno{(7.85)}\nonumber$$                            \label{7.85}        
 where the numerical values of $\delta=21.5^{\circ}\;$; 
$\delta^{'}=72.5^{\circ}\;$; $\lambda=0.22\;$ and 
$\gamma_{9}=0.3$ are used. In (7.75) the phase of
`` $\theta_{9}$ '' is essentially the phase difference    
of strong interaction between external tree and penguin
diagram. It is interesting that :
$\theta_{\rm{ex}}\cong{\theta_{9}}$ from (7.46) and (7.85).     

\section{Conclusions}
\hspace{\parindent}
Construction of our model are carried out rather by 
axiomatic way. First we set three fundamental 
hypotheses : \newline $\;$ \newline
(${\bf 1}$)$\;$`` The space is four 
dimensional : one is time and three are real space. '' 
\par Comment : It is very 
difficult to answer why the space is four dimensional.
So we \par \hspace{2.1cm} are obliged to make this hypothesis.
\newline 
(${\bf 2}$)$\;$`` All gage fields are Cartan 
connections equipped with Soldering Mechanism. ''
\par Comment : With the aid of Soldering Mechanism 
all gauge fields can propagate in \par \hspace{2.1cm}
the real space-time.
\newline
(${\bf 3}$)$\;$`` The primodial matter fields 
({\bf primon} and {\bf anti-primon}) created from 
vaccum \par at the Big Bang of the universe 
are supersymmetric pair of 
${\:}$spin-1/2 fermion and \par spin-0 boson${\:}$, 
both of which have electric charge $|e/6|$. '' \par
Comment : The quantized electric charge of $|e/6|$ is 
explained by that electro mag- \par \hspace{2.1cm}
netic $U(1)$-$\!$gauge group is the unfactorized 
subgroup of $SL(2,C)$ which \par \hspace{2.2cm}have 
6 generators. Therefore `` $e$ '' must be normalized 
by `` 6 ''.
\newline $\;$ \newline
Combining (${\bf 1}$) with (${\bf 2}$), only four types 
of sets of classical groups can be found$\,$, namely : 
$\;F_{1}=SO(1,4)/SO(1,3)\;$(gravity),
$F_{2}=SU(3)/SU(2)\;$(QED),
$\;F_{3}=SL(2,C)/SL(1,C)\;$ (QCD)
and $\;F_{4}=SO(5)/SO(4)\;$ (Weak). It is noticeable 
that $F_{i}\;(i=1\sim4)$ are all $({\rm dimF}=4)$-spaces. 
Soldering Mechanism with 4-dimensional $F_{i}$-spaces 
naturally induces the existence of {\bf six-fields-set} 
:$\; \{$ ${\bf C},\overline{\bf C},
{\bf B}, \overline{\bf B},{\bf G_{1}},{\bf G_{2}}$ $\}$ 
which are massless scalar fields and 
they enable ${\bf g'}$-valued 
gauge fields $\{$gravity, electro magnetic.
strong and weak$\}$ to propagate in the real space-time 
with the aid of the metric-tensers : ${\bf g}^{{\mu}{\nu}}_{a}$ 
($a$ is $gravity$, $electro magnetic$, $strong$ and $weak$).
Our model thinks that the {\bf six-fields-set} not only 
play the role of mathematical tools for 
${\bf BRS}$-$\!${\bf invariance} but also really exist 
at every point in the universe.
In general {\bf massless scalar fields} generate 
{\bf repulsive forces}. Therefore the 
{\bf six-fields-set} generate repulsive forces
at every point of the universe and expand the universe. 
So they might be candidates for {\bf Dark Energy}. 
In our model, four kinds of gauge fields are different 
things in quality and then grand unification does 
not occur even at Plank energy.
\par The hypothesis : (${\bf 3}$) is crucially 
important to build up quarks and leptons.
After the Big Bang, {\bf primons} are composed 
into subquarks,
among which (1,1,1)-state subquarks of $SU(3)_C\otimes
{SU(2)_L^h}\otimes{SU(2)_R^h}$ gauge symmetry are neutral
and they cannot interact with ordinary matters. As it 
could be thought that plenty of (1,1,1)-state 
subquarks have created at the beginning of the universe, 
these neutral subquarks could be the candidate 
for `` {\bf Dark Matter} ''.  
The (3,3,1)-state subquark and the others are composed 
into ordinary matters. In our model weak interacting 
${\bf W^{\pm,}}$-and ${\bf Z^{0}}$-boson are also 
composite objects 
and they  have  scalar partners :$\:$ ${\bf S^{\pm,0}}$. 
Mass difference between them is
supposed to be not so large because its origine is `` 
hyperfine splittig ''. \par
 Our model show that one proton has the configuration
 of $({\bf u}{\bf u}{\bf d}): (2{\bf \alpha}, 
 \overline{\bf \alpha}, 
 3{\bf x}, \overline{\bf x})$; 
 electron: $(\overline{\bf \alpha},2\overline{\bf x})$;
 neutrino: $({\bf \alpha}, \overline{\bf x})$; 
 antineutrino: $(\overline{\bf \alpha}, {\bf x})$ and the dark 
 matters are constructed from the same amount of matters 
 and antimatters because of their neutral charges. 
 Further it is said that  
 the universe contains  almost the same number of protons 
 and electrons. These considerations lead the thought that 
`` {\bf The universe is the matter-antimatter-even object.} '' 
 This idea is different from the current thought that
 the universe is made of matters only. 
 Then in our model the problem 
 about CP violation in the early universe does not occur.
 \par In our model the existence of the 4th generation is, 
in kind, not inhibited because the generation-making 
mechanism is just to add ${\bf y}$-subquarks. 
In fact, if the experimental evidence of    
 $1$-$(|V_{ud}|^{2}+|V_{us}|^{2}+|V_{ub}|^{2})$
 =$0.0017\pm0.0015$ at the $1\sigma$ level[31] is 
taken seriously[30], it cannot be said that 
there is not any possibility of the 4th generation. 
But whether the 4th generation really exists or not may 
depend on the details of the substructure dynamics 
inside quarks, that is, the possibility of the existence 
of the dynamical stable states with the addition of three 
${\bf y}$-subquarks : namely, whether the sum of the 
kinetic energies of the constituent subquarks may balance 
to the binding energy to form the stable states, or not. 
If the non-existence of the 4th generation is finally 
confirmed, that fact will offer one of the clues to solve 
the substructure dynamics. 
\par 
In our model, phenomena of CP violatins and
 Mass differences ($\Delta M_P$)
of $P^{0}$ and $\overline{P^{0}}$ are originated from 
subquark dynamics. Among various subquarks neutral 
${\bf y}$-subquark plays the important role. Namely 
$P^{0}-\overline{P^{0}}$ mixings occur by 
${\bf y}$-subquark exchange between constituent quarks 
in $P^{0}$ or $\overline{P^{0}}$, which generate 
indirect CP violations and $\Delta M_P$. 
Further $({\bf y} \longrightarrow 2{\bf g}_h)$-processes
give complex phases to vertex parts of the flavour mixings,
which generate direct and mixing-induced CP violations. 
There are two parameters of phases : $\delta$ and
$\delta^{'}$, the values of which are evaluated by the
experimental data of Belle and BaBar. Using them some  
predictions are examined by the various data and the 
results are roughly good.\par Concerning 
$\Delta M_{D^{0}}$, our model predicts the value of 
$\rm{O}(10^{-14})$. We hope further accumulation of 
the data by the CLEO detector.
 \par
 To conclude, we have discussed the possibility that
 the subquark dynamics play the essential 
 role in all flavor-changing phenomena.
 \par
{\bf Acknowledgements} \par
We would like to thank A. Takamura for valuable discussions. 
We would also like to thank for the hospitality at Toyota
College of Technology.


\clearpage
  

\setlength{\unitlength}{0.7mm}
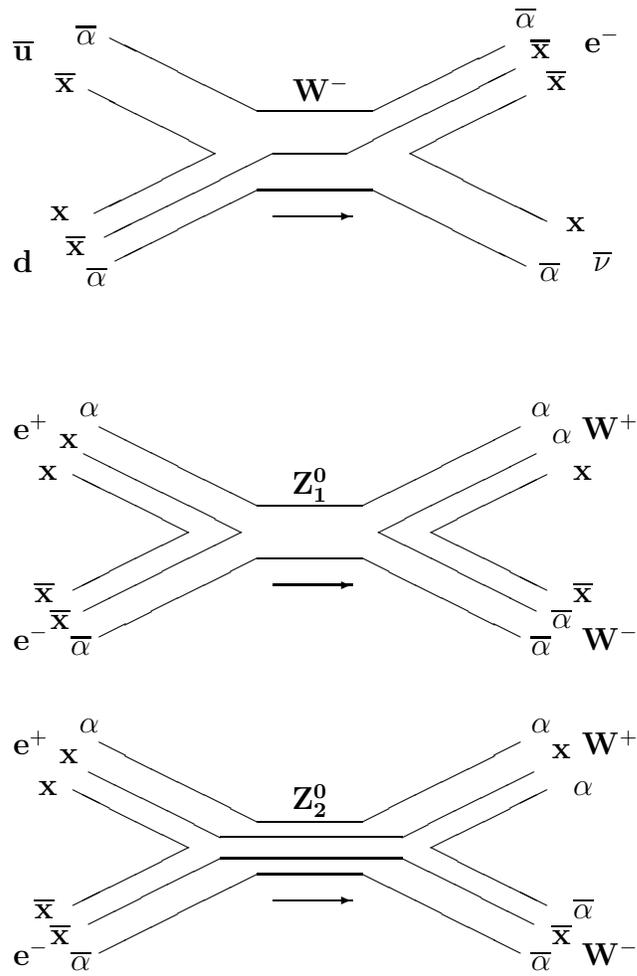
\begin{figure}
\begin{center}
\begin{picture}(180,270)(15,-20)
\put(47,216){\makebox(7,7){${\bf \overline{\alpha}}$}}
\put(43,207){\makebox(7,7){${\bf \overline{x}}$}}
\put(35,213){\makebox(7,7){${\bf \overline u}$}}
\put(49,170){\makebox(7,7){${\bf \overline{\alpha}}$}}
\put(45,176){\makebox(7,7){${\bf \overline x}$}}
\put(42,182){\makebox(7,7){${\bf x}$}}
\put(35,173){\makebox(7,7){${\bf d}$}}
\put(132,221){\makebox(7,7)[bl]{${\bf \overline{\alpha}}$}}
\put(135,215){\makebox(7,7)[bl]{${\bf \overline x}$}}
\put(138,209){\makebox(7,7)[bl]{${\bf \overline x}$}}
\put(145,215){\makebox(7,7){${\bf e^-}$}}
\put(140,180){\makebox(7,7){${\bf x}$}}
\put(135,171){\makebox(7,7){${\bf \overline{\alpha}}$}}
\put(145,173){\makebox(7,7){${\bf \overline{\nu}}$}}
\put(90,207){\makebox(10,10)[b]{${\bf W^-}$}}
\put(83,205){\line(-2,1){28}}
\put(75,197){\line(-2,1){24}}
\put(75,197){\line(-2,-1){23}}
\put(86,197){\line(-2,-1){32}}
\put(83,190){\line(-2,-1){27}}
\put(86,185){\vector(1,0){15}}
\put(83,205){\line(1,0){22}}
\put(86,197){\line(1,0){14}}
\put(83,190){\line(1,0){22}}
\put(105,205){\line(2,1){25}}
\put(100,197){\line(2,1){32}}
\put(112,197){\line(2,1){22}}
\put(112,197){\line(2,-1){26}}
\put(105,190){\line(2,-1){29}}
\put(46,147){\makebox(7,7)[br]{${\bf {\alpha}}$}}
\put(42,141){\makebox(7,7)[br]{${\bf x}$}}
\put(38,135){\makebox(7,7)[br]{${\bf x}$}}
\put(39,110){\makebox(7,7){${\bf \overline{x}}$}}
\put(42,105){\makebox(7,7){${\bf \overline x}$}}
\put(46,100){\makebox(7,7){${\bf \overline{\alpha}}$}}
\put(135,147){\makebox(7,7)[bl]{${\bf \alpha}$}}
\put(139,142){\makebox(7,7)[bl]{${\bf \alpha}$}}
\put(143,135){\makebox(7,7)[bl]{${\bf x}$}}
\put(143,110){\makebox(7,7)[l]{${\bf \overline{x}}$}}
\put(139,105){\makebox(7,7)[l]{${\bf \overline{\alpha}}$}}
\put(135,100){\makebox(7,7)[l]{${\bf \overline{\alpha}}$}}
\put(35,140){\makebox(10,10){${\bf e^+}$}}
\put(35,100){\makebox(10,10){${\bf e^-}$}}
\put(145,140){\makebox(10,10){${\bf W^+}$}}
\put(145,100){\makebox(10,10){${\bf W^-}$}}
\put(88,129){\makebox(10,10){${\bf Z_1^0}$}}
\put(83,130){\line(-2,1){30}}
\put(80,125){\line(-2,1){30}}
\put(70,125){\line(-2,1){22}}
\put(70,125){\line(-2,-1){22}}
\put(80,125){\line(-2,-1){30}}
\put(86,115){\vector(1,0){15}}
\put(83,120){\line(-2,-1){30}}
\put(83,130){\line(1,0){20}}
\put(83,120){\line(1,0){20}}
\put(103,130){\line(2,1){30}}
\put(106,125){\line(2,1){30}}
\put(116,125){\line(2,1){22}}
\put(116,125){\line(2,-1){22}}
\put(106,125){\line(2,-1){30}}
\put(103,120){\line(2,-1){30}}
\put(46,87){\makebox(7,7)[br]{${\bf {\alpha}}$}}
\put(42,81){\makebox(7,7)[br]{${\bf x}$}}
\put(38,75){\makebox(7,7)[br]{${\bf x}$}}
\put(39,50){\makebox(7,7){${\bf \overline{x}}$}}
\put(42,45){\makebox(7,7){${\bf \overline x}$}}
\put(46,40){\makebox(7,7){${\bf \overline{\alpha}}$}}
\put(135,87){\makebox(7,7)[bl]{${\bf \alpha}$}}
\put(139,82){\makebox(7,7)[bl]{${\bf x}$}}
\put(143,75){\makebox(7,7)[bl]{${\bf \alpha}$}}
\put(143,50){\makebox(7,7)[l]{${\bf \overline{\alpha}}$}}
\put(139,45){\makebox(7,7)[l]{${\bf \overline{x}}$}}
\put(135,40){\makebox(7,7)[l]{${\bf \overline{\alpha}}$}}
\put(35,80){\makebox(10,10){${\bf e^+}$}}
\put(35,40){\makebox(10,10){${\bf e^-}$}}
\put(145,80){\makebox(10,10){${\bf W^+}$}}
\put(145,40){\makebox(10,10){${\bf W^-}$}}
\put(88,69){\makebox(10,10){${\bf Z_2^0}$}}
\put(83,70){\line(-2,1){30}}
\put(76,67){\line(-2,1){25}}
\put(70,65){\line(-2,1){22}}
\put(70,65){\line(-2,-1){22}}
\put(76,63){\line(-2,-1){25}}
\put(86,55){\vector(1,0){15}}
\put(83,60){\line(-2,-1){30}}
\put(83,70){\line(1,0){20}}
\put(76,67){\line(1,0){34.5}}
\put(76,63){\line(1,0){34.5}}
\put(83,60){\line(1,0){20}}
\put(103,70){\line(2,1){30}}
\put(110.5,67){\line(2,1){25}}
\put(116,65){\line(2,1){22}}
\put(116,65){\line(2,-1){22}}
\put(110.5,63){\line(2,-1){25}}
\put(103,60){\line(2,-1){30}}
\end{picture}
\end{center}
\caption{Subquark-line diagrams of the weak interactions}
\end{figure}

\begin{figure}
\setlength{\unitlength}{0.7mm}
\begin{center}
\begin{picture}(180,250)(15,-20)
\put(0,115){\makebox(80,80)[l]{\Huge ${\bf y}$}}
\put(5,165){\makebox(30,30)[r]{${\Lambda}
             \hspace{1mm}({\Theta})$}}
\put(5,115){\makebox(30,30)[r]{${\overline{\Lambda}
             \hspace{1mm}(\overline{\Theta})}$}}
\put(40,180){\line(1,0){60}}
\put(40,130){\line(1,0){60}}
\put(100,130){\line(0,1){50}}
\multiput(100,180)(4,0){10}{\line(1,0){3}}
\multiput(100,130)(4,0){10}{\line(1,0){3}}
\put(120,165){\makebox(30,30)[r]{${\bf g}_h$}}
\put(120,115){\makebox(30,30)[r]{${\bf g}_h$}}
\end{picture}
\end{center}
\caption{The (${\bf y} \longrightarrow 2{\bf g}_h)$-process by 
 primon-level diagram}
\end{figure}
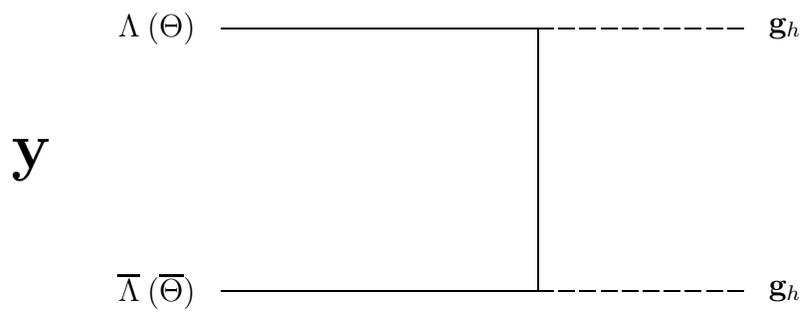

\setlength{\unitlength}{0.8mm}
\begin{figure}
\begin{center}
\begin{picture}(180,250)(15,-20)

\put(140,185){\makebox(30,30)[l]{(A)}}
\put(0,185){\makebox(30,30)[r]{${V}
             \hspace{1mm}({Ps})$}}
\put(60,220){\line(1,0){60}}
\multiput(90,185)(0,4){9}{\line(0,1){3}}
\put(66,225){\makebox(50,10)[b]{
            $+{g\displaystyle{\frac{\lambda_{ik}^{a}}{2}}}$}}
\put(50,216){\makebox(7,7)[r]{${\bf q}_{i}$}}
\put(123,216){\makebox(7,7)[l]{${\bf q}_{k}$}}
\put(50,181){\makebox(7,7)[r]{${\bf \overline{q}}_{j}$}}
\put(60,185){\line(1,0){60}}
\put(123,181){\makebox(7,7)[l]{${\bf \overline{q}}_{l}$}}
\put(66,176){\makebox(50,10)[b]{
            ${-g\displaystyle{\frac{\lambda_{jl}^{b}}{2}}}$}}
   
\put(140,95){\makebox(30,30)[l]{(B)}}
\put(0,95){\makebox(30,30)[r]{${Z^0}
             \hspace{1mm}({S^0})$}}
\put(60,130){\line(1,0){60}}
\multiput(90,95)(0,4){9}{\line(0,1){3}}
\put(66,135){\makebox(50,10)[b]{
            $+{g_{h}\displaystyle{\frac{\tau_{ik}^{a}}{2}}}$}}
\put(50,126){\makebox(7,7)[r]{${\bf \alpha}_{i}$}}
\put(123,126){\makebox(7,7)[l]{${\bf \alpha}_{k}$}}
\put(50,91){\makebox(7,7)[r]{${\bf \overline{\alpha}}_{j}$}}
\put(60,95){\line(1,0){60}}
\put(123,91){\makebox(7,7)[l]{${\bf \overline{\alpha}}_{l}$}}
\put(66,86){\makebox(50,10)[b]{
            $+{g_{h}\displaystyle{\frac{\tau_{jl}^{b}}{2}}}$}}
             
\end{picture}
\end{center}
\caption{
   (A) Gluon exchange in ${\bf q}\overline{\bf q}$ system;
    (B) Hypergluon exchange in
      ${\bf \alpha}\overline{\bf \alpha}$ system.
       }
\end{figure}

\setlength{\unitlength}{0.8mm}
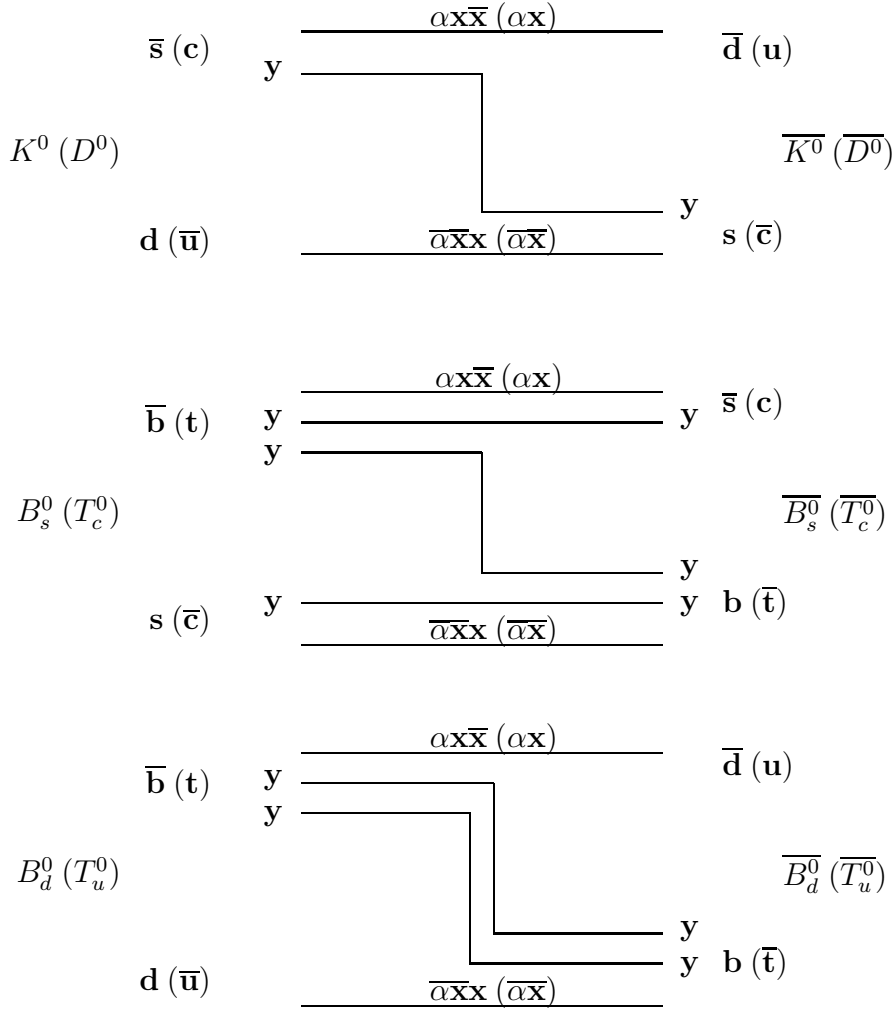
\begin{figure}
\begin{center}
\begin{picture}(180,250)(15,-20)
\put(140,155){\makebox(30,30)[l]{${\overline{K^0}}
             \hspace{1mm}({\overline{D^0}})$}}
\put(0,155){\makebox(30,30)[r]{${K^0}
             \hspace{1mm}({D^0})$}}
\put(30,182){\makebox(15,10)[r]{${\bf {\overline{s}}
             \hspace{1mm}(c)}$}}
\put(60,190){\line(1,0){60}}
\put(67,191){\makebox(50,10)[b]{${\bf {\alpha x \overline{x}
                                 \hspace{1mm}(\alpha x)}}$}}
\put(130,182){\makebox(15,10)[l]{${\bf {\overline {d}}
             \hspace{1mm}(u)}$}}
\put(50,180){\makebox(7,7)[r]{${\bf y}$}}
\put(60,183){\line(1,0){30}}
\put(90,160){\line(0,1){23}}
\put(90,160){\line(1,0){30}}
\put(123,157){\makebox(7,7)[l]{${\bf y}$}}
\put(30,150){\makebox(15,10)[r]{${\bf {d\hspace{1mm}
             (\overline{u})}}$}}
\put(60,153){\line(1,0){60}}
\put(130,151){\makebox(15,10)[l]{${\bf {s\hspace{1mm}
              (\overline{c})}}$}}
\put(67,154){\makebox(50,10)[b]{${\bf {\overline{\alpha} 
    \overline{x} x \hspace{1mm}(\overline{\alpha} 
    \overline{x})}}$}}
\put(140,95){\makebox(30,30)[l]{${\overline{B_s^0}}
             \hspace{1mm}({\overline{T_c^0}})$}}
\put(0,95){\makebox(30,30)[r]{${B_s^0}
             \hspace{1mm}({T_c^0})$}}
\put(30,120){\makebox(15,10)[r]{${\bf {\overline{b}}
             \hspace{1mm}(t)}$}}
\put(60,130){\line(1,0){60}}
\put(67,131){\makebox(50,10)[b]{
            ${\bf {\alpha x \overline{x}
                                \hspace{1mm}(\alpha x)}}$}}
\put(130,123){\makebox(15,10)[l]{${\bf {\overline {s}}
             \hspace{1mm}(c)}$}}
\put(50,122){\makebox(7,7)[r]{${\bf y}$}}
\put(60,125){\line(1,0){60}}
\put(123,122){\makebox(7,7)[l]{${\bf y}$}}
\put(50,116){\makebox(7,7)[r]{${\bf y}$}}
\put(60,120){\line(1,0){30}}
\put(90,100){\line(0,1){20}}
\put(90,100){\line(1,0){30}}
\put(123,97){\makebox(7,7)[l]{${\bf y}$}}
\put(50,91){\makebox(7,7)[r]{${\bf y}$}}
\put(60,95){\line(1,0){60}}
\put(123,91){\makebox(7,7)[l]{${\bf y}$}}
\put(30,87){\makebox(15,10)[r]{${\bf {s
           \hspace{1mm}(\overline{c})}}$}}
\put(60,88){\line(1,0){60}}
\put(67,89){\makebox(50,10)[b]{${\bf {\overline{\alpha} 
    \overline{x} x \hspace{1mm}(\overline{\alpha} 
    \overline{x})}}$}}
\put(130,90){\makebox(15,10)[l]{${\bf {b\hspace{1mm}
             (\overline{t})}}$}}
\put(140,35){\makebox(30,30)[l]{${\overline{B_d^0}}
             \hspace{1mm}({\overline{T_u^0}})$}}
\put(0,35){\makebox(30,30)[r]{${B_d^0}
             \hspace{1mm}({T_u^0})$}}
\put(30,60){\makebox(15,10)[r]{${\bf {\overline{b}}
             \hspace{1mm}(t)}$}}
\put(60,70){\line(1,0){60}}
\put(67,71){\makebox(50,10)[b]{${\bf {\alpha x \overline{x}
                                \hspace{1mm}(\alpha x)}}$}}
\put(130,63){\makebox(15,10)[l]{${\bf {\overline {d}}
             \hspace{1mm}(u)}$}}
\put(50,62){\makebox(7,7)[r]{${\bf y}$}}
\put(60,65){\line(1,0){32}}
\put(92,40){\line(0,1){25}}
\put(50,56){\makebox(7,7)[r]{${\bf y}$}}
\put(60,60){\line(1,0){28}}
\put(88,35){\line(0,1){25}}
\put(92,40){\line(1,0){28}}
\put(123,37){\makebox(7,7)[l]{${\bf y}$}}
\put(88,35){\line(1,0){32}}
\put(123,31){\makebox(7,7)[l]{${\bf y}$}}
\put(30,27){\makebox(15,10)[r]{${\bf {d
           \hspace{1mm}(\overline{u})}}$}}
\put(60,28){\line(1,0){60}}
\put(67,29){\makebox(50,10)[b]{${\bf {\overline{\alpha} 
           \overline{x} x \hspace{1mm}
           (\overline{\alpha} \overline{x})}}$}}
\put(130,30){\makebox(15,10)[l]{${\bf {b\hspace{1mm}
             (\overline{t})}}$}}
\end{picture}
\end{center}
\caption{  Schematic illustrations of $P^0$-$\overline{P^0}$ mixings
  by ${\bf y}$-exchange interactions}
\end{figure}

\end{document}